\documentclass[useAMS,usenatbib,twocolumn]{mn2e}
\usepackage{natbib,graphics,epsfig}
\usepackage{amsmath}
\usepackage{amssymb}
\usepackage{textcomp}
\usepackage{graphicx}
\usepackage{color}
\usepackage[draft]{hyperref}
\usepackage{txfonts}
\usepackage{float,subfig}
  \definecolor{snow}{rgb}{1.000000,0.980392,0.980392}
  \definecolor{ghost white}{rgb}{0.972549,0.972549,1.000000}
  \definecolor{GhostWhite}{rgb}{0.972549,0.972549,1.000000}
  \definecolor{white smoke}{rgb}{0.960784,0.960784,0.960784}
  \definecolor{WhiteSmoke}{rgb}{0.960784,0.960784,0.960784}
  \definecolor{gainsboro}{rgb}{0.862745,0.862745,0.862745}
  \definecolor{floral white}{rgb}{1.000000,0.980392,0.941176}
  \definecolor{FloralWhite}{rgb}{1.000000,0.980392,0.941176}
  \definecolor{old lace}{rgb}{0.992157,0.960784,0.901961}
  \definecolor{OldLace}{rgb}{0.992157,0.960784,0.901961}
  \definecolor{linen}{rgb}{0.980392,0.941176,0.901961}
  \definecolor{antique white}{rgb}{0.980392,0.921569,0.843137}
  \definecolor{AntiqueWhite}{rgb}{0.980392,0.921569,0.843137}
  \definecolor{papaya whip}{rgb}{1.000000,0.937255,0.835294}
  \definecolor{PapayaWhip}{rgb}{1.000000,0.937255,0.835294}
  \definecolor{blanched almond}{rgb}{1.000000,0.921569,0.803922}
  \definecolor{BlanchedAlmond}{rgb}{1.000000,0.921569,0.803922}
  \definecolor{bisque}{rgb}{1.000000,0.894118,0.768627}
  \definecolor{peach puff}{rgb}{1.000000,0.854902,0.725490}
  \definecolor{PeachPuff}{rgb}{1.000000,0.854902,0.725490}
  \definecolor{navajo white}{rgb}{1.000000,0.870588,0.678431}
  \definecolor{NavajoWhite}{rgb}{1.000000,0.870588,0.678431}
  \definecolor{moccasin}{rgb}{1.000000,0.894118,0.709804}
  \definecolor{cornsilk}{rgb}{1.000000,0.972549,0.862745}
  \definecolor{ivory}{rgb}{1.000000,1.000000,0.941176}
  \definecolor{lemon chiffon}{rgb}{1.000000,0.980392,0.803922}
  \definecolor{LemonChiffon}{rgb}{1.000000,0.980392,0.803922}
  \definecolor{seashell}{rgb}{1.000000,0.960784,0.933333}
  \definecolor{honeydew}{rgb}{0.941176,1.000000,0.941176}
  \definecolor{mint cream}{rgb}{0.960784,1.000000,0.980392}
  \definecolor{MintCream}{rgb}{0.960784,1.000000,0.980392}
  \definecolor{azure}{rgb}{0.941176,1.000000,1.000000}
  \definecolor{alice blue}{rgb}{0.941176,0.972549,1.000000}
  \definecolor{AliceBlue}{rgb}{0.941176,0.972549,1.000000}
  \definecolor{lavender}{rgb}{0.901961,0.901961,0.980392}
  \definecolor{lavender blush}{rgb}{1.000000,0.941176,0.960784}
  \definecolor{LavenderBlush}{rgb}{1.000000,0.941176,0.960784}
  \definecolor{misty rose}{rgb}{1.000000,0.894118,0.882353}
  \definecolor{MistyRose}{rgb}{1.000000,0.894118,0.882353}
  \definecolor{white}{rgb}{1.000000,1.000000,1.000000}
  \definecolor{black}{rgb}{0.000000,0.000000,0.000000}
  \definecolor{dark slate gray}{rgb}{0.184314,0.309804,0.309804}
  \definecolor{DarkSlateGray}{rgb}{0.184314,0.309804,0.309804}
  \definecolor{dark slate grey}{rgb}{0.184314,0.309804,0.309804}
  \definecolor{DarkSlateGrey}{rgb}{0.184314,0.309804,0.309804}
  \definecolor{dim gray}{rgb}{0.411765,0.411765,0.411765}
  \definecolor{DimGray}{rgb}{0.411765,0.411765,0.411765}
  \definecolor{dim grey}{rgb}{0.411765,0.411765,0.411765}
  \definecolor{DimGrey}{rgb}{0.411765,0.411765,0.411765}
  \definecolor{slate gray}{rgb}{0.439216,0.501961,0.564706}
  \definecolor{SlateGray}{rgb}{0.439216,0.501961,0.564706}
  \definecolor{slate grey}{rgb}{0.439216,0.501961,0.564706}
  \definecolor{SlateGrey}{rgb}{0.439216,0.501961,0.564706}
  \definecolor{light slate gray}{rgb}{0.466667,0.533333,0.600000}
  \definecolor{LightSlateGray}{rgb}{0.466667,0.533333,0.600000}
  \definecolor{light slate grey}{rgb}{0.466667,0.533333,0.600000}
  \definecolor{LightSlateGrey}{rgb}{0.466667,0.533333,0.600000}
  \definecolor{gray}{rgb}{0.745098,0.745098,0.745098}
  \definecolor{grey}{rgb}{0.745098,0.745098,0.745098}
  \definecolor{light grey}{rgb}{0.827451,0.827451,0.827451}
  \definecolor{LightGrey}{rgb}{0.827451,0.827451,0.827451}
  \definecolor{light gray}{rgb}{0.827451,0.827451,0.827451}
  \definecolor{LightGray}{rgb}{0.827451,0.827451,0.827451}
  \definecolor{midnight blue}{rgb}{0.098039,0.098039,0.439216}
  \definecolor{MidnightBlue}{rgb}{0.098039,0.098039,0.439216}
  \definecolor{navy}{rgb}{0.000000,0.000000,0.501961}
  \definecolor{navy blue}{rgb}{0.000000,0.000000,0.501961}
  \definecolor{NavyBlue}{rgb}{0.000000,0.000000,0.501961}
  \definecolor{cornflower blue}{rgb}{0.392157,0.584314,0.929412}
  \definecolor{CornflowerBlue}{rgb}{0.392157,0.584314,0.929412}
  \definecolor{dark slate blue}{rgb}{0.282353,0.239216,0.545098}
  \definecolor{DarkSlateBlue}{rgb}{0.282353,0.239216,0.545098}
  \definecolor{slate blue}{rgb}{0.415686,0.352941,0.803922}
  \definecolor{SlateBlue}{rgb}{0.415686,0.352941,0.803922}
  \definecolor{medium slate blue}{rgb}{0.482353,0.407843,0.933333}
  \definecolor{MediumSlateBlue}{rgb}{0.482353,0.407843,0.933333}
  \definecolor{light slate blue}{rgb}{0.517647,0.439216,1.000000}
  \definecolor{LightSlateBlue}{rgb}{0.517647,0.439216,1.000000}
  \definecolor{medium blue}{rgb}{0.000000,0.000000,0.803922}
  \definecolor{MediumBlue}{rgb}{0.000000,0.000000,0.803922}
  \definecolor{royal blue}{rgb}{0.254902,0.411765,0.882353}
  \definecolor{RoyalBlue}{rgb}{0.254902,0.411765,0.882353}
  \definecolor{blue}{rgb}{0.000000,0.000000,1.000000}
  \definecolor{dodger blue}{rgb}{0.117647,0.564706,1.000000}
  \definecolor{DodgerBlue}{rgb}{0.117647,0.564706,1.000000}
  \definecolor{deep sky blue}{rgb}{0.000000,0.749020,1.000000}
  \definecolor{DeepSkyBlue}{rgb}{0.000000,0.749020,1.000000}
  \definecolor{sky blue}{rgb}{0.529412,0.807843,0.921569}
  \definecolor{SkyBlue}{rgb}{0.529412,0.807843,0.921569}
  \definecolor{light sky blue}{rgb}{0.529412,0.807843,0.980392}
  \definecolor{LightSkyBlue}{rgb}{0.529412,0.807843,0.980392}
  \definecolor{steel blue}{rgb}{0.274510,0.509804,0.705882}
  \definecolor{SteelBlue}{rgb}{0.274510,0.509804,0.705882}
  \definecolor{light steel blue}{rgb}{0.690196,0.768627,0.870588}
  \definecolor{LightSteelBlue}{rgb}{0.690196,0.768627,0.870588}
  \definecolor{light blue}{rgb}{0.678431,0.847059,0.901961}
  \definecolor{LightBlue}{rgb}{0.678431,0.847059,0.901961}
  \definecolor{powder blue}{rgb}{0.690196,0.878431,0.901961}
  \definecolor{PowderBlue}{rgb}{0.690196,0.878431,0.901961}
  \definecolor{pale turquoise}{rgb}{0.686275,0.933333,0.933333}
  \definecolor{PaleTurquoise}{rgb}{0.686275,0.933333,0.933333}
  \definecolor{dark turquoise}{rgb}{0.000000,0.807843,0.819608}
  \definecolor{DarkTurquoise}{rgb}{0.000000,0.807843,0.819608}
  \definecolor{medium turquoise}{rgb}{0.282353,0.819608,0.800000}
  \definecolor{MediumTurquoise}{rgb}{0.282353,0.819608,0.800000}
  \definecolor{turquoise}{rgb}{0.250980,0.878431,0.815686}
  \definecolor{cyan}{rgb}{0.000000,1.000000,1.000000}
  \definecolor{light cyan}{rgb}{0.878431,1.000000,1.000000}
  \definecolor{LightCyan}{rgb}{0.878431,1.000000,1.000000}
  \definecolor{cadet blue}{rgb}{0.372549,0.619608,0.627451}
  \definecolor{CadetBlue}{rgb}{0.372549,0.619608,0.627451}
  \definecolor{medium aquamarine}{rgb}{0.400000,0.803922,0.666667}
  \definecolor{MediumAquamarine}{rgb}{0.400000,0.803922,0.666667}
  \definecolor{aquamarine}{rgb}{0.498039,1.000000,0.831373}
  \definecolor{dark green}{rgb}{0.000000,0.392157,0.000000}
  \definecolor{DarkGreen}{rgb}{0.000000,0.392157,0.000000}
  \definecolor{dark olive green}{rgb}{0.333333,0.419608,0.184314}
  \definecolor{DarkOliveGreen}{rgb}{0.333333,0.419608,0.184314}
  \definecolor{dark sea green}{rgb}{0.560784,0.737255,0.560784}
  \definecolor{DarkSeaGreen}{rgb}{0.560784,0.737255,0.560784}
  \definecolor{sea green}{rgb}{0.180392,0.545098,0.341176}
  \definecolor{SeaGreen}{rgb}{0.180392,0.545098,0.341176}
  \definecolor{medium sea green}{rgb}{0.235294,0.701961,0.443137}
  \definecolor{MediumSeaGreen}{rgb}{0.235294,0.701961,0.443137}
  \definecolor{light sea green}{rgb}{0.125490,0.698039,0.666667}
  \definecolor{LightSeaGreen}{rgb}{0.125490,0.698039,0.666667}
  \definecolor{pale green}{rgb}{0.596078,0.984314,0.596078}
  \definecolor{PaleGreen}{rgb}{0.596078,0.984314,0.596078}
  \definecolor{spring green}{rgb}{0.000000,1.000000,0.498039}
  \definecolor{SpringGreen}{rgb}{0.000000,1.000000,0.498039}
  \definecolor{lawn green}{rgb}{0.486275,0.988235,0.000000}
  \definecolor{LawnGreen}{rgb}{0.486275,0.988235,0.000000}
  \definecolor{green}{rgb}{0.000000,1.000000,0.000000}
  \definecolor{chartreuse}{rgb}{0.498039,1.000000,0.000000}
  \definecolor{medium spring green}{rgb}{0.000000,0.980392,0.603922}
  \definecolor{MediumSpringGreen}{rgb}{0.000000,0.980392,0.603922}
  \definecolor{green yellow}{rgb}{0.678431,1.000000,0.184314}
  \definecolor{GreenYellow}{rgb}{0.678431,1.000000,0.184314}
  \definecolor{lime green}{rgb}{0.196078,0.803922,0.196078}
  \definecolor{LimeGreen}{rgb}{0.196078,0.803922,0.196078}
  \definecolor{yellow green}{rgb}{0.603922,0.803922,0.196078}
  \definecolor{YellowGreen}{rgb}{0.603922,0.803922,0.196078}
  \definecolor{forest green}{rgb}{0.133333,0.545098,0.133333}
  \definecolor{ForestGreen}{rgb}{0.133333,0.545098,0.133333}
  \definecolor{olive drab}{rgb}{0.419608,0.556863,0.137255}
  \definecolor{OliveDrab}{rgb}{0.419608,0.556863,0.137255}
  \definecolor{dark khaki}{rgb}{0.741176,0.717647,0.419608}
  \definecolor{DarkKhaki}{rgb}{0.741176,0.717647,0.419608}
  \definecolor{khaki}{rgb}{0.941176,0.901961,0.549020}
  \definecolor{pale goldenrod}{rgb}{0.933333,0.909804,0.666667}
  \definecolor{PaleGoldenrod}{rgb}{0.933333,0.909804,0.666667}
  \definecolor{light goldenrod yellow}{rgb}{0.980392,0.980392,0.823529}
  \definecolor{LightGoldenrodYellow}{rgb}{0.980392,0.980392,0.823529}
  \definecolor{light yellow}{rgb}{1.000000,1.000000,0.878431}
  \definecolor{LightYellow}{rgb}{1.000000,1.000000,0.878431}
  \definecolor{yellow}{rgb}{1.000000,1.000000,0.000000}
  \definecolor{gold}{rgb}{1.000000,0.843137,0.000000}
  \definecolor{light goldenrod}{rgb}{0.933333,0.866667,0.509804}
  \definecolor{LightGoldenrod}{rgb}{0.933333,0.866667,0.509804}
  \definecolor{goldenrod}{rgb}{0.854902,0.647059,0.125490}
  \definecolor{dark goldenrod}{rgb}{0.721569,0.525490,0.043137}
  \definecolor{DarkGoldenrod}{rgb}{0.721569,0.525490,0.043137}
  \definecolor{rosy brown}{rgb}{0.737255,0.560784,0.560784}
  \definecolor{RosyBrown}{rgb}{0.737255,0.560784,0.560784}
  \definecolor{indian red}{rgb}{0.803922,0.360784,0.360784}
  \definecolor{IndianRed}{rgb}{0.803922,0.360784,0.360784}
  \definecolor{saddle brown}{rgb}{0.545098,0.270588,0.074510}
  \definecolor{SaddleBrown}{rgb}{0.545098,0.270588,0.074510}
  \definecolor{sienna}{rgb}{0.627451,0.321569,0.176471}
  \definecolor{peru}{rgb}{0.803922,0.521569,0.247059}
  \definecolor{burlywood}{rgb}{0.870588,0.721569,0.529412}
  \definecolor{beige}{rgb}{0.960784,0.960784,0.862745}
  \definecolor{wheat}{rgb}{0.960784,0.870588,0.701961}
  \definecolor{sandy brown}{rgb}{0.956863,0.643137,0.376471}
  \definecolor{SandyBrown}{rgb}{0.956863,0.643137,0.376471}
  \definecolor{tan}{rgb}{0.823529,0.705882,0.549020}
  \definecolor{chocolate}{rgb}{0.823529,0.411765,0.117647}
  \definecolor{firebrick}{rgb}{0.698039,0.133333,0.133333}
  \definecolor{brown}{rgb}{0.647059,0.164706,0.164706}
  \definecolor{dark salmon}{rgb}{0.913725,0.588235,0.478431}
  \definecolor{DarkSalmon}{rgb}{0.913725,0.588235,0.478431}
  \definecolor{salmon}{rgb}{0.980392,0.501961,0.447059}
  \definecolor{light salmon}{rgb}{1.000000,0.627451,0.478431}
  \definecolor{LightSalmon}{rgb}{1.000000,0.627451,0.478431}
  \definecolor{orange}{rgb}{1.000000,0.647059,0.000000}
  \definecolor{dark orange}{rgb}{1.000000,0.549020,0.000000}
  \definecolor{DarkOrange}{rgb}{1.000000,0.549020,0.000000}
  \definecolor{coral}{rgb}{1.000000,0.498039,0.313726}
  \definecolor{light coral}{rgb}{0.941176,0.501961,0.501961}
  \definecolor{LightCoral}{rgb}{0.941176,0.501961,0.501961}
  \definecolor{tomato}{rgb}{1.000000,0.388235,0.278431}
  \definecolor{orange red}{rgb}{1.000000,0.270588,0.000000}
  \definecolor{OrangeRed}{rgb}{1.000000,0.270588,0.000000}
  \definecolor{red}{rgb}{1.000000,0.000000,0.000000}
  \definecolor{hot pink}{rgb}{1.000000,0.411765,0.705882}
  \definecolor{HotPink}{rgb}{1.000000,0.411765,0.705882}
  \definecolor{deep pink}{rgb}{1.000000,0.078431,0.576471}
  \definecolor{DeepPink}{rgb}{1.000000,0.078431,0.576471}
  \definecolor{pink}{rgb}{1.000000,0.752941,0.796078}
  \definecolor{light pink}{rgb}{1.000000,0.713726,0.756863}
  \definecolor{LightPink}{rgb}{1.000000,0.713726,0.756863}
  \definecolor{pale violet red}{rgb}{0.858824,0.439216,0.576471}
  \definecolor{PaleVioletRed}{rgb}{0.858824,0.439216,0.576471}
  \definecolor{maroon}{rgb}{0.690196,0.188235,0.376471}
  \definecolor{medium violet red}{rgb}{0.780392,0.082353,0.521569}
  \definecolor{MediumVioletRed}{rgb}{0.780392,0.082353,0.521569}
  \definecolor{violet red}{rgb}{0.815686,0.125490,0.564706}
  \definecolor{VioletRed}{rgb}{0.815686,0.125490,0.564706}
  \definecolor{magenta}{rgb}{1.000000,0.000000,1.000000}
  \definecolor{violet}{rgb}{0.933333,0.509804,0.933333}
  \definecolor{plum}{rgb}{0.866667,0.627451,0.866667}
  \definecolor{orchid}{rgb}{0.854902,0.439216,0.839216}
  \definecolor{medium orchid}{rgb}{0.729412,0.333333,0.827451}
  \definecolor{MediumOrchid}{rgb}{0.729412,0.333333,0.827451}
  \definecolor{dark orchid}{rgb}{0.600000,0.196078,0.800000}
  \definecolor{DarkOrchid}{rgb}{0.600000,0.196078,0.800000}
  \definecolor{dark violet}{rgb}{0.580392,0.000000,0.827451}
  \definecolor{DarkViolet}{rgb}{0.580392,0.000000,0.827451}
  \definecolor{blue violet}{rgb}{0.541176,0.168627,0.886275}
  \definecolor{BlueViolet}{rgb}{0.541176,0.168627,0.886275}
  \definecolor{purple}{rgb}{0.627451,0.125490,0.941176}
  \definecolor{medium purple}{rgb}{0.576471,0.439216,0.858824}
  \definecolor{MediumPurple}{rgb}{0.576471,0.439216,0.858824}
  \definecolor{thistle}{rgb}{0.847059,0.749020,0.847059}
  \definecolor{snow1}{rgb}{1.000000,0.980392,0.980392}
  \definecolor{snow2}{rgb}{0.933333,0.913725,0.913725}
  \definecolor{snow3}{rgb}{0.803922,0.788235,0.788235}
  \definecolor{snow4}{rgb}{0.545098,0.537255,0.537255}
  \definecolor{seashell1}{rgb}{1.000000,0.960784,0.933333}
  \definecolor{seashell2}{rgb}{0.933333,0.898039,0.870588}
  \definecolor{seashell3}{rgb}{0.803922,0.772549,0.749020}
  \definecolor{seashell4}{rgb}{0.545098,0.525490,0.509804}
  \definecolor{AntiqueWhite1}{rgb}{1.000000,0.937255,0.858824}
  \definecolor{AntiqueWhite2}{rgb}{0.933333,0.874510,0.800000}
  \definecolor{AntiqueWhite3}{rgb}{0.803922,0.752941,0.690196}
  \definecolor{AntiqueWhite4}{rgb}{0.545098,0.513726,0.470588}
  \definecolor{bisque1}{rgb}{1.000000,0.894118,0.768627}
  \definecolor{bisque2}{rgb}{0.933333,0.835294,0.717647}
  \definecolor{bisque3}{rgb}{0.803922,0.717647,0.619608}
  \definecolor{bisque4}{rgb}{0.545098,0.490196,0.419608}
  \definecolor{PeachPuff1}{rgb}{1.000000,0.854902,0.725490}
  \definecolor{PeachPuff2}{rgb}{0.933333,0.796078,0.678431}
  \definecolor{PeachPuff3}{rgb}{0.803922,0.686275,0.584314}
  \definecolor{PeachPuff4}{rgb}{0.545098,0.466667,0.396078}
  \definecolor{NavajoWhite1}{rgb}{1.000000,0.870588,0.678431}
  \definecolor{NavajoWhite2}{rgb}{0.933333,0.811765,0.631373}
  \definecolor{NavajoWhite3}{rgb}{0.803922,0.701961,0.545098}
  \definecolor{NavajoWhite4}{rgb}{0.545098,0.474510,0.368627}
  \definecolor{LemonChiffon1}{rgb}{1.000000,0.980392,0.803922}
  \definecolor{LemonChiffon2}{rgb}{0.933333,0.913725,0.749020}
  \definecolor{LemonChiffon3}{rgb}{0.803922,0.788235,0.647059}
  \definecolor{LemonChiffon4}{rgb}{0.545098,0.537255,0.439216}
  \definecolor{cornsilk1}{rgb}{1.000000,0.972549,0.862745}
  \definecolor{cornsilk2}{rgb}{0.933333,0.909804,0.803922}
  \definecolor{cornsilk3}{rgb}{0.803922,0.784314,0.694118}
  \definecolor{cornsilk4}{rgb}{0.545098,0.533333,0.470588}
  \definecolor{ivory1}{rgb}{1.000000,1.000000,0.941176}
  \definecolor{ivory2}{rgb}{0.933333,0.933333,0.878431}
  \definecolor{ivory3}{rgb}{0.803922,0.803922,0.756863}
  \definecolor{ivory4}{rgb}{0.545098,0.545098,0.513726}
  \definecolor{honeydew1}{rgb}{0.941176,1.000000,0.941176}
  \definecolor{honeydew2}{rgb}{0.878431,0.933333,0.878431}
  \definecolor{honeydew3}{rgb}{0.756863,0.803922,0.756863}
  \definecolor{honeydew4}{rgb}{0.513726,0.545098,0.513726}
  \definecolor{LavenderBlush1}{rgb}{1.000000,0.941176,0.960784}
  \definecolor{LavenderBlush2}{rgb}{0.933333,0.878431,0.898039}
  \definecolor{LavenderBlush3}{rgb}{0.803922,0.756863,0.772549}
  \definecolor{LavenderBlush4}{rgb}{0.545098,0.513726,0.525490}
  \definecolor{MistyRose1}{rgb}{1.000000,0.894118,0.882353}
  \definecolor{MistyRose2}{rgb}{0.933333,0.835294,0.823529}
  \definecolor{MistyRose3}{rgb}{0.803922,0.717647,0.709804}
  \definecolor{MistyRose4}{rgb}{0.545098,0.490196,0.482353}
  \definecolor{azure1}{rgb}{0.941176,1.000000,1.000000}
  \definecolor{azure2}{rgb}{0.878431,0.933333,0.933333}
  \definecolor{azure3}{rgb}{0.756863,0.803922,0.803922}
  \definecolor{azure4}{rgb}{0.513726,0.545098,0.545098}
  \definecolor{SlateBlue1}{rgb}{0.513726,0.435294,1.000000}
  \definecolor{SlateBlue2}{rgb}{0.478431,0.403922,0.933333}
  \definecolor{SlateBlue3}{rgb}{0.411765,0.349020,0.803922}
  \definecolor{SlateBlue4}{rgb}{0.278431,0.235294,0.545098}
  \definecolor{RoyalBlue1}{rgb}{0.282353,0.462745,1.000000}
  \definecolor{RoyalBlue2}{rgb}{0.262745,0.431373,0.933333}
  \definecolor{RoyalBlue3}{rgb}{0.227451,0.372549,0.803922}
  \definecolor{RoyalBlue4}{rgb}{0.152941,0.250980,0.545098}
  \definecolor{blue1}{rgb}{0.000000,0.000000,1.000000}
  \definecolor{blue2}{rgb}{0.000000,0.000000,0.933333}
  \definecolor{blue3}{rgb}{0.000000,0.000000,0.803922}
  \definecolor{blue4}{rgb}{0.000000,0.000000,0.545098}
  \definecolor{DodgerBlue1}{rgb}{0.117647,0.564706,1.000000}
  \definecolor{DodgerBlue2}{rgb}{0.109804,0.525490,0.933333}
  \definecolor{DodgerBlue3}{rgb}{0.094118,0.454902,0.803922}
  \definecolor{DodgerBlue4}{rgb}{0.062745,0.305882,0.545098}
  \definecolor{SteelBlue1}{rgb}{0.388235,0.721569,1.000000}
  \definecolor{SteelBlue2}{rgb}{0.360784,0.674510,0.933333}
  \definecolor{SteelBlue3}{rgb}{0.309804,0.580392,0.803922}
  \definecolor{SteelBlue4}{rgb}{0.211765,0.392157,0.545098}
  \definecolor{DeepSkyBlue1}{rgb}{0.000000,0.749020,1.000000}
  \definecolor{DeepSkyBlue2}{rgb}{0.000000,0.698039,0.933333}
  \definecolor{DeepSkyBlue3}{rgb}{0.000000,0.603922,0.803922}
  \definecolor{DeepSkyBlue4}{rgb}{0.000000,0.407843,0.545098}
  \definecolor{SkyBlue1}{rgb}{0.529412,0.807843,1.000000}
  \definecolor{SkyBlue2}{rgb}{0.494118,0.752941,0.933333}
  \definecolor{SkyBlue3}{rgb}{0.423529,0.650980,0.803922}
  \definecolor{SkyBlue4}{rgb}{0.290196,0.439216,0.545098}
  \definecolor{LightSkyBlue1}{rgb}{0.690196,0.886275,1.000000}
  \definecolor{LightSkyBlue2}{rgb}{0.643137,0.827451,0.933333}
  \definecolor{LightSkyBlue3}{rgb}{0.552941,0.713726,0.803922}
  \definecolor{LightSkyBlue4}{rgb}{0.376471,0.482353,0.545098}
  \definecolor{SlateGray1}{rgb}{0.776471,0.886275,1.000000}
  \definecolor{SlateGray2}{rgb}{0.725490,0.827451,0.933333}
  \definecolor{SlateGray3}{rgb}{0.623529,0.713726,0.803922}
  \definecolor{SlateGray4}{rgb}{0.423529,0.482353,0.545098}
  \definecolor{LightSteelBlue1}{rgb}{0.792157,0.882353,1.000000}
  \definecolor{LightSteelBlue2}{rgb}{0.737255,0.823529,0.933333}
  \definecolor{LightSteelBlue3}{rgb}{0.635294,0.709804,0.803922}
  \definecolor{LightSteelBlue4}{rgb}{0.431373,0.482353,0.545098}
  \definecolor{LightBlue1}{rgb}{0.749020,0.937255,1.000000}
  \definecolor{LightBlue2}{rgb}{0.698039,0.874510,0.933333}
  \definecolor{LightBlue3}{rgb}{0.603922,0.752941,0.803922}
  \definecolor{LightBlue4}{rgb}{0.407843,0.513726,0.545098}
  \definecolor{LightCyan1}{rgb}{0.878431,1.000000,1.000000}
  \definecolor{LightCyan2}{rgb}{0.819608,0.933333,0.933333}
  \definecolor{LightCyan3}{rgb}{0.705882,0.803922,0.803922}
  \definecolor{LightCyan4}{rgb}{0.478431,0.545098,0.545098}
  \definecolor{PaleTurquoise1}{rgb}{0.733333,1.000000,1.000000}
  \definecolor{PaleTurquoise2}{rgb}{0.682353,0.933333,0.933333}
  \definecolor{PaleTurquoise3}{rgb}{0.588235,0.803922,0.803922}
  \definecolor{PaleTurquoise4}{rgb}{0.400000,0.545098,0.545098}
  \definecolor{CadetBlue1}{rgb}{0.596078,0.960784,1.000000}
  \definecolor{CadetBlue2}{rgb}{0.556863,0.898039,0.933333}
  \definecolor{CadetBlue3}{rgb}{0.478431,0.772549,0.803922}
  \definecolor{CadetBlue4}{rgb}{0.325490,0.525490,0.545098}
  \definecolor{turquoise1}{rgb}{0.000000,0.960784,1.000000}
  \definecolor{turquoise2}{rgb}{0.000000,0.898039,0.933333}
  \definecolor{turquoise3}{rgb}{0.000000,0.772549,0.803922}
  \definecolor{turquoise4}{rgb}{0.000000,0.525490,0.545098}
  \definecolor{cyan1}{rgb}{0.000000,1.000000,1.000000}
  \definecolor{cyan2}{rgb}{0.000000,0.933333,0.933333}
  \definecolor{cyan3}{rgb}{0.000000,0.803922,0.803922}
  \definecolor{cyan4}{rgb}{0.000000,0.545098,0.545098}
  \definecolor{DarkSlateGray1}{rgb}{0.592157,1.000000,1.000000}
  \definecolor{DarkSlateGray2}{rgb}{0.552941,0.933333,0.933333}
  \definecolor{DarkSlateGray3}{rgb}{0.474510,0.803922,0.803922}
  \definecolor{DarkSlateGray4}{rgb}{0.321569,0.545098,0.545098}
  \definecolor{aquamarine1}{rgb}{0.498039,1.000000,0.831373}
  \definecolor{aquamarine2}{rgb}{0.462745,0.933333,0.776471}
  \definecolor{aquamarine3}{rgb}{0.400000,0.803922,0.666667}
  \definecolor{aquamarine4}{rgb}{0.270588,0.545098,0.454902}
  \definecolor{DarkSeaGreen1}{rgb}{0.756863,1.000000,0.756863}
  \definecolor{DarkSeaGreen2}{rgb}{0.705882,0.933333,0.705882}
  \definecolor{DarkSeaGreen3}{rgb}{0.607843,0.803922,0.607843}
  \definecolor{DarkSeaGreen4}{rgb}{0.411765,0.545098,0.411765}
  \definecolor{SeaGreen1}{rgb}{0.329412,1.000000,0.623529}
  \definecolor{SeaGreen2}{rgb}{0.305882,0.933333,0.580392}
  \definecolor{SeaGreen3}{rgb}{0.262745,0.803922,0.501961}
  \definecolor{SeaGreen4}{rgb}{0.180392,0.545098,0.341176}
  \definecolor{PaleGreen1}{rgb}{0.603922,1.000000,0.603922}
  \definecolor{PaleGreen2}{rgb}{0.564706,0.933333,0.564706}
  \definecolor{PaleGreen3}{rgb}{0.486275,0.803922,0.486275}
  \definecolor{PaleGreen4}{rgb}{0.329412,0.545098,0.329412}
  \definecolor{SpringGreen1}{rgb}{0.000000,1.000000,0.498039}
  \definecolor{SpringGreen2}{rgb}{0.000000,0.933333,0.462745}
  \definecolor{SpringGreen3}{rgb}{0.000000,0.803922,0.400000}
  \definecolor{SpringGreen4}{rgb}{0.000000,0.545098,0.270588}
  \definecolor{green1}{rgb}{0.000000,1.000000,0.000000}
  \definecolor{green2}{rgb}{0.000000,0.933333,0.000000}
  \definecolor{green3}{rgb}{0.000000,0.803922,0.000000}
  \definecolor{green4}{rgb}{0.000000,0.545098,0.000000}
  \definecolor{chartreuse1}{rgb}{0.498039,1.000000,0.000000}
  \definecolor{chartreuse2}{rgb}{0.462745,0.933333,0.000000}
  \definecolor{chartreuse3}{rgb}{0.400000,0.803922,0.000000}
  \definecolor{chartreuse4}{rgb}{0.270588,0.545098,0.000000}
  \definecolor{OliveDrab1}{rgb}{0.752941,1.000000,0.243137}
  \definecolor{OliveDrab2}{rgb}{0.701961,0.933333,0.227451}
  \definecolor{OliveDrab3}{rgb}{0.603922,0.803922,0.196078}
  \definecolor{OliveDrab4}{rgb}{0.411765,0.545098,0.133333}
  \definecolor{DarkOliveGreen1}{rgb}{0.792157,1.000000,0.439216}
  \definecolor{DarkOliveGreen2}{rgb}{0.737255,0.933333,0.407843}
  \definecolor{DarkOliveGreen3}{rgb}{0.635294,0.803922,0.352941}
  \definecolor{DarkOliveGreen4}{rgb}{0.431373,0.545098,0.239216}
  \definecolor{khaki1}{rgb}{1.000000,0.964706,0.560784}
  \definecolor{khaki2}{rgb}{0.933333,0.901961,0.521569}
  \definecolor{khaki3}{rgb}{0.803922,0.776471,0.450980}
  \definecolor{khaki4}{rgb}{0.545098,0.525490,0.305882}
  \definecolor{LightGoldenrod1}{rgb}{1.000000,0.925490,0.545098}
  \definecolor{LightGoldenrod2}{rgb}{0.933333,0.862745,0.509804}
  \definecolor{LightGoldenrod3}{rgb}{0.803922,0.745098,0.439216}
  \definecolor{LightGoldenrod4}{rgb}{0.545098,0.505882,0.298039}
  \definecolor{LightYellow1}{rgb}{1.000000,1.000000,0.878431}
  \definecolor{LightYellow2}{rgb}{0.933333,0.933333,0.819608}
  \definecolor{LightYellow3}{rgb}{0.803922,0.803922,0.705882}
  \definecolor{LightYellow4}{rgb}{0.545098,0.545098,0.478431}
  \definecolor{yellow1}{rgb}{1.000000,1.000000,0.000000}
  \definecolor{yellow2}{rgb}{0.933333,0.933333,0.000000}
  \definecolor{yellow3}{rgb}{0.803922,0.803922,0.000000}
  \definecolor{yellow4}{rgb}{0.545098,0.545098,0.000000}
  \definecolor{gold1}{rgb}{1.000000,0.843137,0.000000}
  \definecolor{gold2}{rgb}{0.933333,0.788235,0.000000}
  \definecolor{gold3}{rgb}{0.803922,0.678431,0.000000}
  \definecolor{gold4}{rgb}{0.545098,0.458824,0.000000}
  \definecolor{goldenrod1}{rgb}{1.000000,0.756863,0.145098}
  \definecolor{goldenrod2}{rgb}{0.933333,0.705882,0.133333}
  \definecolor{goldenrod3}{rgb}{0.803922,0.607843,0.113725}
  \definecolor{goldenrod4}{rgb}{0.545098,0.411765,0.078431}
  \definecolor{DarkGoldenrod1}{rgb}{1.000000,0.725490,0.058824}
  \definecolor{DarkGoldenrod2}{rgb}{0.933333,0.678431,0.054902}
  \definecolor{DarkGoldenrod3}{rgb}{0.803922,0.584314,0.047059}
  \definecolor{DarkGoldenrod4}{rgb}{0.545098,0.396078,0.031373}
  \definecolor{RosyBrown1}{rgb}{1.000000,0.756863,0.756863}
  \definecolor{RosyBrown2}{rgb}{0.933333,0.705882,0.705882}
  \definecolor{RosyBrown3}{rgb}{0.803922,0.607843,0.607843}
  \definecolor{RosyBrown4}{rgb}{0.545098,0.411765,0.411765}
  \definecolor{IndianRed1}{rgb}{1.000000,0.415686,0.415686}
  \definecolor{IndianRed2}{rgb}{0.933333,0.388235,0.388235}
  \definecolor{IndianRed3}{rgb}{0.803922,0.333333,0.333333}
  \definecolor{IndianRed4}{rgb}{0.545098,0.227451,0.227451}
  \definecolor{sienna1}{rgb}{1.000000,0.509804,0.278431}
  \definecolor{sienna2}{rgb}{0.933333,0.474510,0.258824}
  \definecolor{sienna3}{rgb}{0.803922,0.407843,0.223529}
  \definecolor{sienna4}{rgb}{0.545098,0.278431,0.149020}
  \definecolor{burlywood1}{rgb}{1.000000,0.827451,0.607843}
  \definecolor{burlywood2}{rgb}{0.933333,0.772549,0.568627}
  \definecolor{burlywood3}{rgb}{0.803922,0.666667,0.490196}
  \definecolor{burlywood4}{rgb}{0.545098,0.450980,0.333333}
  \definecolor{wheat1}{rgb}{1.000000,0.905882,0.729412}
  \definecolor{wheat2}{rgb}{0.933333,0.847059,0.682353}
  \definecolor{wheat3}{rgb}{0.803922,0.729412,0.588235}
  \definecolor{wheat4}{rgb}{0.545098,0.494118,0.400000}
  \definecolor{tan1}{rgb}{1.000000,0.647059,0.309804}
  \definecolor{tan2}{rgb}{0.933333,0.603922,0.286275}
  \definecolor{tan3}{rgb}{0.803922,0.521569,0.247059}
  \definecolor{tan4}{rgb}{0.545098,0.352941,0.168627}
  \definecolor{chocolate1}{rgb}{1.000000,0.498039,0.141176}
  \definecolor{chocolate2}{rgb}{0.933333,0.462745,0.129412}
  \definecolor{chocolate3}{rgb}{0.803922,0.400000,0.113725}
  \definecolor{chocolate4}{rgb}{0.545098,0.270588,0.074510}
  \definecolor{firebrick1}{rgb}{1.000000,0.188235,0.188235}
  \definecolor{firebrick2}{rgb}{0.933333,0.172549,0.172549}
  \definecolor{firebrick3}{rgb}{0.803922,0.149020,0.149020}
  \definecolor{firebrick4}{rgb}{0.545098,0.101961,0.101961}
  \definecolor{brown1}{rgb}{1.000000,0.250980,0.250980}
  \definecolor{brown2}{rgb}{0.933333,0.231373,0.231373}
  \definecolor{brown3}{rgb}{0.803922,0.200000,0.200000}
  \definecolor{brown4}{rgb}{0.545098,0.137255,0.137255}
  \definecolor{salmon1}{rgb}{1.000000,0.549020,0.411765}
  \definecolor{salmon2}{rgb}{0.933333,0.509804,0.384314}
  \definecolor{salmon3}{rgb}{0.803922,0.439216,0.329412}
  \definecolor{salmon4}{rgb}{0.545098,0.298039,0.223529}
  \definecolor{LightSalmon1}{rgb}{1.000000,0.627451,0.478431}
  \definecolor{LightSalmon2}{rgb}{0.933333,0.584314,0.447059}
  \definecolor{LightSalmon3}{rgb}{0.803922,0.505882,0.384314}
  \definecolor{LightSalmon4}{rgb}{0.545098,0.341176,0.258824}
  \definecolor{orange1}{rgb}{1.000000,0.647059,0.000000}
  \definecolor{orange2}{rgb}{0.933333,0.603922,0.000000}
  \definecolor{orange3}{rgb}{0.803922,0.521569,0.000000}
  \definecolor{orange4}{rgb}{0.545098,0.352941,0.000000}
  \definecolor{DarkOrange1}{rgb}{1.000000,0.498039,0.000000}
  \definecolor{DarkOrange2}{rgb}{0.933333,0.462745,0.000000}
  \definecolor{DarkOrange3}{rgb}{0.803922,0.400000,0.000000}
  \definecolor{DarkOrange4}{rgb}{0.545098,0.270588,0.000000}
  \definecolor{coral1}{rgb}{1.000000,0.447059,0.337255}
  \definecolor{coral2}{rgb}{0.933333,0.415686,0.313726}
  \definecolor{coral3}{rgb}{0.803922,0.356863,0.270588}
  \definecolor{coral4}{rgb}{0.545098,0.243137,0.184314}
  \definecolor{tomato1}{rgb}{1.000000,0.388235,0.278431}
  \definecolor{tomato2}{rgb}{0.933333,0.360784,0.258824}
  \definecolor{tomato3}{rgb}{0.803922,0.309804,0.223529}
  \definecolor{tomato4}{rgb}{0.545098,0.211765,0.149020}
  \definecolor{OrangeRed1}{rgb}{1.000000,0.270588,0.000000}
  \definecolor{OrangeRed2}{rgb}{0.933333,0.250980,0.000000}
  \definecolor{OrangeRed3}{rgb}{0.803922,0.215686,0.000000}
  \definecolor{OrangeRed4}{rgb}{0.545098,0.145098,0.000000}
  \definecolor{red1}{rgb}{1.000000,0.000000,0.000000}
  \definecolor{red2}{rgb}{0.933333,0.000000,0.000000}
  \definecolor{red3}{rgb}{0.803922,0.000000,0.000000}
  \definecolor{red4}{rgb}{0.545098,0.000000,0.000000}
  \definecolor{DeepPink1}{rgb}{1.000000,0.078431,0.576471}
  \definecolor{DeepPink2}{rgb}{0.933333,0.070588,0.537255}
  \definecolor{DeepPink3}{rgb}{0.803922,0.062745,0.462745}
  \definecolor{DeepPink4}{rgb}{0.545098,0.039216,0.313726}
  \definecolor{HotPink1}{rgb}{1.000000,0.431373,0.705882}
  \definecolor{HotPink2}{rgb}{0.933333,0.415686,0.654902}
  \definecolor{HotPink3}{rgb}{0.803922,0.376471,0.564706}
  \definecolor{HotPink4}{rgb}{0.545098,0.227451,0.384314}
  \definecolor{pink1}{rgb}{1.000000,0.709804,0.772549}
  \definecolor{pink2}{rgb}{0.933333,0.662745,0.721569}
  \definecolor{pink3}{rgb}{0.803922,0.568627,0.619608}
  \definecolor{pink4}{rgb}{0.545098,0.388235,0.423529}
  \definecolor{LightPink1}{rgb}{1.000000,0.682353,0.725490}
  \definecolor{LightPink2}{rgb}{0.933333,0.635294,0.678431}
  \definecolor{LightPink3}{rgb}{0.803922,0.549020,0.584314}
  \definecolor{LightPink4}{rgb}{0.545098,0.372549,0.396078}
  \definecolor{PaleVioletRed1}{rgb}{1.000000,0.509804,0.670588}
  \definecolor{PaleVioletRed2}{rgb}{0.933333,0.474510,0.623529}
  \definecolor{PaleVioletRed3}{rgb}{0.803922,0.407843,0.537255}
  \definecolor{PaleVioletRed4}{rgb}{0.545098,0.278431,0.364706}
  \definecolor{maroon1}{rgb}{1.000000,0.203922,0.701961}
  \definecolor{maroon2}{rgb}{0.933333,0.188235,0.654902}
  \definecolor{maroon3}{rgb}{0.803922,0.160784,0.564706}
  \definecolor{maroon4}{rgb}{0.545098,0.109804,0.384314}
  \definecolor{VioletRed1}{rgb}{1.000000,0.243137,0.588235}
  \definecolor{VioletRed2}{rgb}{0.933333,0.227451,0.549020}
  \definecolor{VioletRed3}{rgb}{0.803922,0.196078,0.470588}
  \definecolor{VioletRed4}{rgb}{0.545098,0.133333,0.321569}
  \definecolor{magenta1}{rgb}{1.000000,0.000000,1.000000}
  \definecolor{magenta2}{rgb}{0.933333,0.000000,0.933333}
  \definecolor{magenta3}{rgb}{0.803922,0.000000,0.803922}
  \definecolor{magenta4}{rgb}{0.545098,0.000000,0.545098}
  \definecolor{orchid1}{rgb}{1.000000,0.513726,0.980392}
  \definecolor{orchid2}{rgb}{0.933333,0.478431,0.913725}
  \definecolor{orchid3}{rgb}{0.803922,0.411765,0.788235}
  \definecolor{orchid4}{rgb}{0.545098,0.278431,0.537255}
  \definecolor{plum1}{rgb}{1.000000,0.733333,1.000000}
  \definecolor{plum2}{rgb}{0.933333,0.682353,0.933333}
  \definecolor{plum3}{rgb}{0.803922,0.588235,0.803922}
  \definecolor{plum4}{rgb}{0.545098,0.400000,0.545098}
  \definecolor{MediumOrchid1}{rgb}{0.878431,0.400000,1.000000}
  \definecolor{MediumOrchid2}{rgb}{0.819608,0.372549,0.933333}
  \definecolor{MediumOrchid3}{rgb}{0.705882,0.321569,0.803922}
  \definecolor{MediumOrchid4}{rgb}{0.478431,0.215686,0.545098}
  \definecolor{DarkOrchid1}{rgb}{0.749020,0.243137,1.000000}
  \definecolor{DarkOrchid2}{rgb}{0.698039,0.227451,0.933333}
  \definecolor{DarkOrchid3}{rgb}{0.603922,0.196078,0.803922}
  \definecolor{DarkOrchid4}{rgb}{0.407843,0.133333,0.545098}
  \definecolor{purple1}{rgb}{0.607843,0.188235,1.000000}
  \definecolor{purple2}{rgb}{0.568627,0.172549,0.933333}
  \definecolor{purple3}{rgb}{0.490196,0.149020,0.803922}
  \definecolor{purple4}{rgb}{0.333333,0.101961,0.545098}
  \definecolor{MediumPurple1}{rgb}{0.670588,0.509804,1.000000}
  \definecolor{MediumPurple2}{rgb}{0.623529,0.474510,0.933333}
  \definecolor{MediumPurple3}{rgb}{0.537255,0.407843,0.803922}
  \definecolor{MediumPurple4}{rgb}{0.364706,0.278431,0.545098}
  \definecolor{thistle1}{rgb}{1.000000,0.882353,1.000000}
  \definecolor{thistle2}{rgb}{0.933333,0.823529,0.933333}
  \definecolor{thistle3}{rgb}{0.803922,0.709804,0.803922}
  \definecolor{thistle4}{rgb}{0.545098,0.482353,0.545098}
  \definecolor{gray0}{rgb}{0.000000,0.000000,0.000000}
  \definecolor{grey0}{rgb}{0.000000,0.000000,0.000000}
  \definecolor{gray1}{rgb}{0.011765,0.011765,0.011765}
  \definecolor{grey1}{rgb}{0.011765,0.011765,0.011765}
  \definecolor{gray2}{rgb}{0.019608,0.019608,0.019608}
  \definecolor{grey2}{rgb}{0.019608,0.019608,0.019608}
  \definecolor{gray3}{rgb}{0.031373,0.031373,0.031373}
  \definecolor{grey3}{rgb}{0.031373,0.031373,0.031373}
  \definecolor{gray4}{rgb}{0.039216,0.039216,0.039216}
  \definecolor{grey4}{rgb}{0.039216,0.039216,0.039216}
  \definecolor{gray5}{rgb}{0.050980,0.050980,0.050980}
  \definecolor{grey5}{rgb}{0.050980,0.050980,0.050980}
  \definecolor{gray6}{rgb}{0.058824,0.058824,0.058824}
  \definecolor{grey6}{rgb}{0.058824,0.058824,0.058824}
  \definecolor{gray7}{rgb}{0.070588,0.070588,0.070588}
  \definecolor{grey7}{rgb}{0.070588,0.070588,0.070588}
  \definecolor{gray8}{rgb}{0.078431,0.078431,0.078431}
  \definecolor{grey8}{rgb}{0.078431,0.078431,0.078431}
  \definecolor{gray9}{rgb}{0.090196,0.090196,0.090196}
  \definecolor{grey9}{rgb}{0.090196,0.090196,0.090196}
  \definecolor{gray10}{rgb}{0.101961,0.101961,0.101961}
  \definecolor{grey10}{rgb}{0.101961,0.101961,0.101961}
  \definecolor{gray11}{rgb}{0.109804,0.109804,0.109804}
  \definecolor{grey11}{rgb}{0.109804,0.109804,0.109804}
  \definecolor{gray12}{rgb}{0.121569,0.121569,0.121569}
  \definecolor{grey12}{rgb}{0.121569,0.121569,0.121569}
  \definecolor{gray13}{rgb}{0.129412,0.129412,0.129412}
  \definecolor{grey13}{rgb}{0.129412,0.129412,0.129412}
  \definecolor{gray14}{rgb}{0.141176,0.141176,0.141176}
  \definecolor{grey14}{rgb}{0.141176,0.141176,0.141176}
  \definecolor{gray15}{rgb}{0.149020,0.149020,0.149020}
  \definecolor{grey15}{rgb}{0.149020,0.149020,0.149020}
  \definecolor{gray16}{rgb}{0.160784,0.160784,0.160784}
  \definecolor{grey16}{rgb}{0.160784,0.160784,0.160784}
  \definecolor{gray17}{rgb}{0.168627,0.168627,0.168627}
  \definecolor{grey17}{rgb}{0.168627,0.168627,0.168627}
  \definecolor{gray18}{rgb}{0.180392,0.180392,0.180392}
  \definecolor{grey18}{rgb}{0.180392,0.180392,0.180392}
  \definecolor{gray19}{rgb}{0.188235,0.188235,0.188235}
  \definecolor{grey19}{rgb}{0.188235,0.188235,0.188235}
  \definecolor{gray20}{rgb}{0.200000,0.200000,0.200000}
  \definecolor{grey20}{rgb}{0.200000,0.200000,0.200000}
  \definecolor{gray21}{rgb}{0.211765,0.211765,0.211765}
  \definecolor{grey21}{rgb}{0.211765,0.211765,0.211765}
  \definecolor{gray22}{rgb}{0.219608,0.219608,0.219608}
  \definecolor{grey22}{rgb}{0.219608,0.219608,0.219608}
  \definecolor{gray23}{rgb}{0.231373,0.231373,0.231373}
  \definecolor{grey23}{rgb}{0.231373,0.231373,0.231373}
  \definecolor{gray24}{rgb}{0.239216,0.239216,0.239216}
  \definecolor{grey24}{rgb}{0.239216,0.239216,0.239216}
  \definecolor{gray25}{rgb}{0.250980,0.250980,0.250980}
  \definecolor{grey25}{rgb}{0.250980,0.250980,0.250980}
  \definecolor{gray26}{rgb}{0.258824,0.258824,0.258824}
  \definecolor{grey26}{rgb}{0.258824,0.258824,0.258824}
  \definecolor{gray27}{rgb}{0.270588,0.270588,0.270588}
  \definecolor{grey27}{rgb}{0.270588,0.270588,0.270588}
  \definecolor{gray28}{rgb}{0.278431,0.278431,0.278431}
  \definecolor{grey28}{rgb}{0.278431,0.278431,0.278431}
  \definecolor{gray29}{rgb}{0.290196,0.290196,0.290196}
  \definecolor{grey29}{rgb}{0.290196,0.290196,0.290196}
  \definecolor{gray30}{rgb}{0.301961,0.301961,0.301961}
  \definecolor{grey30}{rgb}{0.301961,0.301961,0.301961}
  \definecolor{gray31}{rgb}{0.309804,0.309804,0.309804}
  \definecolor{grey31}{rgb}{0.309804,0.309804,0.309804}
  \definecolor{gray32}{rgb}{0.321569,0.321569,0.321569}
  \definecolor{grey32}{rgb}{0.321569,0.321569,0.321569}
  \definecolor{gray33}{rgb}{0.329412,0.329412,0.329412}
  \definecolor{grey33}{rgb}{0.329412,0.329412,0.329412}
  \definecolor{gray34}{rgb}{0.341176,0.341176,0.341176}
  \definecolor{grey34}{rgb}{0.341176,0.341176,0.341176}
  \definecolor{gray35}{rgb}{0.349020,0.349020,0.349020}
  \definecolor{grey35}{rgb}{0.349020,0.349020,0.349020}
  \definecolor{gray36}{rgb}{0.360784,0.360784,0.360784}
  \definecolor{grey36}{rgb}{0.360784,0.360784,0.360784}
  \definecolor{gray37}{rgb}{0.368627,0.368627,0.368627}
  \definecolor{grey37}{rgb}{0.368627,0.368627,0.368627}
  \definecolor{gray38}{rgb}{0.380392,0.380392,0.380392}
  \definecolor{grey38}{rgb}{0.380392,0.380392,0.380392}
  \definecolor{gray39}{rgb}{0.388235,0.388235,0.388235}
  \definecolor{grey39}{rgb}{0.388235,0.388235,0.388235}
  \definecolor{gray40}{rgb}{0.400000,0.400000,0.400000}
  \definecolor{grey40}{rgb}{0.400000,0.400000,0.400000}
  \definecolor{gray41}{rgb}{0.411765,0.411765,0.411765}
  \definecolor{grey41}{rgb}{0.411765,0.411765,0.411765}
  \definecolor{gray42}{rgb}{0.419608,0.419608,0.419608}
  \definecolor{grey42}{rgb}{0.419608,0.419608,0.419608}
  \definecolor{gray43}{rgb}{0.431373,0.431373,0.431373}
  \definecolor{grey43}{rgb}{0.431373,0.431373,0.431373}
  \definecolor{gray44}{rgb}{0.439216,0.439216,0.439216}
  \definecolor{grey44}{rgb}{0.439216,0.439216,0.439216}
  \definecolor{gray45}{rgb}{0.450980,0.450980,0.450980}
  \definecolor{grey45}{rgb}{0.450980,0.450980,0.450980}
  \definecolor{gray46}{rgb}{0.458824,0.458824,0.458824}
  \definecolor{grey46}{rgb}{0.458824,0.458824,0.458824}
  \definecolor{gray47}{rgb}{0.470588,0.470588,0.470588}
  \definecolor{grey47}{rgb}{0.470588,0.470588,0.470588}
  \definecolor{gray48}{rgb}{0.478431,0.478431,0.478431}
  \definecolor{grey48}{rgb}{0.478431,0.478431,0.478431}
  \definecolor{gray49}{rgb}{0.490196,0.490196,0.490196}
  \definecolor{grey49}{rgb}{0.490196,0.490196,0.490196}
  \definecolor{gray50}{rgb}{0.498039,0.498039,0.498039}
  \definecolor{grey50}{rgb}{0.498039,0.498039,0.498039}
  \definecolor{gray51}{rgb}{0.509804,0.509804,0.509804}
  \definecolor{grey51}{rgb}{0.509804,0.509804,0.509804}
  \definecolor{gray52}{rgb}{0.521569,0.521569,0.521569}
  \definecolor{grey52}{rgb}{0.521569,0.521569,0.521569}
  \definecolor{gray53}{rgb}{0.529412,0.529412,0.529412}
  \definecolor{grey53}{rgb}{0.529412,0.529412,0.529412}
  \definecolor{gray54}{rgb}{0.541176,0.541176,0.541176}
  \definecolor{grey54}{rgb}{0.541176,0.541176,0.541176}
  \definecolor{gray55}{rgb}{0.549020,0.549020,0.549020}
  \definecolor{grey55}{rgb}{0.549020,0.549020,0.549020}
  \definecolor{gray56}{rgb}{0.560784,0.560784,0.560784}
  \definecolor{grey56}{rgb}{0.560784,0.560784,0.560784}
  \definecolor{gray57}{rgb}{0.568627,0.568627,0.568627}
  \definecolor{grey57}{rgb}{0.568627,0.568627,0.568627}
  \definecolor{gray58}{rgb}{0.580392,0.580392,0.580392}
  \definecolor{grey58}{rgb}{0.580392,0.580392,0.580392}
  \definecolor{gray59}{rgb}{0.588235,0.588235,0.588235}
  \definecolor{grey59}{rgb}{0.588235,0.588235,0.588235}
  \definecolor{gray60}{rgb}{0.600000,0.600000,0.600000}
  \definecolor{grey60}{rgb}{0.600000,0.600000,0.600000}
  \definecolor{gray61}{rgb}{0.611765,0.611765,0.611765}
  \definecolor{grey61}{rgb}{0.611765,0.611765,0.611765}
  \definecolor{gray62}{rgb}{0.619608,0.619608,0.619608}
  \definecolor{grey62}{rgb}{0.619608,0.619608,0.619608}
  \definecolor{gray63}{rgb}{0.631373,0.631373,0.631373}
  \definecolor{grey63}{rgb}{0.631373,0.631373,0.631373}
  \definecolor{gray64}{rgb}{0.639216,0.639216,0.639216}
  \definecolor{grey64}{rgb}{0.639216,0.639216,0.639216}
  \definecolor{gray65}{rgb}{0.650980,0.650980,0.650980}
  \definecolor{grey65}{rgb}{0.650980,0.650980,0.650980}
  \definecolor{gray66}{rgb}{0.658824,0.658824,0.658824}
  \definecolor{grey66}{rgb}{0.658824,0.658824,0.658824}
  \definecolor{gray67}{rgb}{0.670588,0.670588,0.670588}
  \definecolor{grey67}{rgb}{0.670588,0.670588,0.670588}
  \definecolor{gray68}{rgb}{0.678431,0.678431,0.678431}
  \definecolor{grey68}{rgb}{0.678431,0.678431,0.678431}
  \definecolor{gray69}{rgb}{0.690196,0.690196,0.690196}
  \definecolor{grey69}{rgb}{0.690196,0.690196,0.690196}
  \definecolor{gray70}{rgb}{0.701961,0.701961,0.701961}
  \definecolor{grey70}{rgb}{0.701961,0.701961,0.701961}
  \definecolor{gray71}{rgb}{0.709804,0.709804,0.709804}
  \definecolor{grey71}{rgb}{0.709804,0.709804,0.709804}
  \definecolor{gray72}{rgb}{0.721569,0.721569,0.721569}
  \definecolor{grey72}{rgb}{0.721569,0.721569,0.721569}
  \definecolor{gray73}{rgb}{0.729412,0.729412,0.729412}
  \definecolor{grey73}{rgb}{0.729412,0.729412,0.729412}
  \definecolor{gray74}{rgb}{0.741176,0.741176,0.741176}
  \definecolor{grey74}{rgb}{0.741176,0.741176,0.741176}
  \definecolor{gray75}{rgb}{0.749020,0.749020,0.749020}
  \definecolor{grey75}{rgb}{0.749020,0.749020,0.749020}
  \definecolor{gray76}{rgb}{0.760784,0.760784,0.760784}
  \definecolor{grey76}{rgb}{0.760784,0.760784,0.760784}
  \definecolor{gray77}{rgb}{0.768627,0.768627,0.768627}
  \definecolor{grey77}{rgb}{0.768627,0.768627,0.768627}
  \definecolor{gray78}{rgb}{0.780392,0.780392,0.780392}
  \definecolor{grey78}{rgb}{0.780392,0.780392,0.780392}
  \definecolor{gray79}{rgb}{0.788235,0.788235,0.788235}
  \definecolor{grey79}{rgb}{0.788235,0.788235,0.788235}
  \definecolor{gray80}{rgb}{0.800000,0.800000,0.800000}
  \definecolor{grey80}{rgb}{0.800000,0.800000,0.800000}
  \definecolor{gray81}{rgb}{0.811765,0.811765,0.811765}
  \definecolor{grey81}{rgb}{0.811765,0.811765,0.811765}
  \definecolor{gray82}{rgb}{0.819608,0.819608,0.819608}
  \definecolor{grey82}{rgb}{0.819608,0.819608,0.819608}
  \definecolor{gray83}{rgb}{0.831373,0.831373,0.831373}
  \definecolor{grey83}{rgb}{0.831373,0.831373,0.831373}
  \definecolor{gray84}{rgb}{0.839216,0.839216,0.839216}
  \definecolor{grey84}{rgb}{0.839216,0.839216,0.839216}
  \definecolor{gray85}{rgb}{0.850980,0.850980,0.850980}
  \definecolor{grey85}{rgb}{0.850980,0.850980,0.850980}
  \definecolor{gray86}{rgb}{0.858824,0.858824,0.858824}
  \definecolor{grey86}{rgb}{0.858824,0.858824,0.858824}
  \definecolor{gray87}{rgb}{0.870588,0.870588,0.870588}
  \definecolor{grey87}{rgb}{0.870588,0.870588,0.870588}
  \definecolor{gray88}{rgb}{0.878431,0.878431,0.878431}
  \definecolor{grey88}{rgb}{0.878431,0.878431,0.878431}
  \definecolor{gray89}{rgb}{0.890196,0.890196,0.890196}
  \definecolor{grey89}{rgb}{0.890196,0.890196,0.890196}
  \definecolor{gray90}{rgb}{0.898039,0.898039,0.898039}
  \definecolor{grey90}{rgb}{0.898039,0.898039,0.898039}
  \definecolor{gray91}{rgb}{0.909804,0.909804,0.909804}
  \definecolor{grey91}{rgb}{0.909804,0.909804,0.909804}
  \definecolor{gray92}{rgb}{0.921569,0.921569,0.921569}
  \definecolor{grey92}{rgb}{0.921569,0.921569,0.921569}
  \definecolor{gray93}{rgb}{0.929412,0.929412,0.929412}
  \definecolor{grey93}{rgb}{0.929412,0.929412,0.929412}
  \definecolor{gray94}{rgb}{0.941176,0.941176,0.941176}
  \definecolor{grey94}{rgb}{0.941176,0.941176,0.941176}
  \definecolor{gray95}{rgb}{0.949020,0.949020,0.949020}
  \definecolor{grey95}{rgb}{0.949020,0.949020,0.949020}
  \definecolor{gray96}{rgb}{0.960784,0.960784,0.960784}
  \definecolor{grey96}{rgb}{0.960784,0.960784,0.960784}
  \definecolor{gray97}{rgb}{0.968627,0.968627,0.968627}
  \definecolor{grey97}{rgb}{0.968627,0.968627,0.968627}
  \definecolor{gray98}{rgb}{0.980392,0.980392,0.980392}
  \definecolor{grey98}{rgb}{0.980392,0.980392,0.980392}
  \definecolor{gray99}{rgb}{0.988235,0.988235,0.988235}
  \definecolor{grey99}{rgb}{0.988235,0.988235,0.988235}
  \definecolor{gray100}{rgb}{1.000000,1.000000,1.000000}
  \definecolor{grey100}{rgb}{1.000000,1.000000,1.000000}
  \definecolor{dark grey}{rgb}{0.662745,0.662745,0.662745}
  \definecolor{DarkGrey}{rgb}{0.662745,0.662745,0.662745}
  \definecolor{dark gray}{rgb}{0.662745,0.662745,0.662745}
  \definecolor{DarkGray}{rgb}{0.662745,0.662745,0.662745}
  \definecolor{dark blue}{rgb}{0.000000,0.000000,0.545098}
  \definecolor{DarkBlue}{rgb}{0.000000,0.000000,0.545098}
  \definecolor{dark cyan}{rgb}{0.000000,0.545098,0.545098}
  \definecolor{DarkCyan}{rgb}{0.000000,0.545098,0.545098}
  \definecolor{dark magenta}{rgb}{0.545098,0.000000,0.545098}
  \definecolor{DarkMagenta}{rgb}{0.545098,0.000000,0.545098}
  \definecolor{dark red}{rgb}{0.545098,0.000000,0.000000}
  \definecolor{DarkRed}{rgb}{0.545098,0.000000,0.000000}
  \definecolor{light green}{rgb}{0.564706,0.933333,0.564706}
  \definecolor{LightGreen}{rgb}{0.564706,0.933333,0.564706}

\usepackage{graphicx}
\usepackage{color}

\usepackage[T1]{fontenc}
\usepackage{aecompl}

\begin{document}

\title[The RomulusC Simulation]{Introducing RomulusC: A Cosmological Simulation of a Galaxy Cluster with Unprecedented Resolution}
\author[M. Tremmel et al.]{M. ~Tremmel$^{1}$\thanks{email: michael.tremmel@yale.edu},
T.~R.~Quinn$^{2}$, 
A.~Ricarte$^{3}$,
A.~Babul$^{4}$,
U.~Chadayammuri$^{3}$,\newauthor
P.~Natarajan$^{3}$,
D.~Nagai$^{1,3,5}$,
A. ~Pontzen$^{6}$,
M. ~Volonteri$^{7}$\\
$^1$Yale Center for Astronomy \& Astrophysics, Physics Department, P.O. Box 208120, New Haven, CT 06520, USA\\
$^2$Astronomy Department, University of Washington, Box 351580, Seattle, WA, 98195-1580\\
$^3$Department of Astronomy, Yale University, New Haven, CT 06511, USA\\
$^4$Department of Physics and Astronomy, University of Victoria, 3800 Finnerty Road, Victoria, BC, V8P 1A1, Canada\\
$^5$Department of Physics, Yale University, New Haven, CT 06520, USA\\
$^6$Department of Physics and Astronomy, University College London, Gower Street, London WC1E 6BT, UK\\
$^7$Sorbonne Universit\`{e}s, UPMC Univ Paris 6 et CNRS, UMR 7095, Institut d`Astrophysique de Paris, 98 bis bd Arago, 75014 Paris, France}

\pagerange{\pageref{firstpage}--\pageref{lastpage}} \pubyear{2015}

\maketitle

\label{firstpage}

\begin{abstract}

We present the first results from {\sc RomulusC}, the highest resolution cosmological hydrodynamic simulation of a galaxy cluster run to date. {\sc RomulusC}, a zoom-in simulation of a halo with $z=0$ mass $10^{14}$ M$_{\odot}$, is run with the same sub-grid physics and resolution as {\sc Romulus25} \citep{tremmel17}. With unprecedented mass and spatial resolution, {\sc RomulusC} represents a unique opportunity to study the evolution of galaxies in dense environments down to dwarf masses. We demonstrate that {\sc RomulusC} results in an intracluster medium (ICM) consistent with observations. The star formation history and stellar mass of the brightest cluster galaxy (BCG) is consistent with observations and abundance matching results, indicating that our sub-grid models, optimized only to reproduce observations of field dwarf and Milky Way mass galaxies, are able to produce reasonable galaxy masses and star formation histories in much higher mass systems. Feedback from supermassive black holes (SMBHs) regulates star formation by driving large-scale, collimated outflows that coexist with a low entropy core. We find that non-BCG cluster member galaxies are substantially quenched compared to the field down to dwarf galaxy masses and, at low masses, quenching is seen to have no dependence on mass or distance from the cluster center. This enhanced quenched population extends beyond $R_{200}$ and is in place at high redshift. Similarly, we predict that SMBH activity is significantly suppressed within clusters outside of the BCG, but show how the effect could be lost when only focusing on the brightest AGN in the most massive galaxies.
\end{abstract}


\section{Introduction}

Galaxy clusters are the most massive, and recently assembled structures in the Universe and their dense environments offer an interesting astrophysical laboratory to test our current theories of galaxy evolution. The setting of a cluster in particular enables us to probe the dynamical processes that transform galaxies, and the interactions with the environment that shape their evolution. The modulation of star formation in cluster galaxies has been a subject of great interest as current observations suggest that star formation in these galaxies is quenched more often compared to isolated galaxies in the field \citep[e.g.][]{weinmann06,geha12,wetzel12,wetzel13, haines13, haines15}. The truncation of star formation in cluster galaxies is thought be a consequence of interactions with the intracluster medium (ICM) via processes such as ram pressure stripping \citep[e.g.][]{chung07, merluzzi13}, `strangulation' \citep[e.g.][]{vandenbosch08,maier16}, whereby the galaxy is starved of its gas reservoir that is not replenished due to the hot environment, or a combination of both \citep{murakami99,bahe15}. Additionally, interactions with other cluster member galaxies through either mergers or fly-by events \citep{moore96,moorelake98,moore99b} may also play an important role in altering the star formation history and structural evolution of cluster member galaxies. 

The central regions of galaxy clusters are also important as they often host the most massive galaxies in the Universe. Cluster samples selected via the Sunyaev-Zel'dovich effect show that $\sim30\%$ of clusters have cool cores \citep{andrade-santos17}, meaning the central regions of their ICM have low entropy and short, sub-Gyr cooling times. The brightest cluster galaxies (BCGs) in these clusters, often located near the center, have relatively low star formation rates compared with estimates based on gas cooling rates which predict upwards of 100s M$_{\odot}$/yr cooling flows \citep[for reviews see][]{fabian94,peterson06}. For BCGs, feedback from active galactic nuclei (AGN) powered by accreting supermassive black holes (SMBHs) is thought to be the primary means of counteracting radiative cooling within the central region of the halo, thereby limiting cooling flows and star formation \citep[e.g.][]{blanton01,babul02,mccarthy08,gaspari2012,babul13,yang16,prasad17,cielo18,guo18}. This premise for BCGs is supported by observations of radio lobes and X-ray cavities \citep[e.g.][]{Boehringer93,mcnamara00,rafferty06,osullivan12}, often associated with jets \citep[e.g.][]{fabian02,croston11}, which can both control the supply of gas in the inner regions of the halo as well as its ability to cool and form stars within the central galaxy. Thus, AGN feedback seems to be crucial not only for the evolution of the most massive galaxies in the Universe, but they may also affect the structure and properties of gas in the centers of clusters.

Multi-wavelength observations of clusters, as well as gravitational lensing studies, are now able to extract more information about the structure of gas within clusters and the evolutionary history of their host galaxies \citep[e.g.][]{mahdavi08,natarajan09,zhang10,mahdavi13,jauzac15,hitomi16, smith16}. Numerical simulations are critical to understand the physical processes that shape the ICM and properties of cluster galaxies that are falling into assembling clusters. Modeling the physical scale of clusters presents a challenge for simulations due to the large dynamic range in the problem; the virial radii of clusters are on the scales of Mpc, while the physical processes of star formation, SMBH accretion, and feedback, as well as various hydrodynamic interactions, operate on kpc scales and far lower. Idealized simulations of galaxy clusters allow for very high resolution and can be useful for understanding the detailed interactions between SMBH feedback and the ICM \citep[e.g.][]{ruszkowski10,li14,li15,li17, gaspari14,prasad15, prasad17, prasad18, cielo18,cielo18b}. However, these simulations lack the cosmological context, the history of hierarchical merging and gas accretion that will shape the more detailed structures of the ICM and transform the population of in-falling galaxies. Cosmological simulations can self consistently model the evolution of a galaxy cluster within a larger-scale, evolving environment starting from realistic initial conditions predicated by the standard cold dark matter model \citep[e.g.][]{lewis00}. Gas and dark matter accumulate realistically and a population of  galaxies bound to the larger cluster halo result naturally from these simulations. Due to their relative rarity, only large volume simulations have any galaxy clusters within them and, due to their size, even zoom-in simulations of clusters are computationally expensive. Recent large-scale cosmological simulations (see Table 1 for a summary of most recent efforts), often calibrated to reproduce observations of high mass galaxies, have shown varying degrees of success in reproducing and studying (among other things) BCG masses and star formation histories  \citep[e.g.][]{bahe17,barnes17b, mccarthy17, pillepich18, ragone-figueroa18}, baryonic content and properties of the ICM \citep[e.g.][]{wu15, liang16, barnes17b, lau17}, AGN feedback and SMBH growth \citep[e.g.][]{mccarthy17, bogdan18}, and the properties and evolution of cluster member galaxies \citep[e.g.][]{pascal16, bahe17, zinger18, pillepich18}. While some simulations are now able to produce realistic BCG masses \citep[e.g.][]{ragone-figueroa18}, quenching star formation at later times often requires sub-grid models for AGN feedback that are explicitly tuned to increase feedback efficiency at higher masses \citep[e.g.][]{weinberger17}. Such models can be very useful and are indeed necessary at lower resolution, but their nature makes it so the interaction between SMBH feedback, halo gas, and star formation in BCGs cannot be considered a fully emergent property of the simulation. For studying cluster member galaxies, suites of dozens or hundreds of simulated clusters provide a large sample of massive cluster galaxies, but the limited resolution means that they will lack well resolved galaxies with masses below that of the Milky Way.


In this Paper, we present initial results from the {\sc RomulusC} zoom-in cosmological simulation of a $1.5\times10^{14}$ M$_{\odot}$ galaxy cluster which is, to our knowledge, the highest resolution cosmological hydrodynamic simulation of a cluster run to $z=0$ to date. Table 1 shows a list of other cosmological simulations that include at least one galaxy cluster of M$_{vir}(z=0)>10^{14}$ M$_{\odot}$ to demonstrate how {\sc RomulusC} represents a significant improvement in both spatial and mass resolution. It is important to note that {\sc RomulusC} (as well as TNG50), only includes a single low mass galaxy cluster while many of the other cosmological simulations produce dozens to hundreds of more massive halos. While such a small sample size is certainly an important limitation, in this paper we focus on the interactions between the ICM and SMBH feedback and the evolution of cluster member galaxies, both of which require high resolution to study in detail. We show how {\sc RomulusC} compares with various observational benchmarks in terms of several key cluster properties including those of the cluster gas and  the brightest cluster galaxy. The high resolution of this simulation  allows us to resolve halo masses as low as $3\times10^{9}$ M$_{\odot}$ with $>10^4$ particles and stellar masses as low as $10^7$ M$_{\odot}$ with $>100$ star particles (though in this paper we make the conservative choice to focus on galaxy masses above $10^8$ M$_{\odot}$). {\sc RomulusC} attains a maximum resolution for hydrodynamics of 70 pc, which has been shown in grid codes to be sufficient to resolve ram pressure stripping \citep{roedinger15}, although a thorough convergence test was not done. This simulation represents the first in a planned suite of high mass, high resolution zoom-in simulations and provides a critical proof of concept for our sub-grid models of star formation and, in particular, supermassive black hole physics, which have in no way been tuned to reproduce realistic galaxy clusters or BCGs. 

In \S2 we discuss the properties of the simulation, including relevant sub-grid physics models. In \S3 we compare bulk properties of the gas in {\sc RomulusC} to various observations. We examine the evolution of both the BCG and other cluster member galaxies in \S4 and discuss our results in \S5. A summary is presented in \S6 and future work pertaining to this simulation and others planned is discussed in \S7.

\begin{table}
\centering
\caption{Comparison with other recent cosmological hydrodynamic simulations that include halos with M$_{vir}>10^{14}$ M$_{\odot}$ as well as full radiative hydrodynamics. M$_{DM}$ and M$_{gas}$ are dark matter and (average) gas particle masses respectively. Note that many of the lower resolution simulations include dozens or hundreds of clusters while high resolution simulations like {\sc RomulusC} and TNG50 only have one.}
\label{symbols}
\begin{tabular}{@{}cccccc}
\hline
Name  & Spatial Res.$^{a}$ & M$_{DM}$ & M$_{gas}$ \\
     &  kpc & $M_{\odot}$ & $M_{\odot}$\\
\hline
{\bf RomulusC} & 0.25 & $3.4\times 10^5$ & $2.1 \times 10^5$\\\\
TNG300$^b$  & 1.5 & $7.9\times10^7$ & $7.4\times10^6$\\\\
TNG100$^b$  & 0.75 & $5.1\times10^6$ & $9.4\times10^5$\\\\
TNG50  & 0.3 & $4.4\times10^5$ & $8.5\times10^4$\\
(in progress$^c$) & & &\\\\
Horizon-AGN$^d$ & 1 & $8.0\times10^7$ & $1.0\times10^7$\\\\
Magneticum$^{e,f}$ & 10 & $1.3\times 10^{10}$ & $2.9\times 10^{9}$\\\\
Magneticum$^{e,f}$ & 3.75 & $6.9\times 10^8$ & $1.4\times 10^8$ \\
high res  &  &  &\\\\
Magneticum$^{e,f}$  & 1.4 & $3.6\times10^7$  & $7.3\times10^6$ \\
ultra high res  &  &  &\\\\
C-EAGLE$^{g,h}$   & 0.7 & $9.6\times10^6$ & $1.8\times10^6$\\\\
EAGLE$^i$      & 0.7 & $9.6\times10^6$ & $1.8\times10^6$\\
(50, 100 Mpc)     &  &  &\\\\
Omega500$^{j,k}$ & 5.4 & $1.6 \times 10^9$ & $2.7\times10^8$\\\\
MACSIS$^l$ & 5.9 & $5.7\times10^9$ & $1.0\times10^9$ \\\\
BAHAMAS$^m$ & 5.9 & $5.7\times10^9$ & $1.0\times10^9$ \\\\
Rhapsody-G$^n$ & 5.0 & $1.0 \times 10^9$ & $1.9\times10^8$ \\
\hline
\end{tabular}
\medskip
\\
$a$. Plummer equivalent values for gravitational softening presented if multiple given. If it varies with redshift, lowest values for low-z are presented. If it varies with particle type, dark matter values are used. When values are presented relative to $h^{-1}$, they are converted to $kpc$ using the value of $h$ corresponding to that simulation's cosmology.\\
$b$. Highest resolution versions are shown. \citet{marinacci17}
$c$. As of the publishing of this paper.
$d$. \citet{dubois14_horizon}
$e$. \citet{bocquet16}
$f$. \citet{dolag16}
$g$. \citet{barnes17b}
$h$. \citet{bahe17}\\
$i$. \citet{eagle15}
$j$. \citet{shirasaki17}
$k$. \citet{nelson14}\\
$l$. \citet{barnes17}
$m$. \citet{mccarthy17}
$n$. \citet{wu15}
\end{table}

\begin{figure*}
\centering
\includegraphics[trim=60mm 25mm 55mm 25mm, clip, width=45mm]{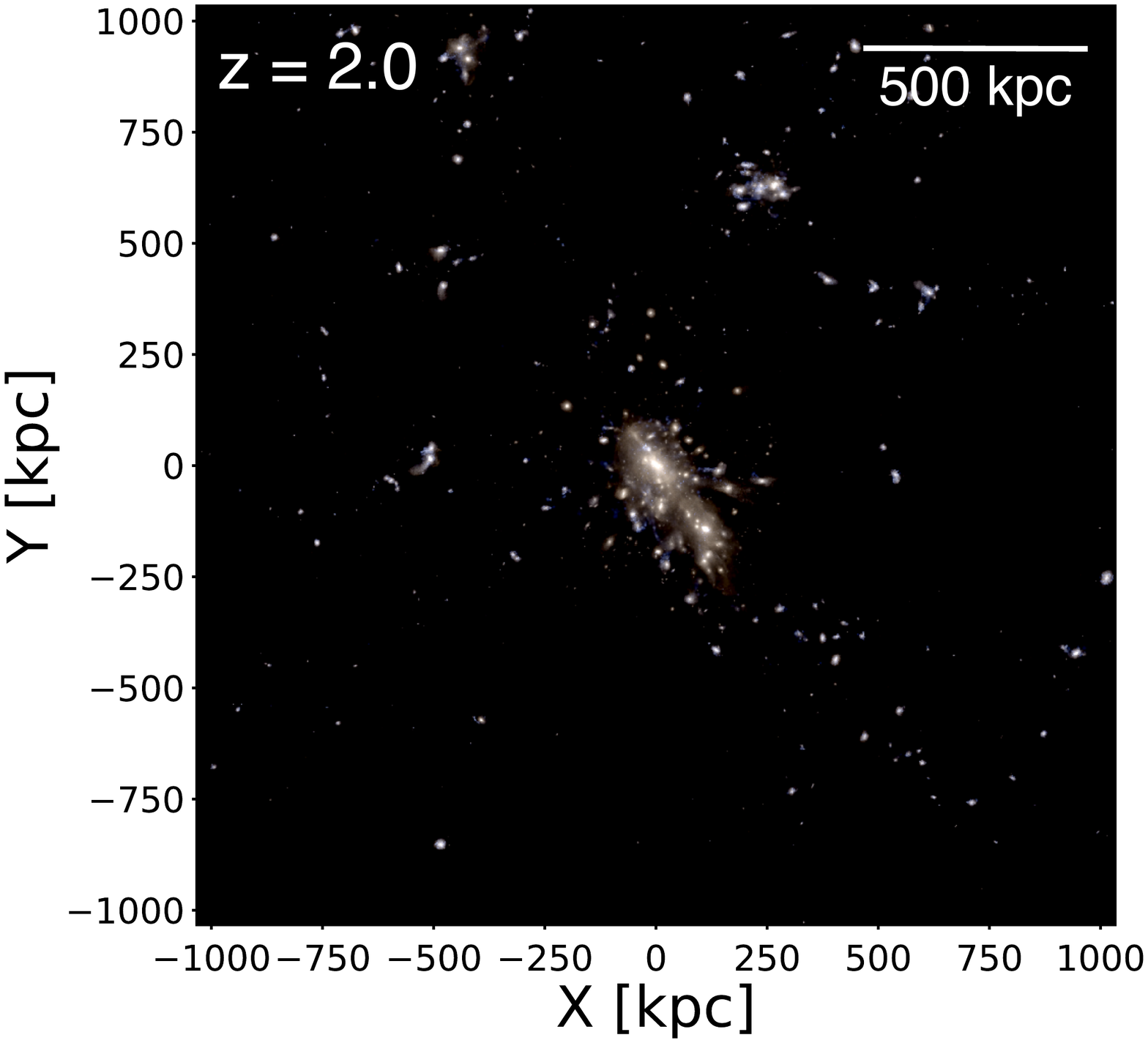}
\includegraphics[trim=60mm 25mm 55mm 25mm, clip, width=45mm]{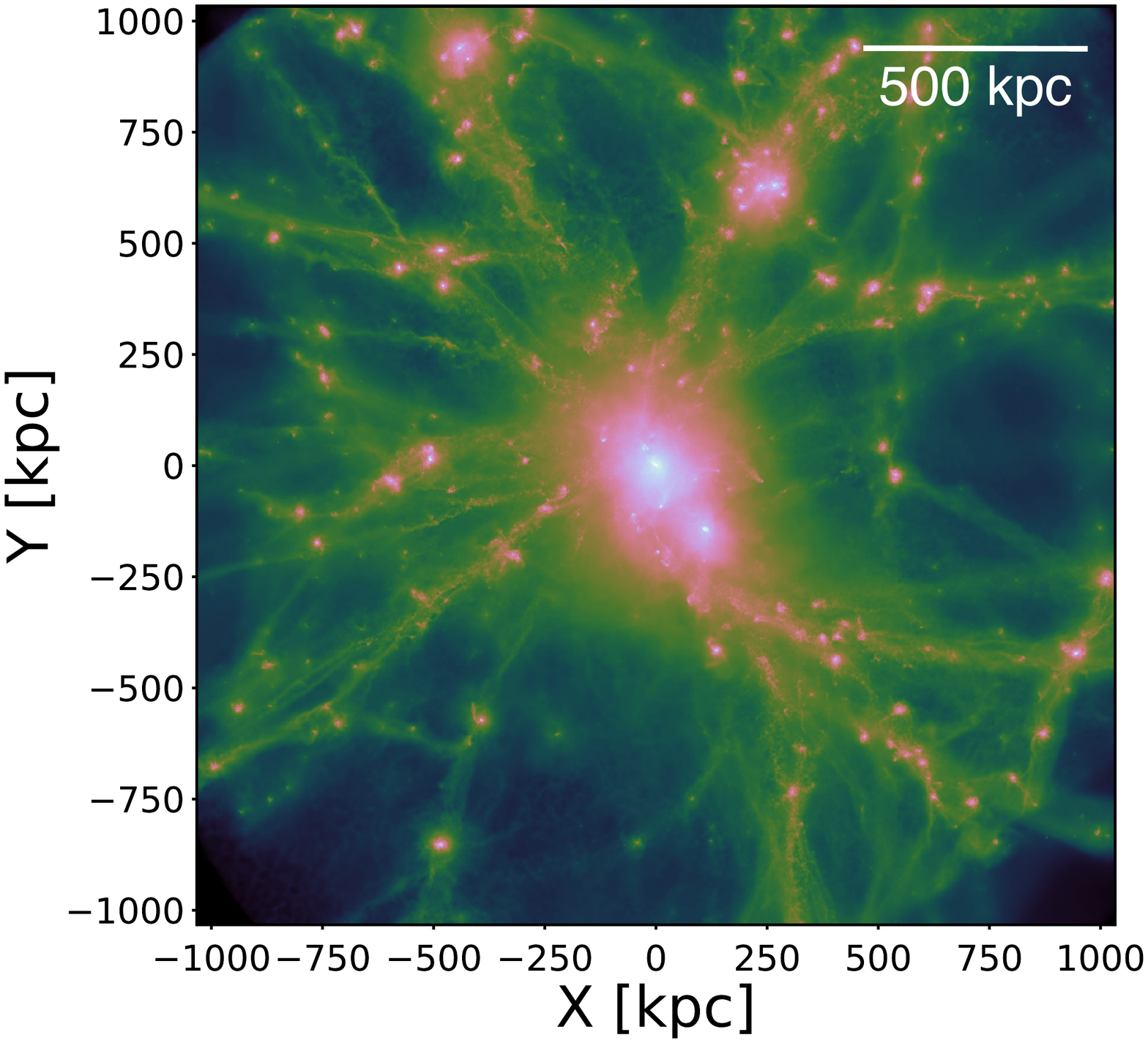}
\includegraphics[trim=60mm 25mm 55mm 25mm, clip, width=45mm]{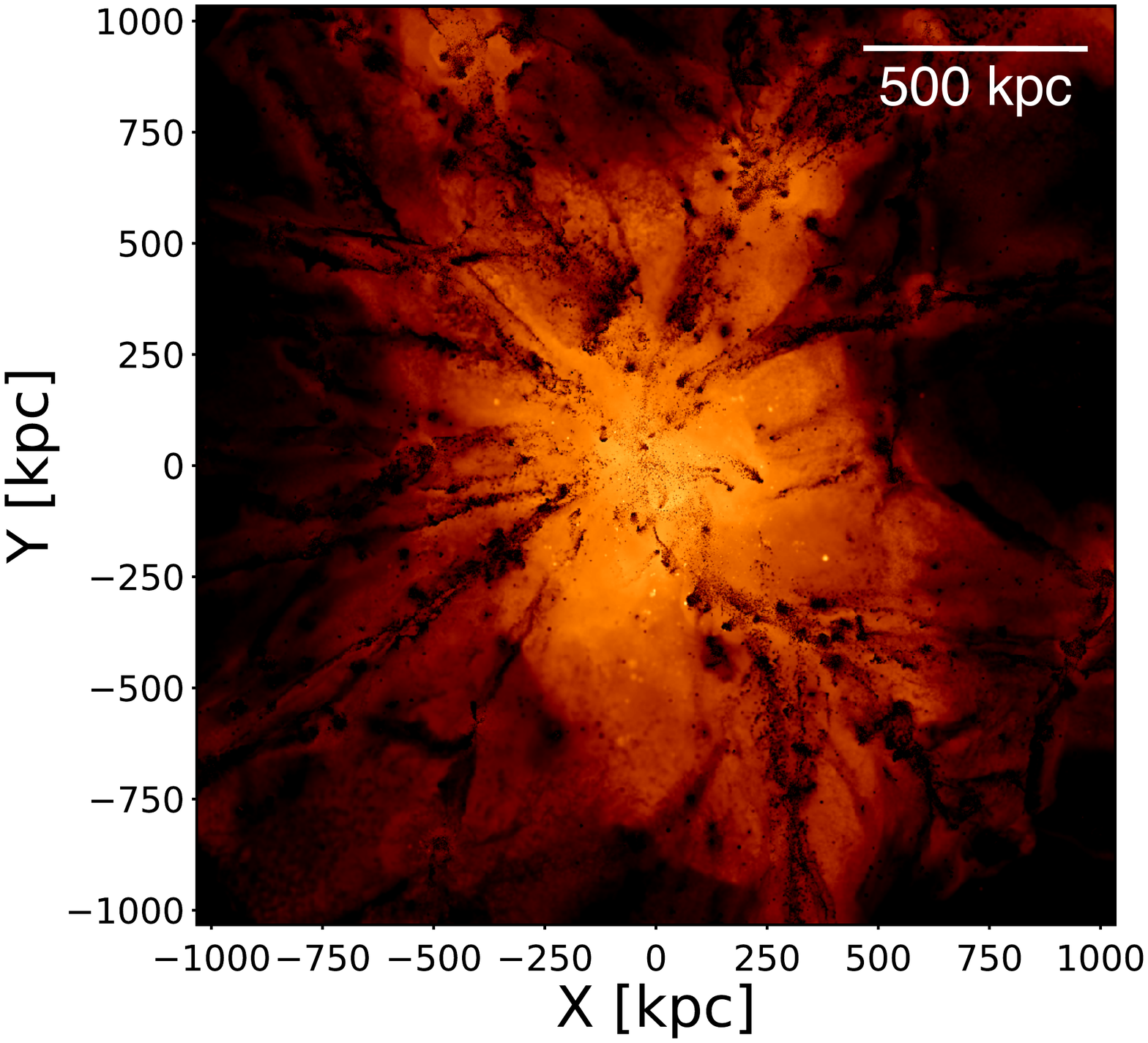}

\includegraphics[trim=60mm 25mm 55mm 25mm, clip, width=45mm]{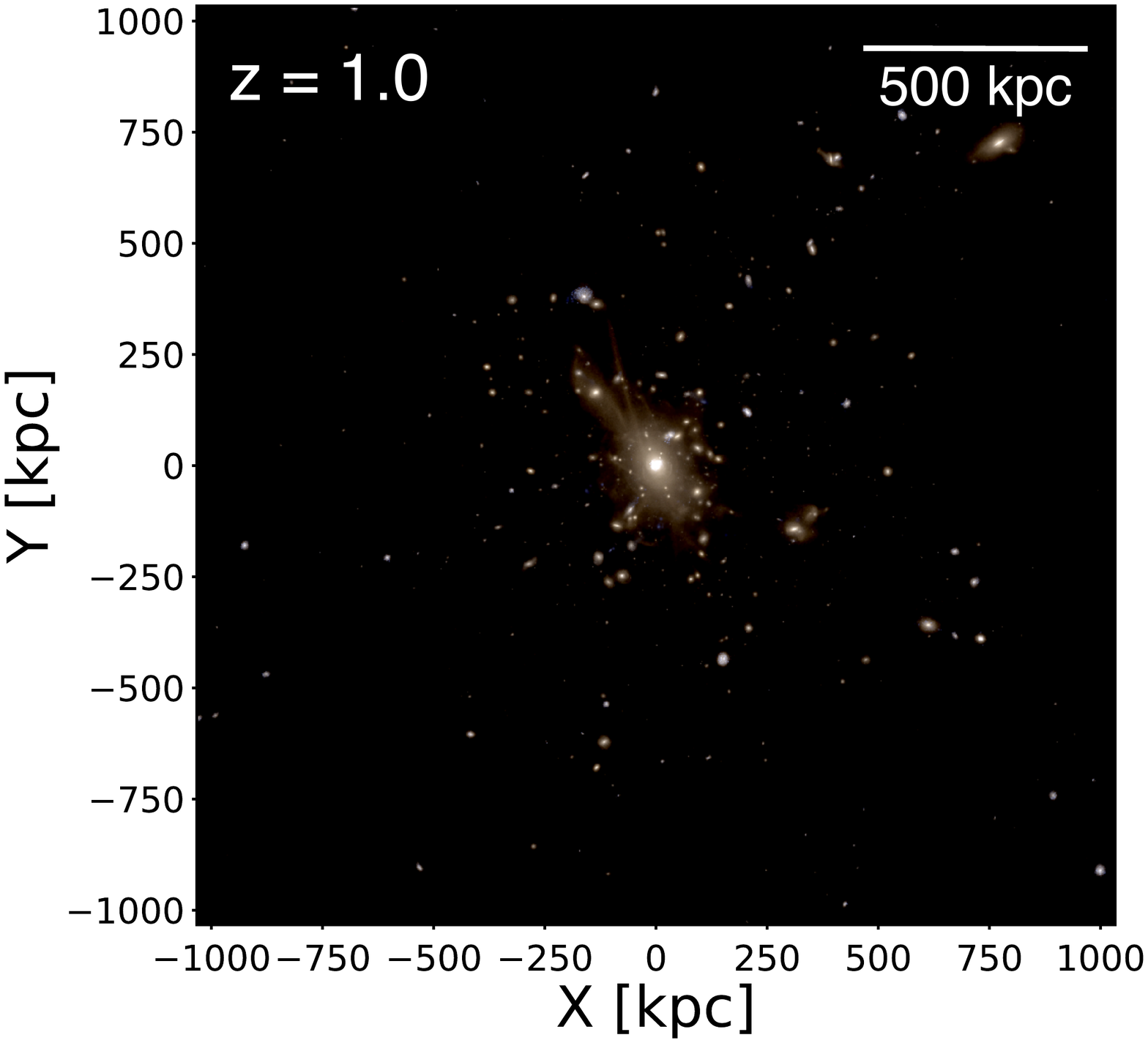}
\includegraphics[trim=60mm 25mm 55mm 25mm, clip, width=45mm]{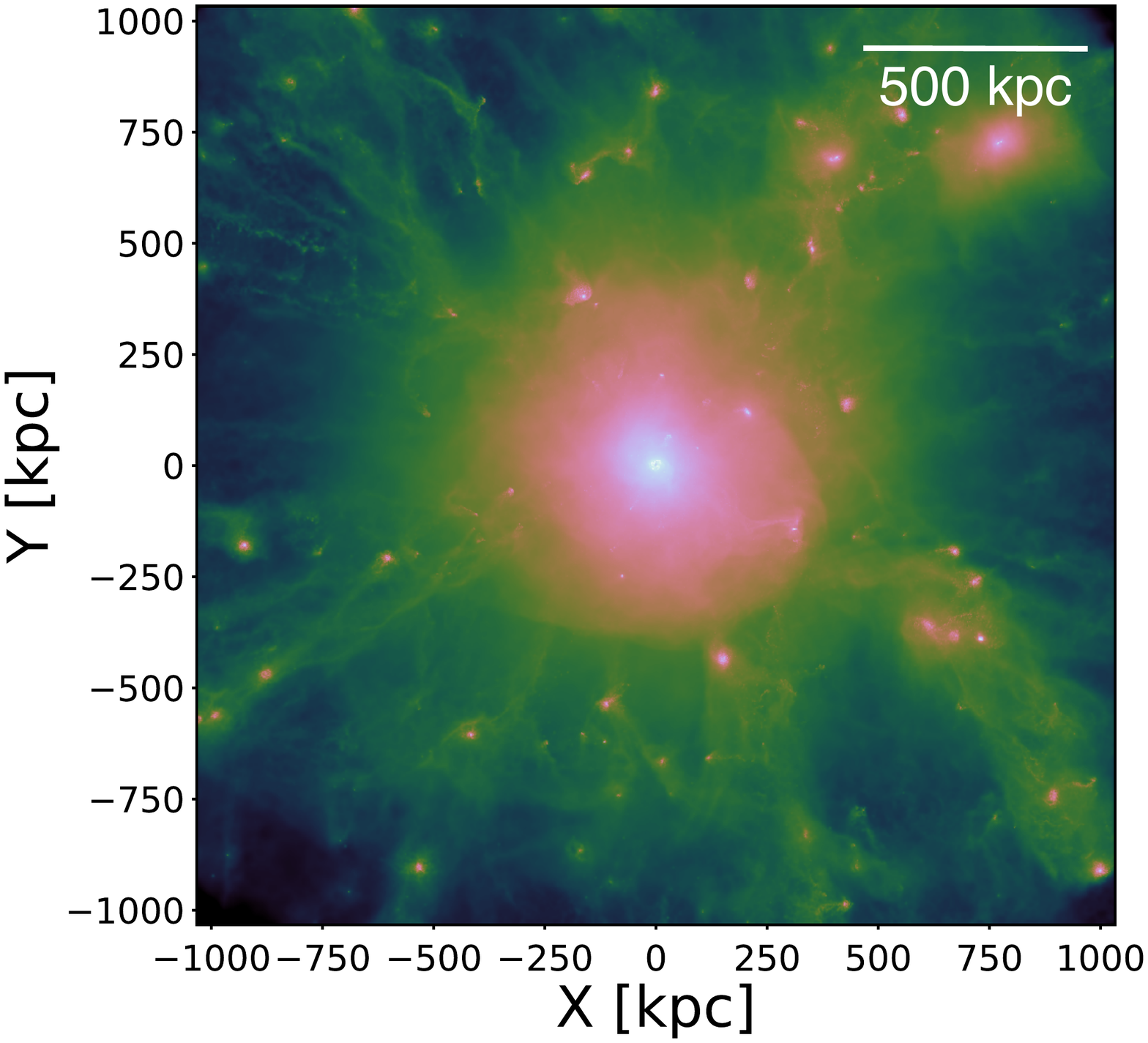}
\includegraphics[trim=60mm 25mm 55mm 25mm, clip, width=45mm]{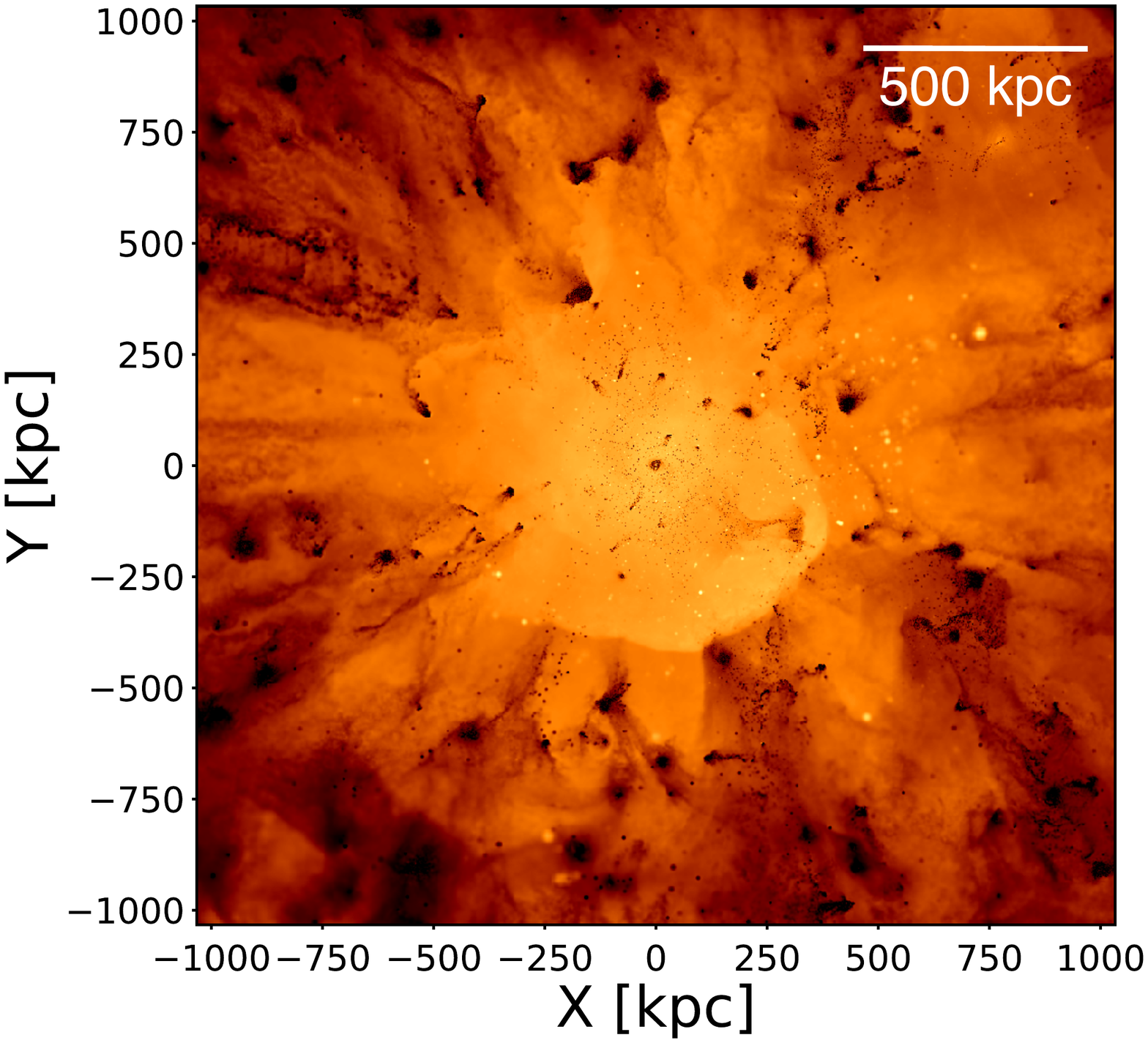}

\includegraphics[trim=60mm 25mm 55mm 25mm, clip, width=45mm]{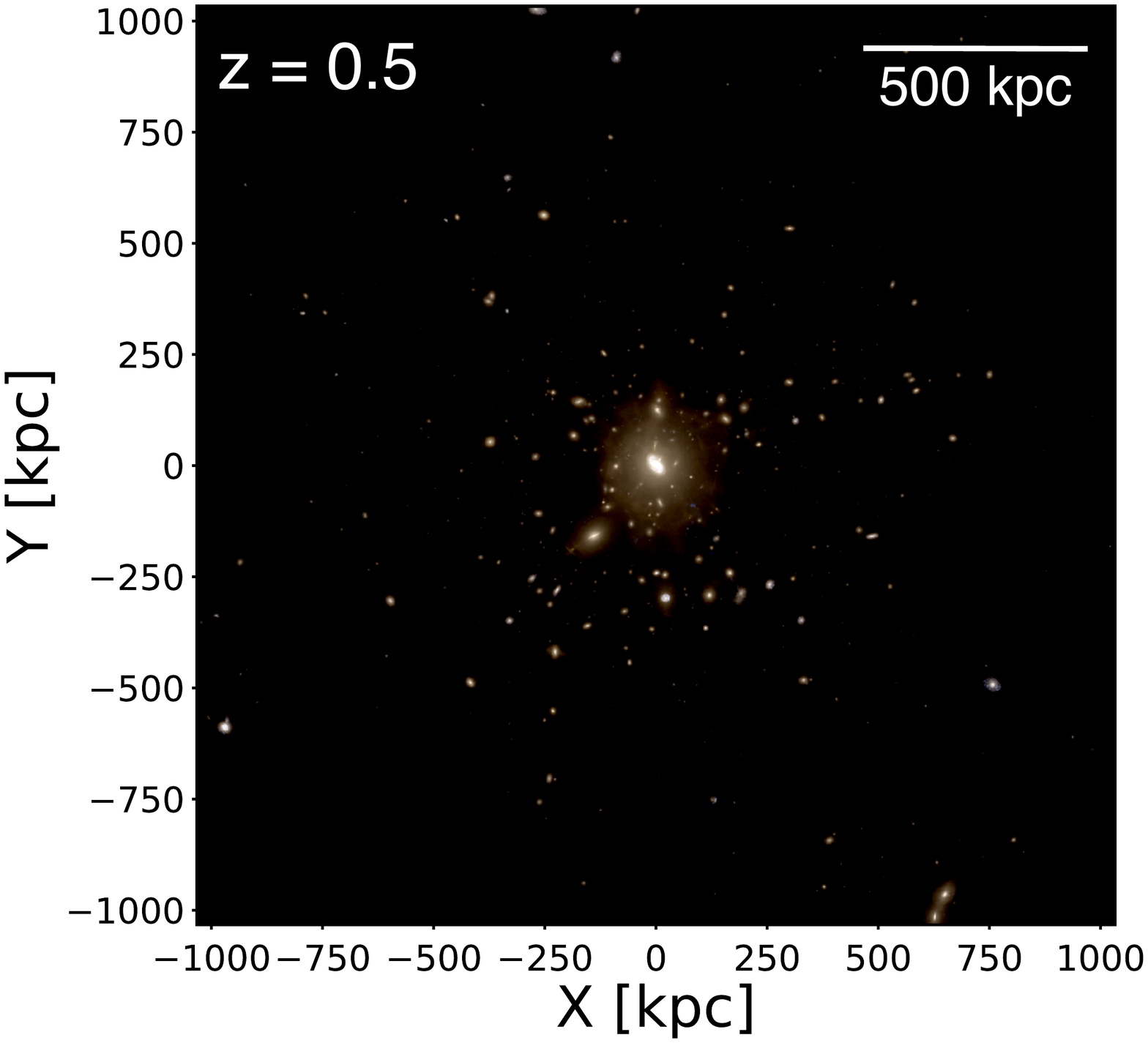}
\includegraphics[trim=60mm 25mm 55mm 25mm, clip, width=45mm]{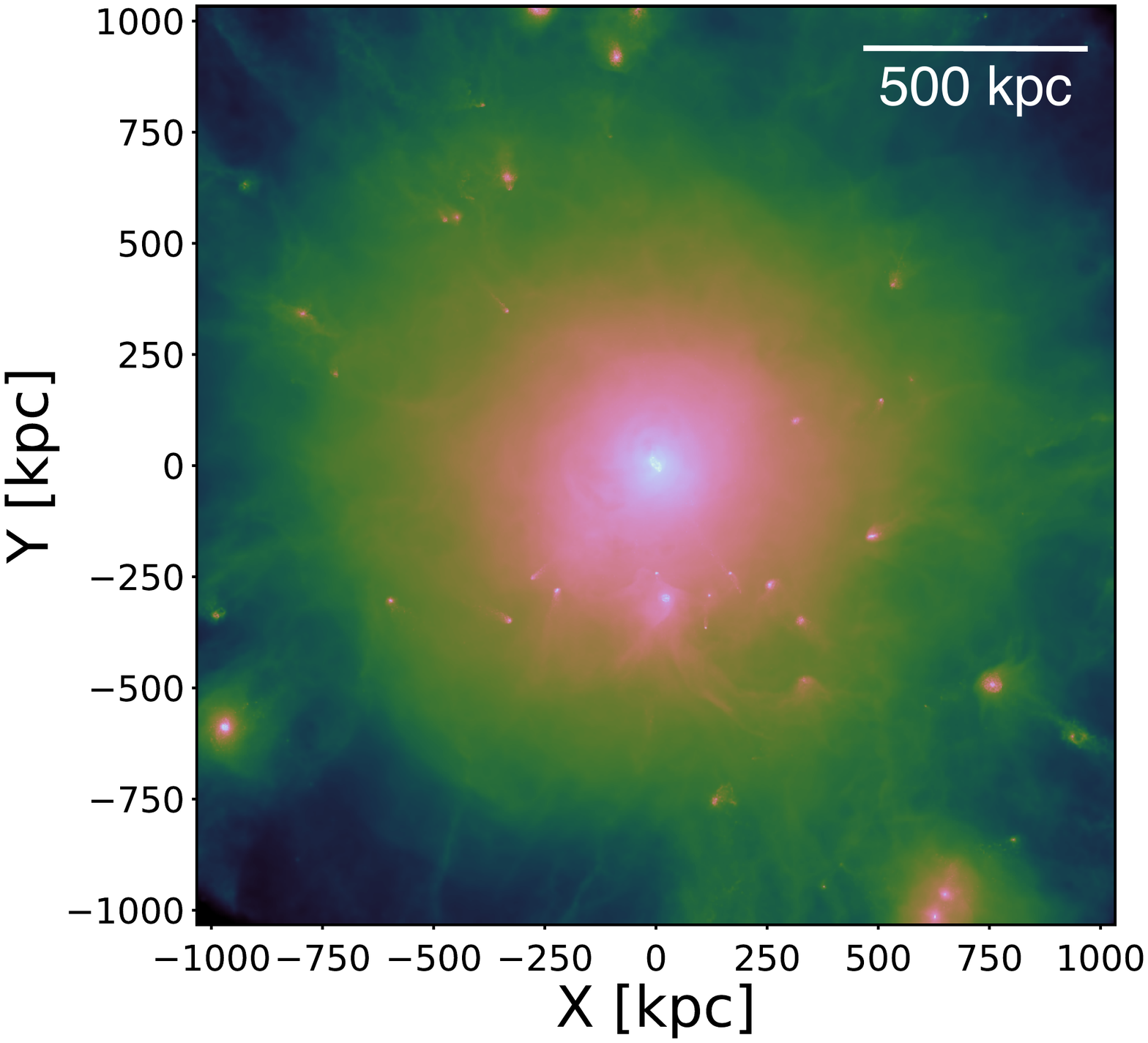}
\includegraphics[trim=60mm 25mm 55mm 25mm, clip, width=45.mm]{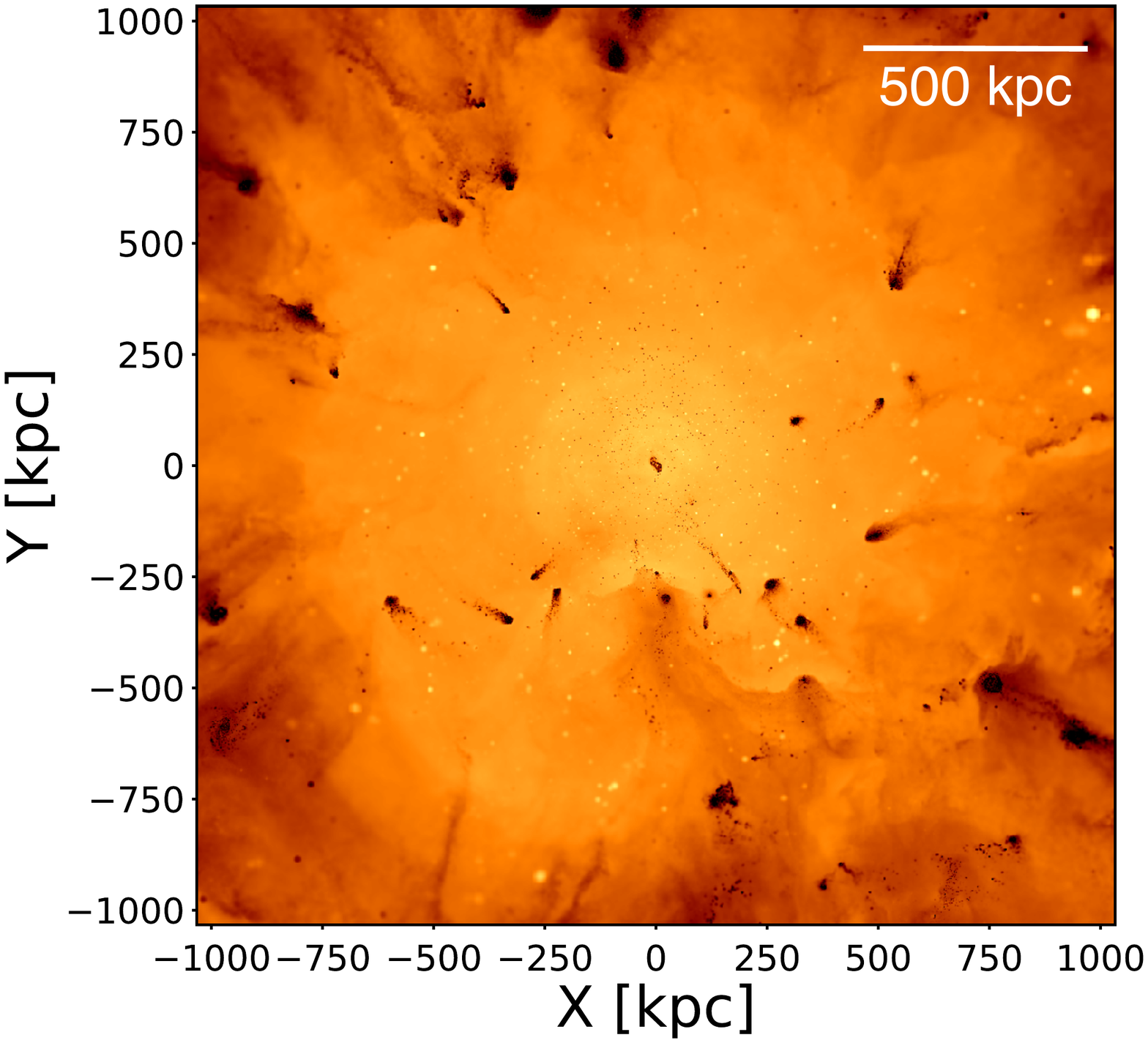}

\includegraphics[trim=60mm 26mm 54mm 25.5mm, clip, width=45.15mm]{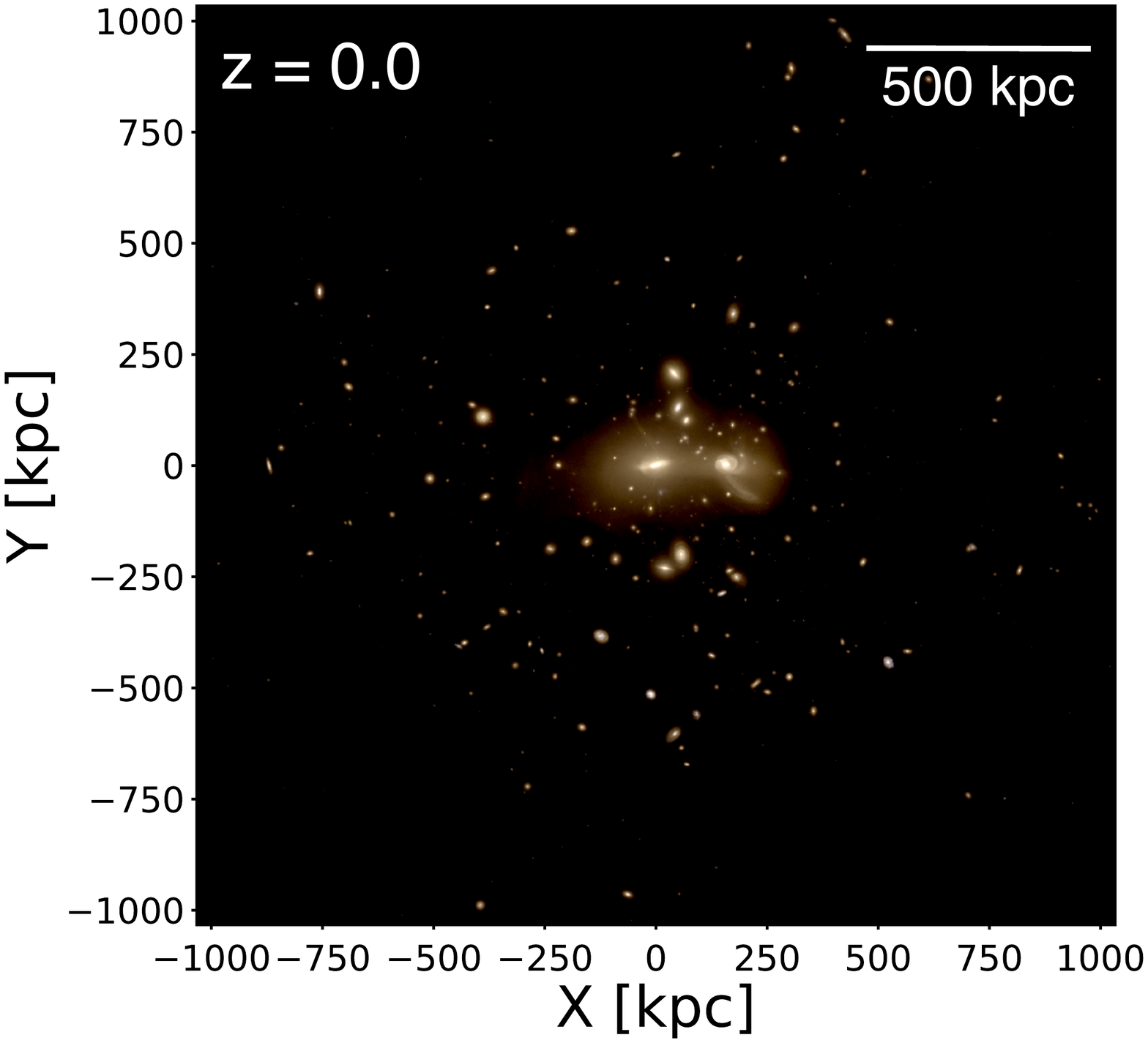}
\includegraphics[trim=60mm 26mm 54mm 25.5mm, clip, width=45.15mm]{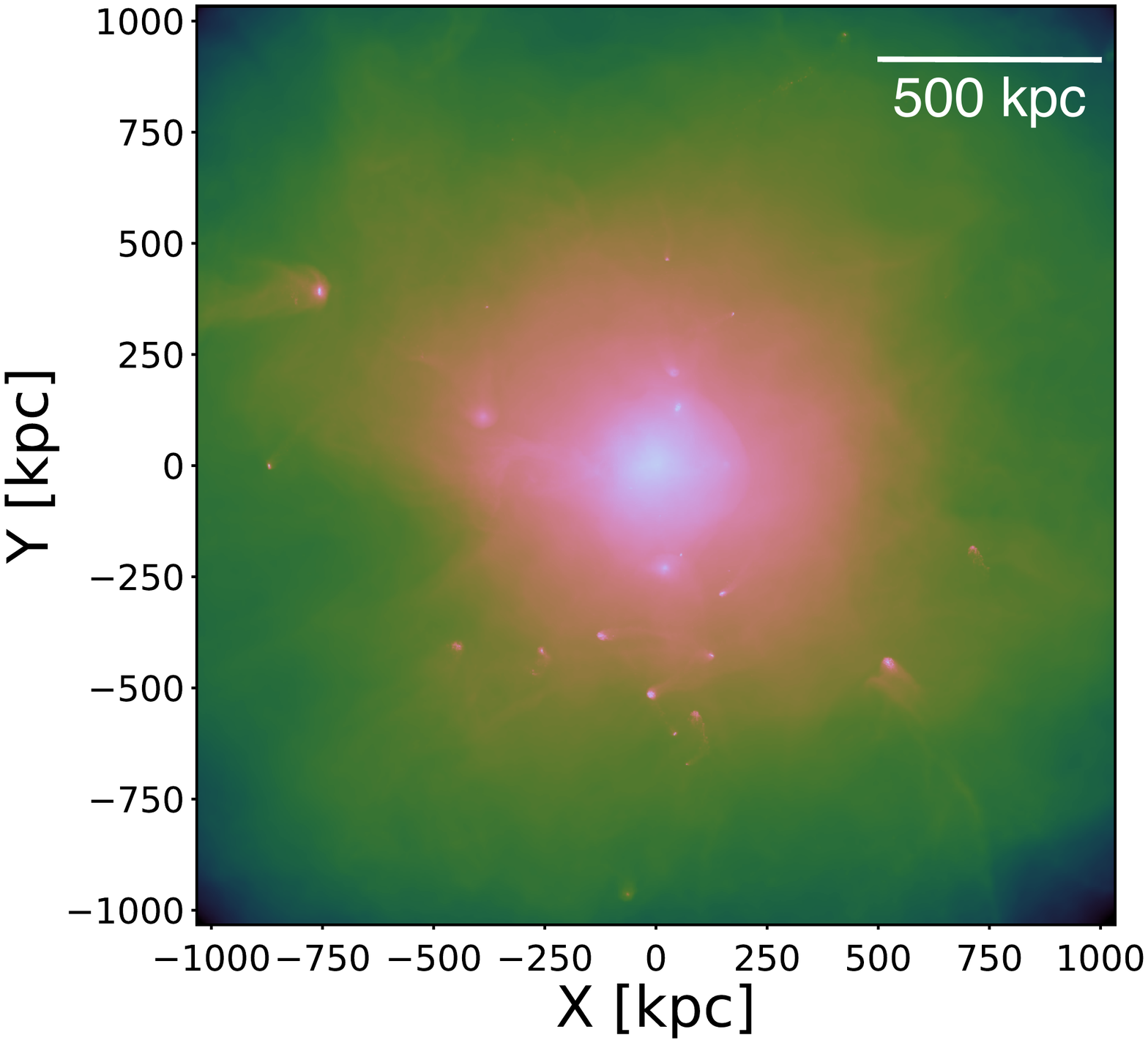}
\includegraphics[trim=60mm 26mm 56mm 24mm, clip, width=44.75mm]{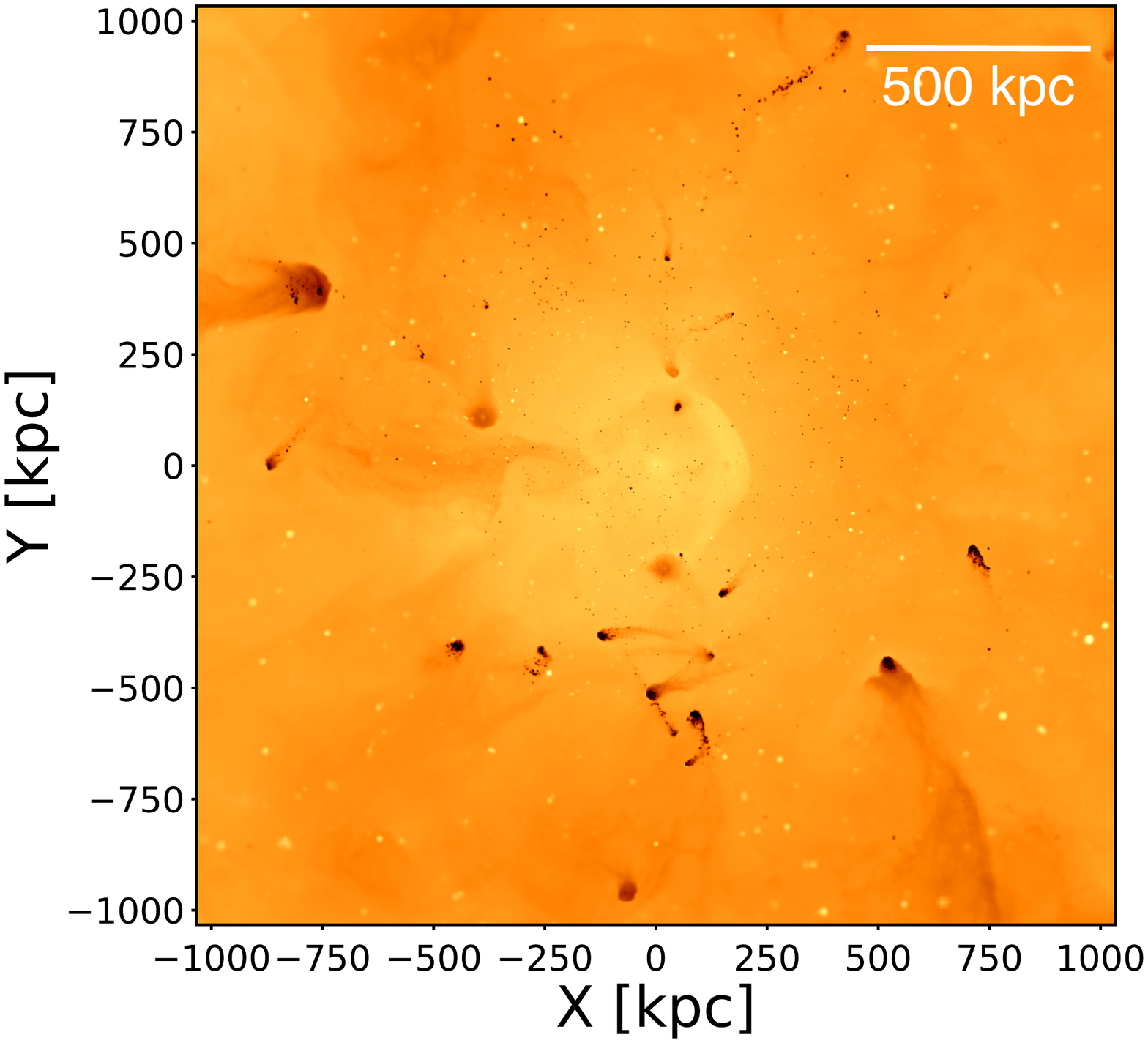}
\caption{{\sc Evolution of Gas and Stars in RomulusC}. Four snapshots of {\sc RomulusC} taken (from the top) at $z = 2$, 1, 0.5, and 0 showing stars (left), gas column density (middle) and gas temperature (right). The temperature is averaged along the line  of sight and weighted by $\rho^2$. All of the plots show the same physical region encompassing what will be $R_{200}$ at $z=0$ (the scales are also in physical units). The stars are shown in UVJ colors assumining a Kroupa IMF. The density ranges from $1\times10^{-5}$ g/cm$^{-2}$ (blue/black) to 0.1  g/cm$^{-2}$ (pink/white). The temperature ranges from $5\times10^4$ K (black) to $5\times10^8$ K (yellow). At early times there are a lot of cold gas filaments penetrating the halo, which triggers more star formation in cluster galaxies. At later times, the halo is massive enough to prevent such filaments from penetrating far into the halo. Ram pressure stripping is evident in the extended tails of cold gas in the cluster galaxies. At $z=0$ there are two galaxies interacting near the center. The less massive of the two, which makes its first pericenter passage into the center of the cluster around $z=0.17$, is half the stellar mass of the BCG and comes from a halo one eighth the mass of the main halo.}
\label{time_evol}
\end{figure*}

\section{The Simulation}


{\sc RomulusC} is a cosmological zoom-in simulation of a small galaxy cluster with $z=0$ total mass of $1.5 \times 10^{14}$ M$_{\odot}$. Halos of this mass are generally only found in uniform volume simulations of at least 50$^3$ Mpc$^3$. Compared with all other modern cosmological simulations that include halos of this mass, {\sc RomulusC} has significantly higher mass and spatial resolution (see table 1) and includes, by $z=0$, 227 galaxies within the virial radius with M$_{\star}>10^8$ M$_{\odot}$, all of which are well resolved with more than $10^4$ particles per halo or more.

The initial conditions for {\sc RomulusC} are extracted from a $50^3$ Mpc$^3$ uniform volume, DM-only simulation. The `zoom-in` volume renormalization technique of \citet{katz93} is used to resimulate at higher resolution the Lagrangian region associated with the most massive halo in the DM-only volume with full hydrodynamic treatment using the new Tree+SPH code {\sc ChaNGa} \citep{changa15}. Due to its ability to scale efficiently up to 500,000 cores, {\sc ChaNGa} is uniquely well suited to run such a large-scale, high resolution simulation. {\sc ChaNGa} includes standard physics modules and physically motivated ingredients previously used in { \sc GASOLINE} \citep{wadsley04,wadsley08,wadsley17} such as a cosmic UV background that includes self-shielding \citep{pontzen08}, star formation, `blastwave' SN feedback \citep{Stinson06}, and low temperature metal cooling. {\sc ChaNGa} includes an updated SPH implementation that eliminates artificial gas surface tension through the use of a geometric mean density in the SPH force expression  
\citep{ritchie01,changa15,governato15}, allowing for the accurate simulation of shearing flows with Kelvin-Helmholtz instabilities. Critical to this work, the simulation includes an updated implementation of turbulent diffusion \citep{wadsley17}, shown to be an important physical process for attaining realistic entropy profiles in galaxy cluster cores \citep{wadsley08} as well as metal distributions in galaxies \citep{shen10}. Further, the simulations include a gradient-based shock detector, a time-dependent artificial viscosity, and an on-the-fly time-step adjustment system, the combination of which allows for a more realistic treatment of both weak and strong shocks \citep{wadsley17}.

{\sc RomulusC} is run with the same hydrodynamics, sub-grid physics, resolution, and cosmology as the {\sc Romulus25} simulation \citep{tremmel17}. The cosmology is $\Lambda$CDM with cosmological parameter values following the most recent results from Planck \citep[$\Omega_0=0.3086$, $\Lambda=0.6914$, h$=0.6777$, $\sigma_8=0.8288$;][]{planck16}. The simulation has a Plummer equivalent force softening, $\epsilon_g$, of $250$ pc (a spline softening of 350 pc is used, which converges to a Newtonian force at $2\epsilon_g$). Unlike many similar cosmological runs, the dark matter distribution is {\it oversampled}, such that we
simulate $3.375$ times more dark matter particles than gas particles, resulting in a dark matter particle mass of $3.39 \times 10^5$M$_{\odot}$ and gas particle mass of $2.12 \times 10^5$M$_{\odot}$. This allows us to decrease numerical noise to more accurately track black hole dynamics \citep{tremmel15}. At this resolution, we confidently resolve halos as small as $\sim3\times10^9$ M$_{\odot}$ with at least $10^4$ particles. 

\begin{figure}
\centering
\includegraphics[trim=15mm 2mm 20mm 0mm, clip, width=90mm]{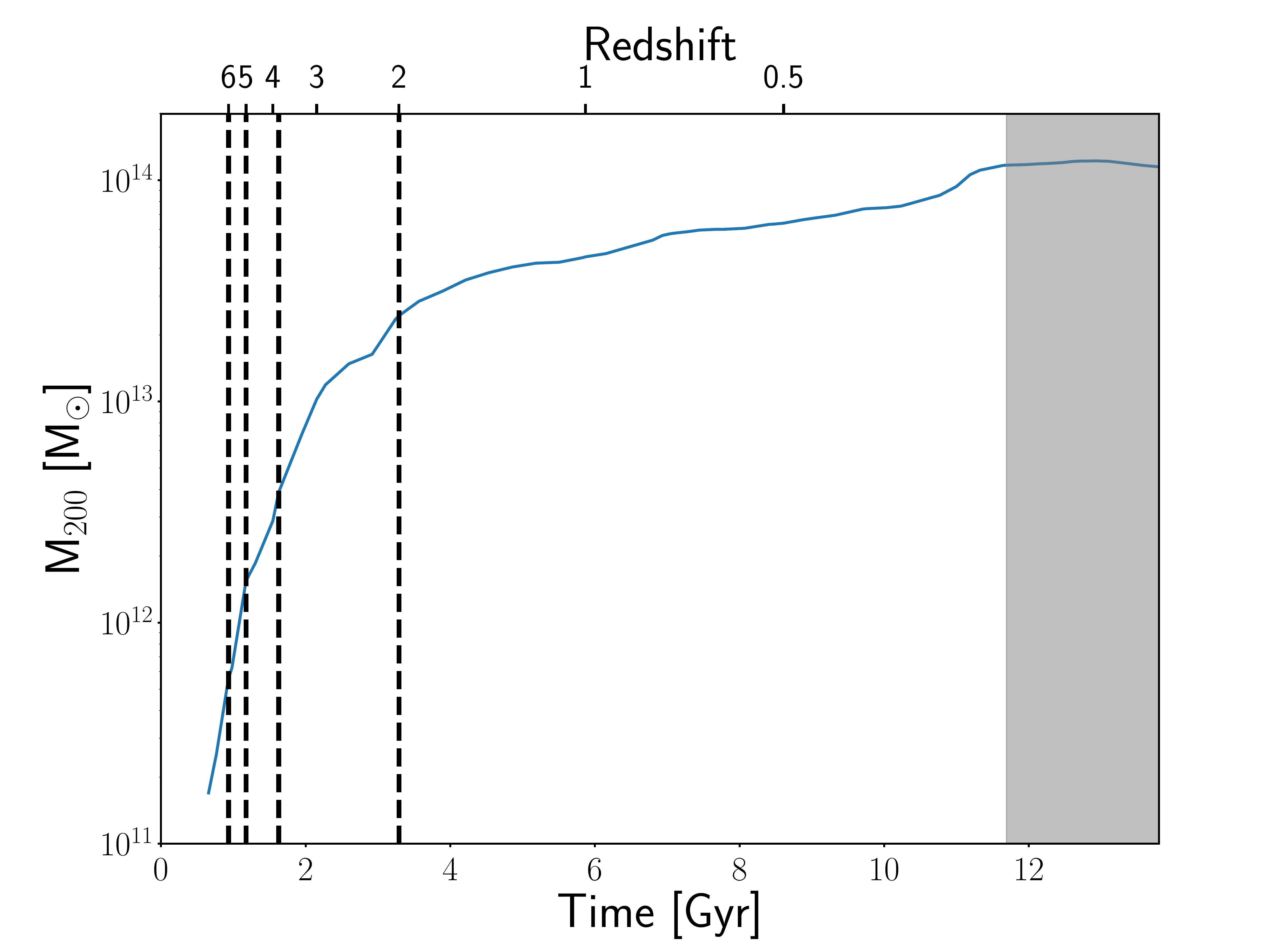}
\caption{{\sc Growth of the RomulusC Cluster}. The Growth in halo mass of {\sc RomulusC}'s main progenitor over cosmic time. By $z\sim2$ the main progenitor to {\sc RomulusC} is as massive as the most massive halo at $z=0$ in our uniform volume simulation, {\sc Romulus25}. Thus, {\sc RomulusC} represents a unique environment in which to examine galaxy evolution and an important addition to our simulation suite.  Mergers of total mass ratio 1:10 or higher are shown as vertical dashed lines. The grey band represents the ongoing merger starting around $z=0.17$, which has a 1:8 total mass ratio and a 1:2 stellar mass ratio. In some of our analysis we will ignore the time period associated with this merger.}
\label{mass_growth}
\end{figure}

Figure~\ref{time_evol} shows the evolution of the stellar and gaseous components of the cluster over time. Figure~\ref{mass_growth} shows the mass evolution of the cluster's main progenitor halo, which reaches $10^{13}$ M$_{\odot}$ by $z = 2$. Figure~\ref{mass_growth} also shows the times for mergers (vertical dashed lines) of total mass ratio at least 1:10. The grey region represents an on-going merger of total mass ratio approximately 1:8 that starts at $z\sim0.2$ and continues through $z=0$. The galaxy involved in this merger can clearly be seen in the bottom left panel of Figure~\ref{time_evol}. The galaxy has a stellar mass that is about half of that of the central brightest cluster galaxy. We note that it may be possible that the lower mass galaxy could be mistaken as the brightest cluster galaxy, though for simplicity we only consider the most massive galaxy to be the BCG in the following analysis. This is also the galaxy located at the potential minimum of the main progenitor branch of the main halo throughout the simulation. We also choose to ignore much of the evolution during this merger and instead focus on the cluster and BCG evolution when the cluster environment is comparatively relaxed prior to the merger event. As briefly discussed in \S3 and \S4.1.2, the cluster would be classified as a cool core cluster for much of the simulation, with a declining entropy profile down to $\sim0.01R_{200}$ and sub-Gyr cooling times in the core, but the merger destroys this structure, leaving the cluster in a non-cool-core state from $z\sim0.2-0$. Because the merger is ongoing at $z=0$, we cannot be sure if this is a permanent transformation or a transient state due to the perturbation away from equilibrium  \citep{poole06,poole08}. This merger event and the ensuing cool core disruption will be explored in more detail in future work.

\subsection{Star Formation and Feedback}

Star formation and associated feedback from supernovae are crucial processes that require sub-grid models in cosmological simulations like {\sc RomulusC}.  As in previous work \citep{Stinson06} for runs at this resolution, star formation (SF) is regulated with parameters that encode star formation efficiency in dense gas, the coupling of SNe to the ISM, and the physical conditions required for star formation:

\begin{enumerate}
\setlength\itemsep{1em}
\item the normalization of the SF efficiency, c$_{\star} = 0.15$, and formation timescale, $\Delta t = 10^6$yr, are both used to calculate the probability of creating a star particle from a gas particle that has a dynamical time $t_{dyn}$

\begin{equation}
p =\frac{m_{gas}}{m_{star}}(1 - e^{-c_{\star} \Delta t /t_{dyn}}), 
\end{equation}
 
\item The fraction of SNe energy that is coupled to the ISM, $\epsilon_{SN} = 0.75$

\item the minimum density, n$_\star = 0.2$ cm$^{-3}$, and maximum temperature, T$_\star = 10^4$ K, thresholds beyond which cold gas is allowed to form stars.
 
 \end{enumerate}




SN feedback is included via the `blastwave' implementation \citep{Stinson06}, and gas cooling is regulated by metal abundance as in \citet{eris11} as well as SPH hydrodynamics and both thermal and metal diffusion as described in \citet{shen10} and \citet{governato15}. We assume a Kroupa IMF \citep{kroupa2001} with associated metal yeields.

An important limitation of the {\sc Romulus} simulations in modeling higher mass halos is that they only include low temperature metal cooling rather than a full implementation of metal line cooling, a major coolant for higher temperature gas in groups and clusters. Metal line cooling has been shown to affect the accretion of gas onto galaxies in the centers of massive halos, though feedback from AGN is also critically important in regulating this process \citep{vandevoort11}. The reason to exclude metal line cooling was made based on results from \citet{christensen14b}, who found that the inclusion of metal cooling without molecular hydrogen physics and more detailed models of star formation resulted in overcooling in galaxies.  The higher temperature metal cooling requires that a true multi-phase interstellar medium (ISM) is maintained to keep the cooling at intermediate temperatures from running away. This means higher resolution and clumpy star formation from molecular hydrogen is needed. Due to the scale of the {\sc Romulus} simulations such resolution criteria are not met and so metal cooling is not implemented.


One possible solution to the overcooling problem would be to boost supernova feedback efficiency \citep[e.g.][]{shen12,dallavecchia12,eagle15, sokolowska16,sokolowska17}, but this will not necessarily provide a realistic ISM or CGM/ICM. \citet{sokolowska16} showed for zoom-in simulations of Milky Way-like galaxies that the inclusion of metal-line cooling and enhanced SN feedback, while producing a reasonable stellar mass, resulted in unrealistic ISM and CGM properties compared to simulations without high temperature metal cooling. \citet{christensen14b} find  that ISM models that include both metal lines and H$_2$ physics result in galaxies with star formation histories and outflow rates more similar to primordial cooling runs than to simulations with metal lines and no H$_2$. Another solution would be to only allow metal cooling in diffuse gas that is less likely to be multiphase, but determining an arbitrary threshold where unresolved multiphase structure exists is difficult, particularly at the boundary of the ISM and the CGM/ICM where the effects will be most important for galaxy evolution, a major focus of the {\sc Romulus} simulations. Given the severe flaws in both of these solutions, we choose to not include metal line cooling in {\sc Romulus}.


\subsection{Black Hole Physics}

One of the major improvements in the {\sc Romulus} Simulations compared to previous work is the more sophisticated modeling of the seeding, fueling, feedback, and dynamics of black holes. Below we summarize the model used in this work focusing particularly on the most relevant aspect: feedback from growing SMBHs. For more details regarding the model and the parameter optimization used to set their free parameters the reader is referred to \citet{tremmel17}.

\subsubsection{SMBH Seeding}

SMBHs are seeded in the simulation based on gas properties, forming in rapidly collapsing, low metallicity regions in the early Universe. The goal is to better approximate theoretical models of SMBH formation where they form at high redshift in lower mass, atomic cooling halos \citep[e.g.][]{schneider02,BrommLoeb2003,LN2006BH}. We isolate pristine gas particles (metalicity $<3\times10^{-4}$) that have reached densities 15 times higher than what is required by our star formation prescription without forming a star or cooling beyond $9.5 \times 10^3$ K (just below the temperature threshold used for star formation, $10^{4}$ K). These regions are collapsing on timescales much shorter than the cooling and star formation timescales and are meant to approximate the regions that would result in SMBHs of mass $>10^6 M_{\odot}$, regardless of the details of their formation mechanism. The initial seed SMBH mass is set to $10^6$ M$_{\odot}$ and is justified by our choice of formation criteria, which would produce black holes that are able to attain higher masses quickly, as there is a lot of dense, collapsing gas nearby that is unlikely to form stars. \citet{tremmel17} show how this method forms most SMBHs within the first Gyr of the simulation, compared with the later seeding times inherent to more common approaches that seed SMBHs based on halo mass thresholds \citep[e.g.][]{springel05b,diMatteo2008,genel14,eagle15}. Unlike these other approaches, our model produces an evolving occupation fraction. At early times, small halos (M$_{vir} \sim 10^{8-10}$ M$_{\odot}$) host newly seeded SMBHs and the occupation fraction then evolves due to hierarchical merging. For example, in {\sc Romulus25} at $z=5$ the SMBH occupation fraction of $\sim10^{10}$ M$_{\odot}$ halos is $\sim60\%$ but at $z=0$ the occupation fraction for halos of this mass drops to $\sim10\%$.

\subsubsection{SMBH Dynamics}

We incorporate the dynamical friction sub-grid model presented in \citet{tremmel15} that permits the more accurate tracking of SMBH orbits within galaxies. The model approximates the unresolved dynamical friction that should act on SMBHs by integrating the Chandrasekar formula from the 90 degree deflection radius out to the gravitational softening length. Close encounters that should occur at this scale are important for dynamical friction, but are unresolved due to both the gravitational softening and limited mass resolution. This approach has been shown to produce a more realistic dynamical evolution of SMBHs at the resolution of {\sc RomulusC} \citep{tremmel15,tremmel18}. 

SMBH dynamics can be important in galaxy clusters since many mergers at high redshift will eventually make up the central galaxy. There are also many glancing encounters between cluster member galaxies \citep{moore96,moorelake98,moore99b}. How SMBHs within galaxies respond to such perturbations can have an effect on their growth and feedback history and so it is important to follow their dynamics realistically. This includes their gradual sinking to halo center resulting from a merger as well as becoming temporarily perturbed away from halo center due to a merger or fly-by interaction. This dynamical evolution can also affect the ability for SMBHs to grow via SMBH mergers or accretion \citep{dicintio17,tremmel18b}. We refer the reader to \citet{tremmel15} for more details about the dynamical friction sub-grid model.

\subsubsection{Accretion and Feedback}

Accretion of gas onto SMBHs is governed by a modified Bondi-Hoyle prescription, re-derived using the same energy balance argument as Bondi-Hoyle but including the additional angular momentum support present for rotating gas. Despite the fact that {\sc RomulusC} has significantly higher mass and spatial resolution compared with other cosmological cluster simulations, the Bondi radius of even the most massive SMBHs remains unresolved. In the simulation, the properties of gas particles are defined by the average values smoothed over a kernel of typical size at least 10\% the gravitational softening, $\epsilon_g$. However, because in reality dense, cool gas has a multiphase structure on scales well below our resolution limit, this will result in gas densities and temperatures that are systematically under and over estimated respectively compared with what they should be nearby the SMBH. This will lead to artificially lower accretion rate estimates due to the failure of the simulation to resolve the multiphase nature of dense gas. We therefore follow \citet{BoothBH2009} and employ a density dependent multiplicative boost factor to our modified Bondi-Hoyle accretion rate calculation in order to account for this unresolved multiphase structure and its impact on gas accretion onto SMBHs.

Taking both angular momentum and unresolved gas structure into account, the accretion equation has the following form:

\begin{equation}
\begin{aligned}
\dot{M}  = \alpha & \pi(GM)^2 \rho \,\times\, \begin{cases}
\frac{1}{(v_{\mathrm{bulk}}^2+c_s^2)^{3/2}} & \text{ if } v_{\mathrm{bulk}}>v_{\theta} \\ \\
\frac{c_s}{(v_{\theta}^2+c_s^2)^{2}} & \text{ if }  v_{\mathrm{bulk}}<v_{\theta}
\end{cases} \\ \\
\alpha & = \begin{cases} (\frac{n}{n_{th,*}} )^\beta & \text{ if } n > n_{*} \\ \\
1 & \text{ if } n < n_{*}
\end{cases}.
\end{aligned}
\end{equation}

\noindent Values for density ($\rho$), number density ($n$), and sound speed ($c_s$) of gas near the SMBH are estimated from smoothing over the 32 nearest gas particles and accretion is not allowed to occur from gas particles farther than $4\epsilon_g$ (1.4 kpc). The tangential velocity, $v_{\theta}$, is derived from the \textit{resolved} kinematics of nearby gas particles and compared to $v_{\mathrm{bulk}}$, the overall bulk motion of the gas that already enters into the Bondi-Hoyle model. This bulk motion is taken to be the minimum relative velocity of the 32 nearest gas particles. When either the bulk motion or internal energy of the gas dominates over rotational motion, the accretion model reverts to the normal Bondi-Hoyle prescription. In both cases, we add the boost factor, $\alpha$, calculated by comparing the number density of nearby gas particles to the threshold for star formation, $n_*$, defined in \S2.1. For lower densities, we assume that the gas is not sufficiently multiphase to require such a boost, as in \citet{BoothBH2009}. How much this boost increases with density is governed by $\beta$, a free parameter, which is set to 2.


An accreting SMBH converts a fraction of the accreted mass into energy that is transferred to nearby gas particles. This feedback efficiency is determined by two separate parameters: $\epsilon_r$, the radiative efficiency, and $\epsilon_f$, the efficiency of energy coupling to nearby gas. The overall feedback efficiency of the SMBH is the product of these two values. For the purposes of optimizing free parameters, we assume that $\epsilon_r$ is 0.1 and treat $\epsilon_f$ as a free parameter set to $0.02$. It should be noted that $\epsilon_r$ and $\epsilon_f$ are not totally degenerate. The radiative efficiency is used to determine the Eddington limit, which we assume is the highest accretion rate attainable by any SMBH in the simulation.

Accretion and feedback are calculated during each SMBH timestep and are meant to represent the total amount of mass and feedback imparted during that time. While the SMBH is growing, thermal energy is transferred to the 32 nearest gas particles and they are not allowed to undergo cooling for a time equal to the SMBH's timestep. \textit{SMBHs are continuously placed on the lowest global time-step in the simulation.} For the majority of the simulation, this ends up being at most $10^5$ years and is more typically $10^3-10^4$ years. Further, gas particles within each SMBH's smoothing region (i.e. the 32 closest particles) are forced to be on a time-step within a factor of 2 of the SMBH. These time-step criteria, as well as the brief cooling shut-off period, ensure 1) a more continuous sampling of accretion and feedback processes and 2) that the gas does not artificially radiate away the energy transferred by the SMBH due to limited spatial and time resolution.

\subsubsection{Comparison with other implementations of feedback}

Often, the choice to impart feedback through kinetic rather than thermal coupling is made due to numerical effects. Thermal energy is radiated away very easily by the dense gas that is generally near growing SMBHs. This is not physical, but rather the result of limited spatial, mass, and time resolution. When massive, dense gas particles are given a large amount of energy instantaneously and then allowed to cool based on their temperature and density properties over the course of their next timestep, they will invariably lose that energy well before their internal properties are able to respond to the energy injection \citep{katz92}. By forcing the SMBHs and surrounding gas particles to have small time-steps and by turning off cooling for a short time, we are able to better approximate a continuous transfer of energy between an accreting SMBH and the surrounding gas. This allows gas that is receiving feedback to expand and become buoyant, driving large-scale, thermally driven, collimated outflows (see \S4.1).

Our adopted feedback model is empirically supported by velocity profiles of observed AGN-driven outflows, which are consistent with being energy conserving, rather than momentum conserving \citep{gaspari17}. Gas that receives feedback on the smallest resolved scales of the simulation is really gas that has been entrained in an outflow initially driven by unresolved processes. Because energy is conserved in this outflow, it makes sense to inject it via thermal energy along with a cooling shutoff to ensure energy conservation. The outflow is then naturally driven by the hot, expanding gas affected by the feedback, similar to `blastwave' SN feedback \citep{Stinson06}.

Many simulations employ a two-mode feedback prescription \citep[e.g.][]{sijacki07,dubois12,vogelsberger2013,weinberger17} whereby a transition occurs between a thermally driven mode of feedback during a radiatively efficient `quasar' mode of SMBH activity to a radiatively inefficient `radio' mode of SMBH activity, which captures momentum-driven feedback and is needed to reproduce the radio lobes observed in massive galaxies. In such models, the efficiency of feedback increases for high mass galaxies/SMBHs, which is needed in order to effectively prevent over-cooling. In effect, the main difference between these models and our own is 1) our overall feedback efficiency is held constant, 2) we always transfer energy thermally, and 3) feedback is implemented in the same way for all galaxies and does not change with redshift, SMBH mass, or galaxy mass.

The reasoning to impart kinetic feedback is supported by the existence of relativistic jets, often associated with radio `bubbles' and X-ray cavities \citep[e.g.][]{fabian02,croston11}. The kinetic power in such structures is thought to play a crucial role in balancing cooling flows in the centers of galaxy clusters \citep{mcnamara12}.  This process is generally modeled in idealized simulations by imparting momentum to gas within a region (usually a cylinder or bipolar cone) centered on the SMBH to approximate the directional momentum coupling of gas due to the presence of a jet \citep[e.g.][]{li14,cielo14,prasad15}. Not only does such a method directly prescribe the momentum and morphology of the outflow on rather large scales, but it also assumes a direction of that momentum transfer. Cosmological simulations (and even idealized cluster simulations) do not have the resolution required to directly follow the spin of SMBHs and {\sc RomulusC} is no exception, so the direction of a jet is purely a result of the sub-grid model assumptions. It has been shown that fixed direction jet heating is unable to reproduce observed cluster gas properties and cannot solve the overcooling problem \citep{vernaleo06,oneill10,babul13}. For this reason, simulations that attempt to model this jet process often incorporate a changing direction, generally modeled as a precessing/re-orienting jet \citep{cielo17,cielo18b} or, as is often the case in cosmological simulations with relatively long time-steps, momentum transfer where nearby gas is given a radial kick equally in all directions \citep{weinberger17}.

How jets couple to the ISM and ICM is still an open question. While constraining the power within these bubbles/lobes from observations can be challenging, the fact that these lobes often encapsulate a power 10-1000 times that of the observed synchrotron radiation seems a relatively robust result (but see \citet{snios18} for a potential exception), implying that much of the energy causing this structure likely resides in more massive particles \citep{deyoung06,birzan08,hardcastle10}. It is possible, therefore, that even with a relativistic jet the majority of feedback energy comes from gas entrained on smaller scales. The practical effect of a jet then would be to modulate the efficiency of feedback coupling to gas on $\sim100$ pc scales. Over long periods of time, this would change how accretion itself regulates in systems dominated by radiatively inefficient accretion processes. This would affect the amount of SMBH accretion, but should only be a secondary effect in terms of the large scale environmental impact of the feedback. Because it is not clear when a jet should or should not be active, given that the detailed physics of accretion are not followed in the simulation, we choose to ignore this effect and maintain a constant feedback efficiency rather than introduce additional free parameters. However, as is briefly discussed in \S 5.1, we do find evidence that higher feedback efficiencies may be required for high mass galaxies/SMBHs. 

In \S4.2 we discuss how the interaction between AGN driven outflows in our simulation and gas on both small (1-10 kpc) and large (10s-100s kpc) scales naturally causes an evolution in wind structure that overcomes the problems seen in fixed direction outflows implemented in idealized simulations. The fact that our model results in highly collimated outflows extending out to large scales shows that a kinetic feedback prescription is \textit{not required} to produce such structures. 

\begin{figure*}
\centering
\includegraphics[trim=15mm 7mm 5mm 5mm, clip, width=150mm]{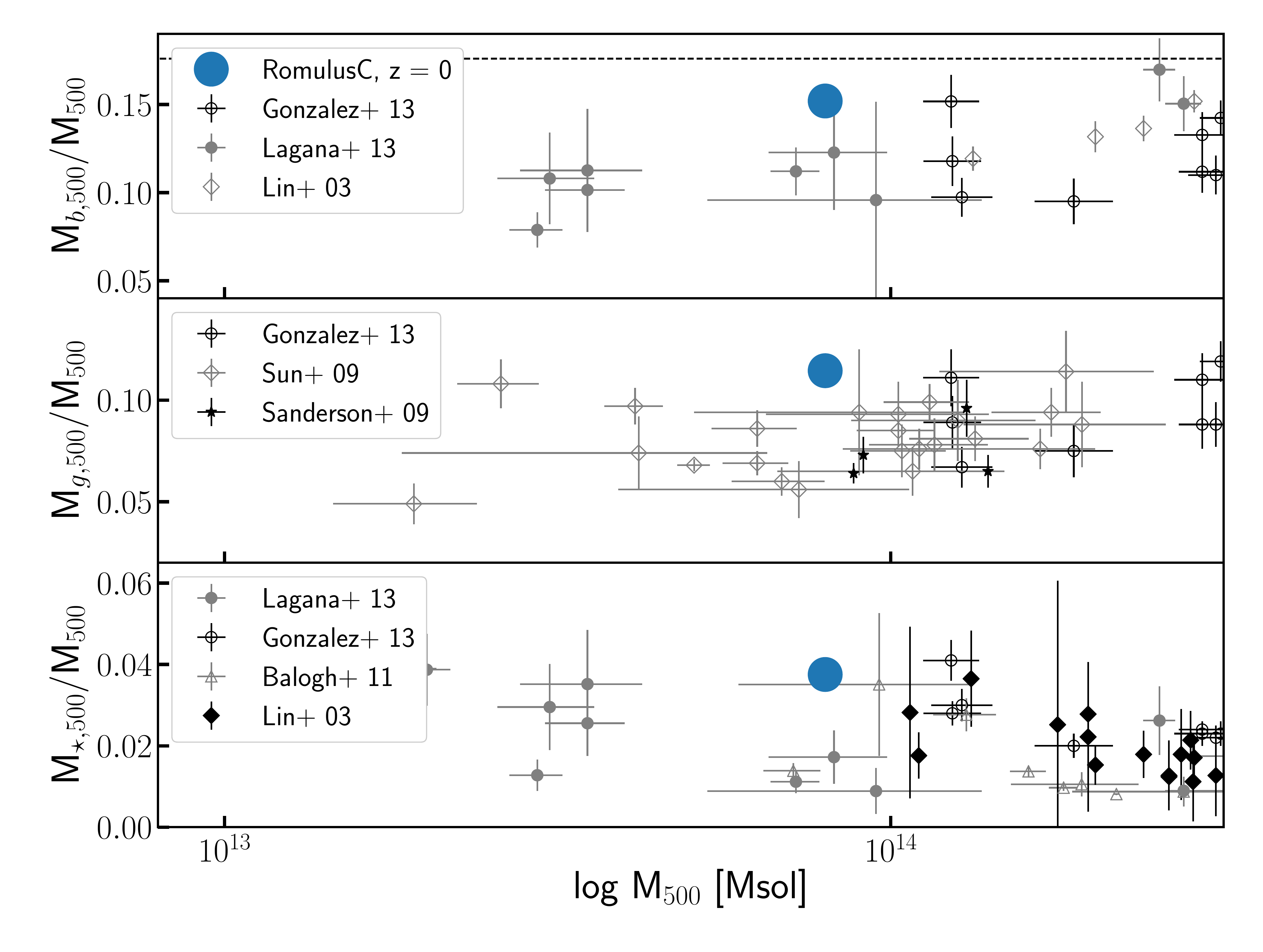}
\caption{{\sc The Baryon Content of Clusters}. The total baryon fraction (top), hot gas fraction (middle), and stellar fraction (bottom) by mass in both observed clusters \citep{lin03, sun09, sanderson09, balogh11,lagana13, gonzalez13}  and {\sc RomulusC} (blue points). All mass fractions for {\sc RomulusC} are within a factor of two of the mean observations at similar masses and within the scatter. We only show $z=0$ values for {\sc RomulusC}, but confirm that there is negligible evolution in this respect from $z=0.3$. All observations are for local (redshift below $\sim0.1$) clusters, though the \citet{lagana13} data extend out to higher redshifts.}
\label{mass_frac}
\end{figure*}

\subsection{Free Parameter Optimization}

As described in \citet{tremmel17}, we use a novel approach for optimizing the free parameters involved in our sub-grid models for both stars and SMBHs and their respective feedback processes. To do this we ran a large set of zoom-in cosmological simulations of halos with masses $10^{10.5}$, $10^{11.5}$, and $10^{12}$ M$_{\odot}$ including full hydrodynamics, star formation, and SMBH physics. The simulations were all run at the same resolution of {\sc RomulusC} and {\sc Romulus25}. Each set of simulations was run using different parameters and graded against different $z=0$ empirical scaling relations related to star formation efficiency, gas content, angular momentum, and black hole growth. From a total of 39 parameter realizations tested using these zoom-in simulations we utilized an adapted Gaussian process Kriging technique to pinpoint regions in parameter space that create galaxies that most closely resemble the mean population at $z=0$ and to determine when we had converged to the optimal choice (see appendix A of \citet{tremmel17} for more details). This resulted in a complete set of sub-grid models governing star formation, stellar feedback, and SMBH accretion and feedback that are optimized to provide realistic $z = 0$ galaxies while maintaining predictive power at higher redshifts and high mass (M$_{vir} > 10^{12}$ M$_{\odot}$). \citet{tremmel17} demonstrated how the parameters result in realistic SMBH and stellar masses for galaxies in halos up to $10^{13}$ M$_{\odot}$ as well as a realistic cosmic star formation and SMBH accretion histories out to high redshift for field galaxy populations. Importantly, this means that the parameters used in {\sc RomulusC} were \textit{in no way constrained to provide realistic results in terms of galaxy evolution in cluster environments}. The results presented in this Paper are, therefore, purely a prediction of our model.

\subsection{Halo and Galaxy Extraction}

For all {\sc Romulus} simulations described in this work, halos are extracted and catalogued using the Amiga Halo Finder \citep{knollmann09}. Halos are defined based on all types of particles (dark matter, gas, and stars) and gravitational unbinding is performed. The centers of halos are defined using a shrinking spheres approach \citep{power03}, which also consistently traces the centers of the central galaxies within each halo.

\section{Properties and Structure of the ICM in RomulusC}

\begin{figure*}
\centering
\includegraphics[trim=10mm 0mm -15mm 0mm, clip, width=150mm]{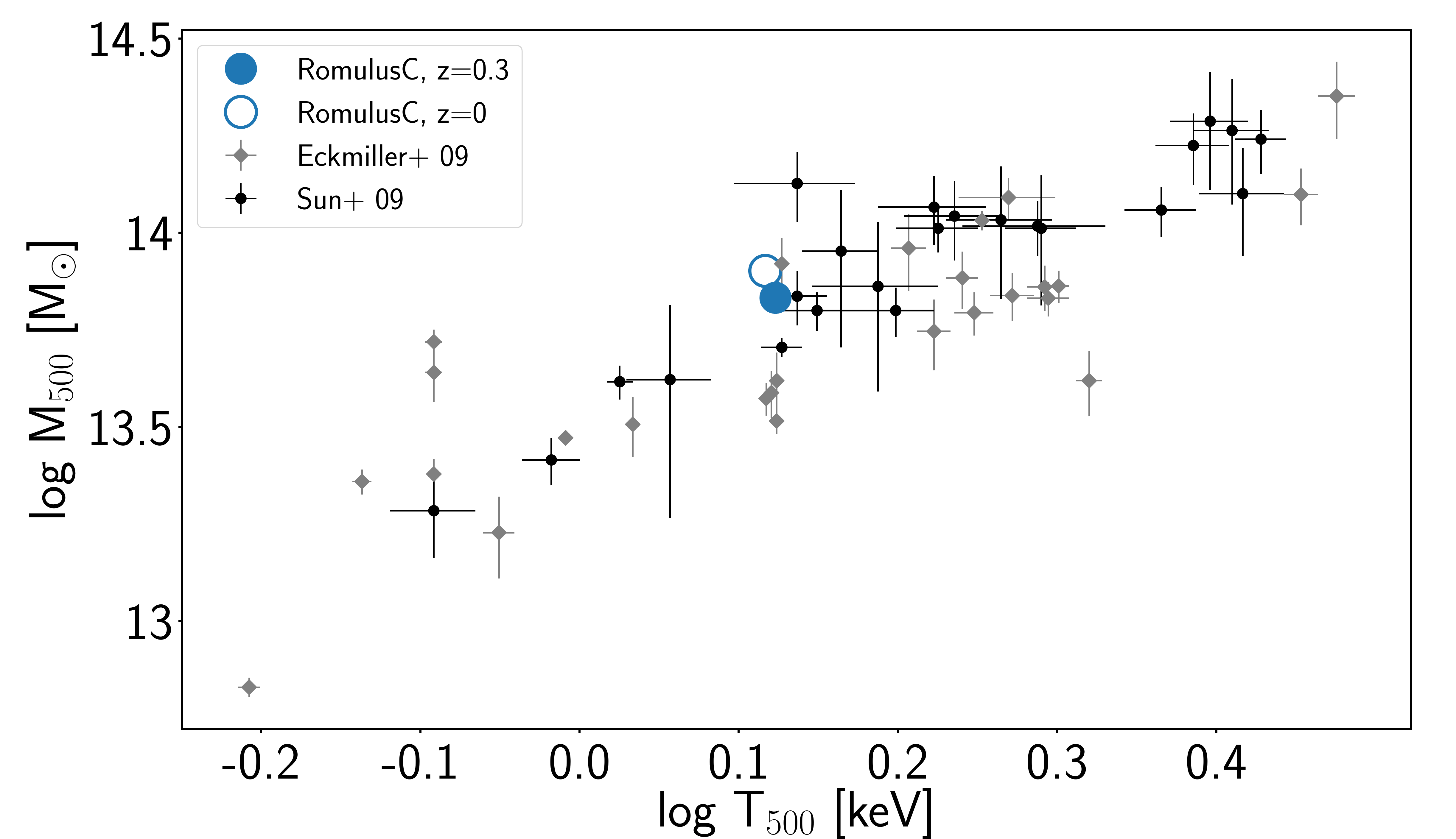}
\caption{{\sc The Temperature of the ICM}. The observed relationship between group and cluster masses (M$_{500}$) and the core excised temperature, T$_{500}$. The open blue circle represents {\sc RomulusC} at $z = 0$ and the solid circle at $z=0.3$, when the cluster is still in equilibrium prior to the in-fall of a galaxy group and what will be an on-going merger at $z=0$. To calculate the temperature we use an average between the emission weighted and mass weighted results for diffuse ($\rho<500 \rho_{crit}$), hot gas detectable by X-ray observatories ($T > 0.1$ keV). This decision is based on results from \citet{liang16} showing that this average better approximates results from more detailed spectroscopic models at this mass scale and are therefore closer to what an observer would see. Our results are insensitive to this choice.}
\label{M_T_rel}
\end{figure*}

In this section, we study the properties and structure of the gas within {\sc RomulusC} and compare these with various observations. The temperature and entropy of cluster gas is determined by the structure of gas that collapses into the cluster to make up the ICM as well as further heating and cooling processes that take place within the cluster (e.g. radiative cooling, heating from AGN and stellar feedback). It is also shaped by the processes of hierarchical merging and galaxy evolution occurring prior to and during cluster formation. Only a cosmological simulation naturally captures these different phases of evolution. Comparing the global properties and structure of the ICM in {\sc RomulusC} is an important benchmark for determining how well our model for feedback, particularly that of AGN, is able to correctly predict ICM properties observed in cluster environments. We show that {\sc RomulusC} reproduces empirical scaling relations, baryonic content, and average ICM profiles in observed clusters. This successful match to key physical properties is crucial and will allow us to use {\sc RomulusC} and future simulations of this type to better understand the origin and evolution of the ICM as well as galaxy evolution within cluster environments. 

In the following analysis, R$_{\Delta}$ is defined as the radius within which the average density is $\Delta\times\rho_{crit}$. A property such as M$_{\Delta}$ is then the mass within R$_{\Delta}$. We also will show some results at both $z=0$ and $z=0.3$. This is because {\sc RomulusC} undergoes a merger with a group at $z\sim0.2$ that is still on going at $z=0$, causing the cluster to be out of equilibrium. The detailed dynamical and hydrodynamical evolution of the cluster during this period will be the topic of future work and is beyond the scope of the current paper. All results for {\sc RomulsuC} are given using 3D radii, though we confirm that this choice makes little difference were we to use projected quantities instead.

\begin{figure*}
\centering
\includegraphics[trim=10mm 0mm -15mm 0mm, clip, width=150mm]{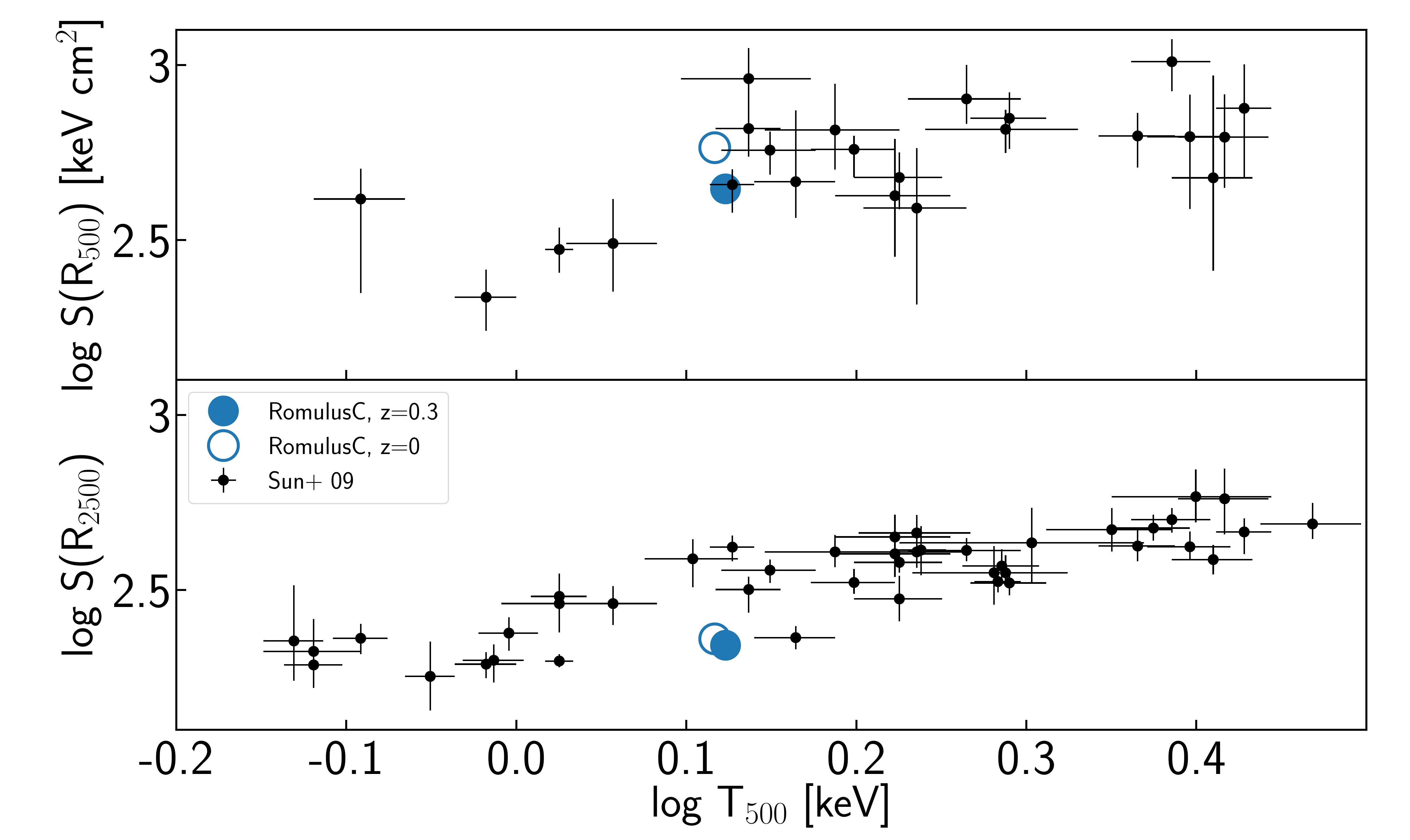}
\caption{{\sc The Entropy of the ICM}. The observed relationship between entropy and overall cluster temperature. The blue points represent results from {\sc RomulusC} at $z=0$ (open) and $z=0.3$ (solid). The entropy is measured within 1 kpc thick annuli at R$_{2500}$ (bottom) and R$_{500}$ (top). Like in Figure~\ref{M_T_rel}, we take the average between mass and emission weighted values for temperature to calculate entropy. Again, only diffuse, hot gas is included in the analysis.}
\label{S_T_rel}
\end{figure*}

Current observations of gas in clusters rely on emission detectable in X-rays, so in this section we only consider gas from {\sc RomulusC} that has temperatures above $10^6$ K, or $\sim0.1$ keV, consistent with the rough lower limits of Chandra's observational sensitivity. Emission weighting is done by weighting each particle by its luminosity in the X-ray. To estimate this, we calculate the volume emissivity, $\epsilon_X$, given by the following equations from \citet{balogh99}:

\begin{equation}
\begin{aligned}
\epsilon_X = & \frac{3}{2} \frac{\rho k_b T}{\mu m_h t_{cool}}\\\\
t_{cool} = & 3.88\times10^{11} \mu m_h \frac{T^{0.5}}{\rho\,(1+5\times10^7 f_m/T)}.
\end{aligned}
\end{equation}

\noindent The metalicity dependent factor, $f_m$, is taken to be 1, consistent with solor metalicity. The results we present are not sensitive to this choice. The other factors, $m_h, k_b$, and $T$, are the mass of hydrogen, the Boltzmann constant, and temperature of the gas respectively. We derive the value of $\mu$ directly for individual gas particles in the simulation based on their tracked metal abundances and ion content. We confirm this choice does not affect our results were we to assume $\mu = 0.59$ for all gas particles. To get the luminosity of each particle, we multiply by its volume, where the volume of the i-th particle is given by $V_i = m_i/\rho_i$, where $m_i$ is the mass of the particle and $\rho_i$ is its density.

Temperatures calculated within a given radius (i.e. $T_{500}$) are calculated excising  the inner $0.15 R_{500}$, consistent with observations \citep[see][and references therein]{liang16}. Specific entropy is calculated at a given radius using the widely accepted proxy, $S(r)$, related to the thermodynamic specific entropy by $ds \propto dlnS$ \citep{balogh99}. This value, which we shall refer to as `entropy' hereafter, is given by

\begin{equation}
S(r) = \frac{k_b T_e(r)}{n_e r^{2/3}}.
\end{equation}


The masses for the observed clusters are typically calculated using X-ray observations and assuming hydrostatic equilibrium. This has been found to underestimate the halo mass using both hydrodynamic simulations \citep{nagai07b} and lensing observations \citep{hoekstra15}. When presenting observations with X-ray derived halo masses, we multiply the published masses by a factor of 1.3 to account for this bias. Nowhere are our conclusions sensitive to the inclusion of this correction value. 


\begin{figure*}
\centering
\includegraphics[trim=-3mm 15mm 0mm 4mm, clip, width=88mm]{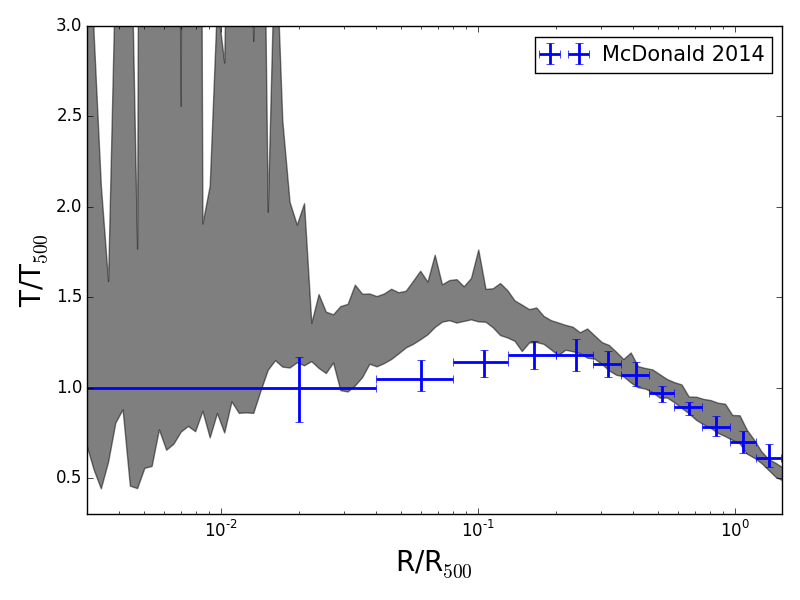}
\includegraphics[trim=3mm 15mm 0mm 0mm, clip, width=85mm]{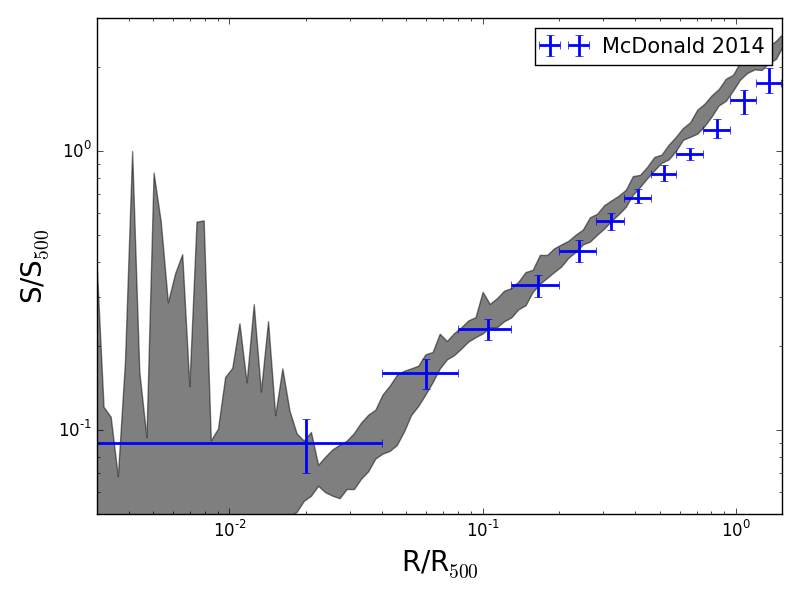}
\includegraphics[trim=0mm 0mm -2mm 4mm, clip, width=89mm]{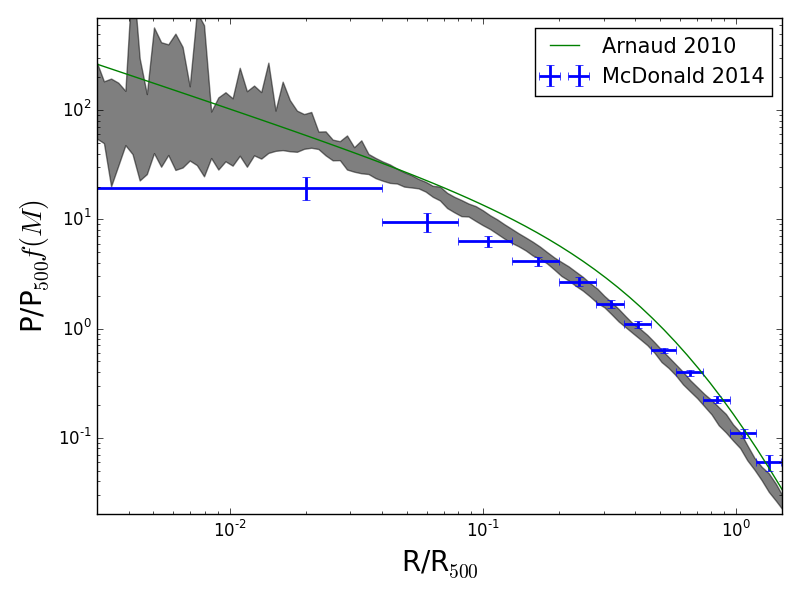}
\includegraphics[trim=6mm -5mm 16mm 4mm, clip, width=84mm]{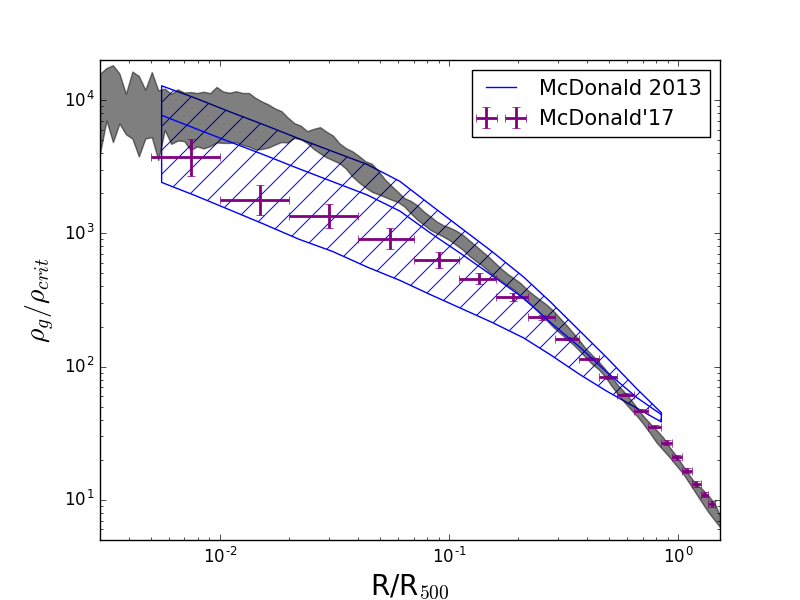}
\caption{{\sc The Structure of the ICM.} Radial profiles of the temperature, entropy, pressure, and density of the ICM gas in {\sc RomulusC} (Grey bands). The grey bands correspond to the range of profiles from $z=0.3-0.5$. This redshift range was chosen because it corresponds to a period in the simulation prior to the late time major merger where the cluster still has a cool core. It is also a similar redshift range as the clusters examined in \citet{mcdonald14}. At $z<0.3$ the in-falling group causes increasing disturbance away from equilibrium. As stated in the text, we will examine the effect of this merger on the ICM properties in future work. Overplotted are the results from \citet{mcdonald13,mcdonald14,mcdonald17}. For pressure we also compare with results from \citet{arnaud10} which are at lower redshifts, but include more similar mass clusters to {\sc RomulusC}. All observed datasets are selected to be cool core clusters, except for the mean density profile from \citet{mcdonald17}. Overall, {\sc RomulusC} fits well with average profiles of cool-core clusters. Compared to \citet{mcdonald14} {\sc RomulusC} has slightly high temperature and pressure within the inner $0.2R_{500}$, but matches well with the \citet{arnaud10} results for pressure within $0.2R_{500}$. The density profile matches well with the  average profile for all clusters from \citet{mcdonald17}, but deviates from self-similarity on small scales where it matches much better median density profile for clusters selected to have low entropy cores \citep{mcdonald13}.}
\label{profiles}
\end{figure*}

Figure~\ref{mass_frac} shows the baryonic, hot gas, and stellar mass fractions within R$_{500}$ for {\sc RomulusC} plotted against a large sample of observed clusters and groups. In all cases, {\sc RomulusC} is consistent with observations, though slightly on the high end of the scatter. This is an important result. As discussed in \S4.1, the central galaxy is quenched by the presence of large-scale outflows driven by a central AGN. Such outflows can affect the ability for gas to cool onto the central BCG by either balancing the radiative cooling of the gas or by expelling it entirely. The fact that {\sc RomulusC} results in realistic stellar and gas mass fractions means that star formation is suppressed without unrealistically evacuating large amounts of gas. This has been an issue in some cosmological simulations \citep{genel14}, where the solution has been to modify the nature of AGN feedback in massive systems \citep{weinberger17}. Our results shown here are similar to what has been found in other recent cosmological cluster simulations \citep[e.g.][]{barnes17b, pillepich18}.

Figure~\ref{M_T_rel} plots the relationship between mass and the ICM temperature within R$_{500}$ for observed groups and clusters, with the results from {\sc RomulusC} overplotted as blue points. Here the temperature for {\sc RomulusC} is taken as an average between the mass weighted and emission weighted temperatures, shown by \citet{liang16} to be a better approximation for the spectroscopic temperature that an observer would derive \citep{mazzotta04,vikhlinin06}. We confirm that our results are insensitive to this choice. The results from {\sc RomulusC} match the observed relationship well at both $z=0$ and $z=0.3$. In order to not include denser gas within substructure \citep[see][]{zhuravleva13}, we only include gas with densities less than $500\rho_{crit}$. This value was arrived at empirically, as more strict cuts lower temperature estimates because they remove hot cluster gas outside $0.15R_{500}$ and less strict cuts also lower temperature estimates because they miss dense substructure. Our results are insensitive to the exact choice of density threshold, in part because emission weighting is only used when the values are averaged with mass weighted values. For future, more detailed analysis of the ICM a more careful approach will be used to remove substructure gas. However, for the purposes of this Paper the simple, single density cut approach is sufficient.

Figure~\ref{S_T_rel} plots the entropy calculated within 1 kpc wide annuli at both R$_{500}$ and R$_{2500}$ for observed groups and clusters with the results from {\sc RomulusC} plotted as blue points at both $z=0.3$ and $z=0$. Again, the average between mass and emission weighted temperature values is used, following \citet{liang16}. The electron density is calculated as a volume weighted average within each annulus. At R$_{500}$, {\sc RomulusC} lies comfortably among the observed low mass clusters. Closer to the center, the entropy is slightly lower in {\sc RomulusC} compared with the average of observations but it is still within the lower edge of observed clusters. The results are insensitive to the width of the annuli. Once again, we exclude gas with density greater than $500\rho_{crit}$.




Figure~\ref{profiles} plots gas temperature, entropy, density, and pressure profiles for {\sc RomulusC} at $z=0.3-0.5$ and compares each to the average profiles observed in cool-core clusters \citep{arnaud10,mcdonald13,mcdonald14} as well as the  average density profile for all clusters \citep{mcdonald17}. While we only include hot ($>10^6$ K) gas in our analysis, no density criterion is used to generate these profiles. For entropy and pressure we derive P$_{500}$ and S$_{500}$ in the same way as described in \citet{mcdonald14} using T$_{500}$ and the average density. The average density is calculated from the electron density, following \citet{mcdonald14,mcdonald17}. For pressure we also normalize by 
$f(M) = (M_{500}/3 \times 10^{14} h^{-1}_{70} M_{\odot})^{0.12}$ in order to compare effectively with clusters of different
mass and assuming the universal pressure profile derived in \citet{nagai07}. 

In the density profile, we see that {\sc RomulusC} matches well with the overall cluster population down to $\sim0.2R_{500}$, which is where deviations from self-similarity are seen in observed clusters \citep{mcdonald13,mcdonald17}. Below this scale, {\sc RomulusC} fits well to the median density profile for cool core clusters from \citet{mcdonald13}. The average observed entropy, pressure, and temperature profiles are all calculated from cool-core selected clusters, so {\sc RomulusC} matches them well down to small scales, although the temperature inside $0.2 R_{500}$ is biased high. The error bars for these average fits are standard errors from the mean (standard deviation divided by $\sqrt{N}$) and there is actually a wide range of temperatures from individual observations spanning 0.5-2 T$_{500}$ within $0.2R_{500}$ (see figure 13 in \citet{mcdonald14}).

We stress that the mass of {\sc RomulusC} is significantly lower than the clusters examined in \citet{mcdonald14,mcdonald17}, which may also affect this comparison. Although the pressure profiles are normalized accordingly assuming a universal profile \citep{nagai07}, the fact that the data from \citet{arnaud10} is for clusters of mass more similar to {\sc RomulusC} might be why we match those results better within $0.2R_{500}$, where self similarity is no longer valid. We use 3D profiles while the observations are not deprojected, though as shown in \citep{mcdonald14} the deprojected profiles are nearly identical to the projected ones for observed clusters.

Matching the observed structure and baryonic content of the ICM is a particularly important result because, as we will explore further in \S4.1, by $z=0.3-0.5$ the central AGN has already been (and continues to be) very active in the cluster. The AGN feedback in {\sc RomulusC} is able to suppress cooling without disrupting the cool-core structure, similar to observed clusters which maintain a stable entropy profile over long timescales \citep{mcdonald14}. In observations and in {\sc RomulusC}, AGN feedback is able to provide a long-term balance to ICM cooling without disrupting the ICM structure. The effect of feedback is also why we see such strong fluctuations in the temperature and entropy profiles at very small radii. 

\begin{figure}
\centering
\includegraphics[trim=0mm 0mm -10mm 5mm, clip, width=80mm]{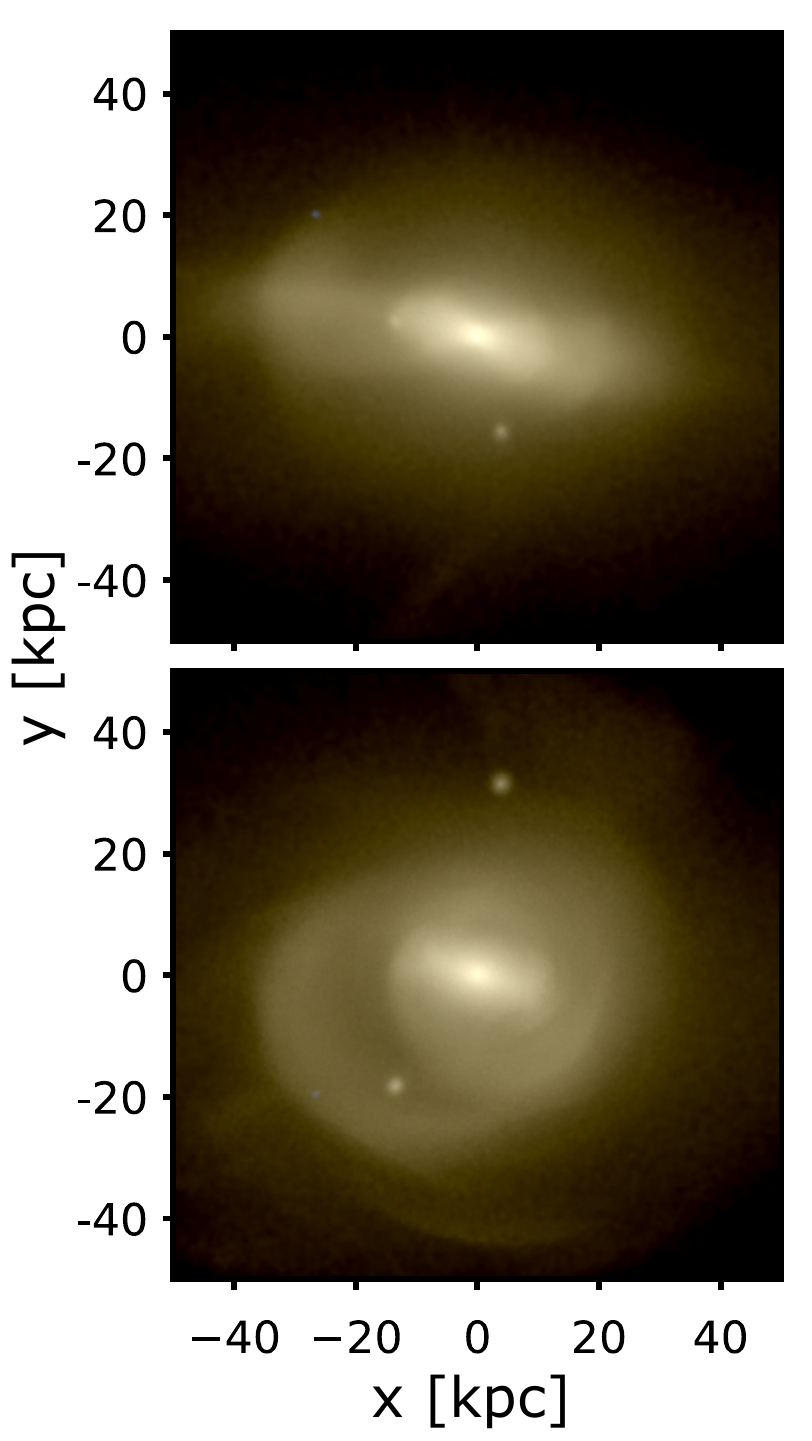}
\caption{{\sc The Brightest Cluster Galaxy}. A uvj image of the brightest cluster galaxy in {\sc RomulusC} at $z=0$ down to 26 mag arcsec$^{-2}$ to show structure on larger scales. The lack of recent star formation has made the galaxy appear very red. Two orthogonal views are shown. There is no longer any stellar disk structure.}
\label{stars}
\end{figure}

\section{Galaxy Evolution in Clusters}

Now that we have established that {\sc RomulusC} results in a cluster with realistic ICM properties, we turn our focus to the evolution of galaxies within the cluster environment. {\sc RomulusC} represents a relatively rare, dense environment. For reference, the cluster attains M$_{200} = 2\times10^{13}$M$_{\odot}$ by $z=2$, a mass equivalent to the highest mass halo in the {\sc Romulus25} $25^3$ Mpc$^3$ uniform volume simulation at $z=0$ (see Figure~\ref{mass_growth}). {\sc RomulusC} therefore traces galaxy evolution within a very dense environment out to high redshift and, in this sense, represents an important addition to the galaxies followed in {\sc Romulus25}. The fact that {\sc Romulus25} has the same resolution and sub-grid physics as {\sc RomulusC} means that we can self consistently compare galaxy properties and evolution between the two simulations in order to examine the effects of a dense environment ({\sc RomulusC}) compared to more isolated galaxies in the field ({\sc Romulus25}). The resolution of the {\sc Romulus} simulations means that we can resolve the evolution of dwarf galaxies in cluster environments better than ever before. While in the following analysis we focus on the population of galaxies and their bulk properties, in future work we will study the evolution of the internal structures of cluster galaxies over a wide range of stellar masses.

\begin{figure}
\centering
\includegraphics[trim=10mm 5mm 20mm 35mm, clip, width=90mm]{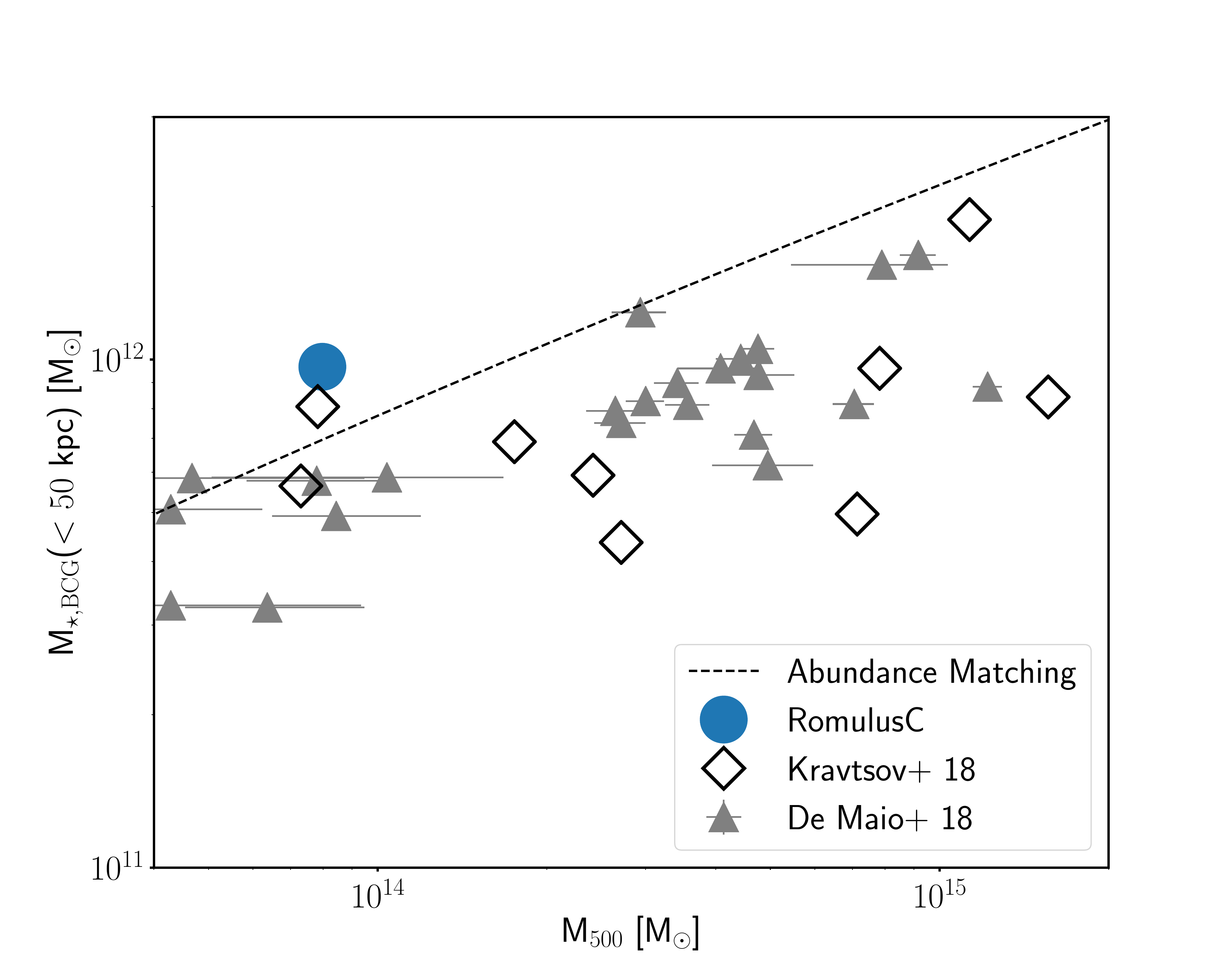}
\caption{{\sc Stellar Mass Halo Mass Relation}. The relationship between the mass of the central BCG and M$_{500}$ for {\sc RomulusC} (blue point) as well as recent observations from \citet{kravtsov18} and \citet{demaio18} (black diamonds and grey triangles). Following the observations we compare with here, we consider the stellar mass within 50 kpc from halo center. Again, we do not plot the $z=0.3$ values because there is negligible change from $z=0$. The BCG in {\sc RomulusC} is within a factor of $\sim2$ of the median observed BCG stellar mass and matches well with abundance matching results derived from updated luminosity functions \citep{kravtsov18}.}
\label{smhm}
\end{figure}

\begin{figure*}
\centering
\includegraphics[trim=0mm 0mm 0mm 0mm, clip, width=180mm]{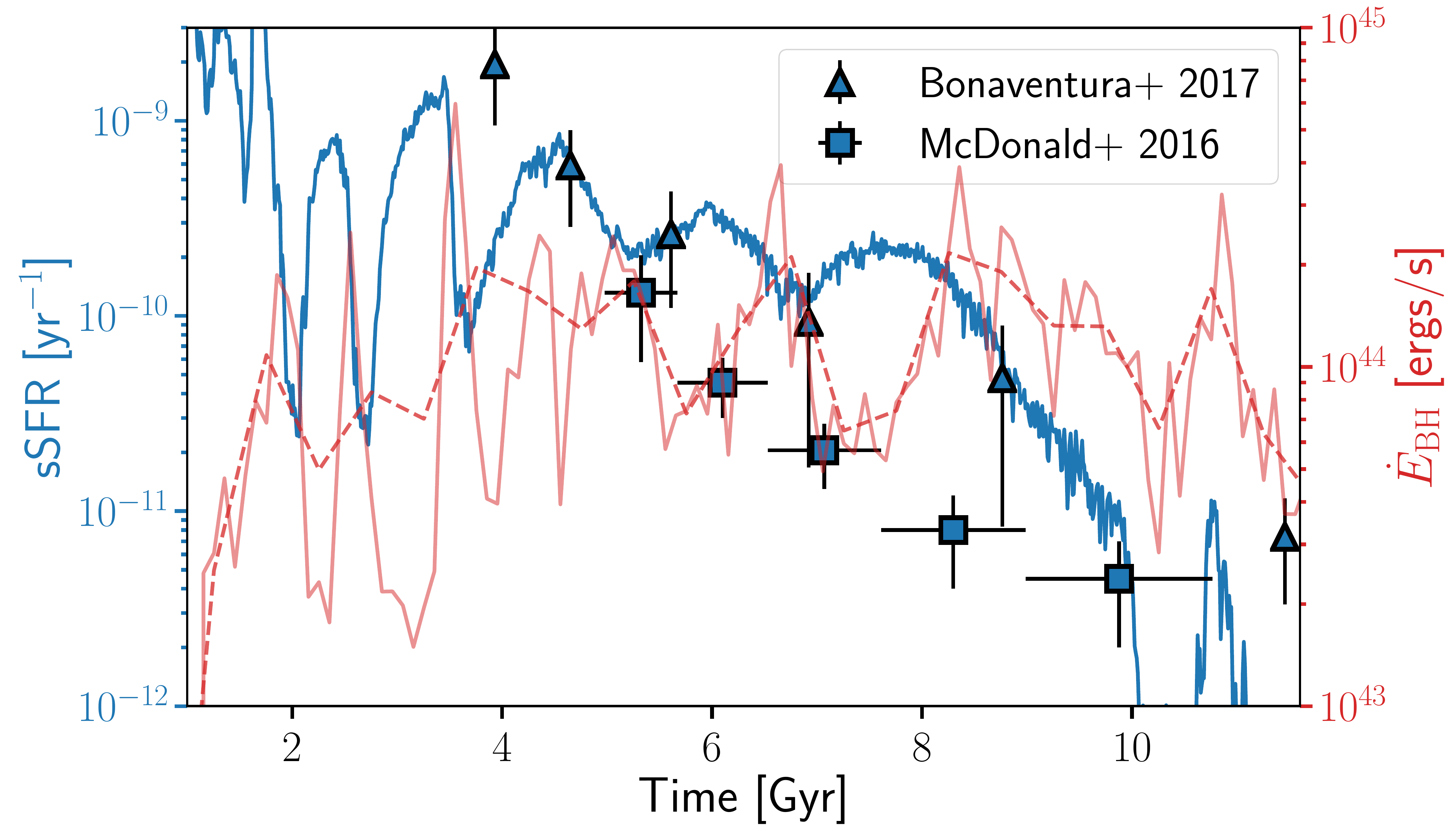}
\caption{{\sc Star Formation and SMBH Feedback History}. The specific star formation rate (blue, averaged over 10 Myr bins) and feedback rate of the central SMBH (red, averaged over 100 and 500 Myr bins for solid and dashed lines respectively) for the BCG in {\sc RomulusC}. The SMBH feedback traces both accretion and feedback energy imparted by the SMBH onto nearby gas. Dips in the sSFR are often associated with peaks of SMBH activity. The peak of activity beginning around 8 Gyr and persisting through 10 Gyr is associated with the final quenching of the BCG. The sSFR history of the {\sc RomulusC} BCG is remarkably close to the average evolution observed in clusters from \citet{bonaventura17}, but slightly high compared to results from \citet{mcdonald16}. The range in time shown is cut off just prior to the infall of the group seen in Figure~\ref{stars}. The merger destroys the cool core, a topic to be explored in future work.}
\label{sfh_bhacc}
\end{figure*}

\subsection{The Brightest Cluster Galaxy}

The most unique cluster galaxy is, of course, the most massive, brightest cluster galaxy which, in {\sc RomulusC}, lies in the center of the halo. This is the first time a galaxy of this size has been simulated at such high resolution in a fully cosmological simulation. Not only will examining this galaxy help us to better understand the interaction between AGN feedback, the central ICM, and the evolution of the BCG, it also represents an important test of our sub-grid physics. As stated in \S2.3, the sub-grid recipes for star formation, SN feedback, and SMBH physics were calibrated to reproduce observed lower mass galaxies (MW mass and below) and, as demonstrated by \citet{tremmel17}, have shown success in reproducing observed properties of galaxies in halos as massive as $10^{13}$ M$_{\odot}$. The ability of such a model to extend over two orders of magnitude in halo mass and still produce realistic central galaxies is a testimony to the success of our optimization routine. It also means that our results are purely a prediction of our simulation with no tuning for cluster environments.

Figure~\ref{stars} shows a synthetic image of the stars in the BCG from two different angles at $z=0$. The galaxy is being disturbed by an ongoing merger, causing a shell-like structure in the diffuse stars. The morphology is that of a dense stellar core with an extended stellar halo and little recent star formation. The stars associated with the cluster halo not inside of substructure extends out to large radii and are difficult to observe. Recent observations have been able to examine the stellar mass of the central galaxy and its extended stellar halo with unprecedented detail \citep{gonzalez13,kravtsov18,demaio18}. In Figure~\ref{smhm} we plot the stellar mass within 50 kpc of halo center in {\sc RomulusC} to compare with recent observations by \citet{demaio18} and \citet{kravtsov18} as a function of M$_{500}$. We also show the results of abundance matching presented in \citet{kravtsov18}. The stellar mass of the {\sc RomulusC} BCG is slightly high relative to abundance matching results and a factor of $\sim2$ higher than the median observed value. While in the simulation we are able to take a three dimensional stellar mass profile, we confirm that projection effects along different lines of sight make no difference in our results. While these results are similar to BCG masses found in lower resolution cosmological simulations \citep[e.g.][]{pillepich18}, recent work presented in \citet{ragone-figueroa18} has resulted in more realistic BCG masses, though higher star formation at low redshift compared to observations.

A slightly high stellar mass for the BCG may imply that the efficiency for SMBH feedback coupling should be increased from our fiducial values. This would allow more energetic feedback for less SMBH growth. Currently, in order to attain highly energetic feedback, the SMBH needs to grow rapidly (given our 2\% coupling efficiency and 10\% radiative efficiency). Such rapid growth requires a lot of relatively dense gas in the cluster center, which would also lead to star formation. Additional physical processes seem to be common in higher mass systems, such as relativistic jets and associated radio lobes \citep{dunn06}. These additional processes associated with radiatively inefficient accretion could effectively increase the coupling efficiency of SMBHs in high mass systems, relative to the value we implement here which is held constant across all black holes in the simulation. 

While we only have a sample of one simulated BCG, the fact that the stellar mass is within even a factor of $\sim2$ of the median of observed BCGs (and near the upper end of the observed scatter) and eventually ceases any substantial star formation is an important result. While outflows from SMBHs that reach out to large radii are important, {\sc RomulusC} demonstrates that modeling such outflows through mechanical prescriptions, which require additional assumptions and free parameters, is not necessarily required. Our implementation of thermal AGN feedback, which is implemented the same way for all black holes in all halos, appears adequate to match current observations.

\begin{figure*}
\centering
\includegraphics[trim=20mm 13mm 13mm 10mm, clip, width=180mm]{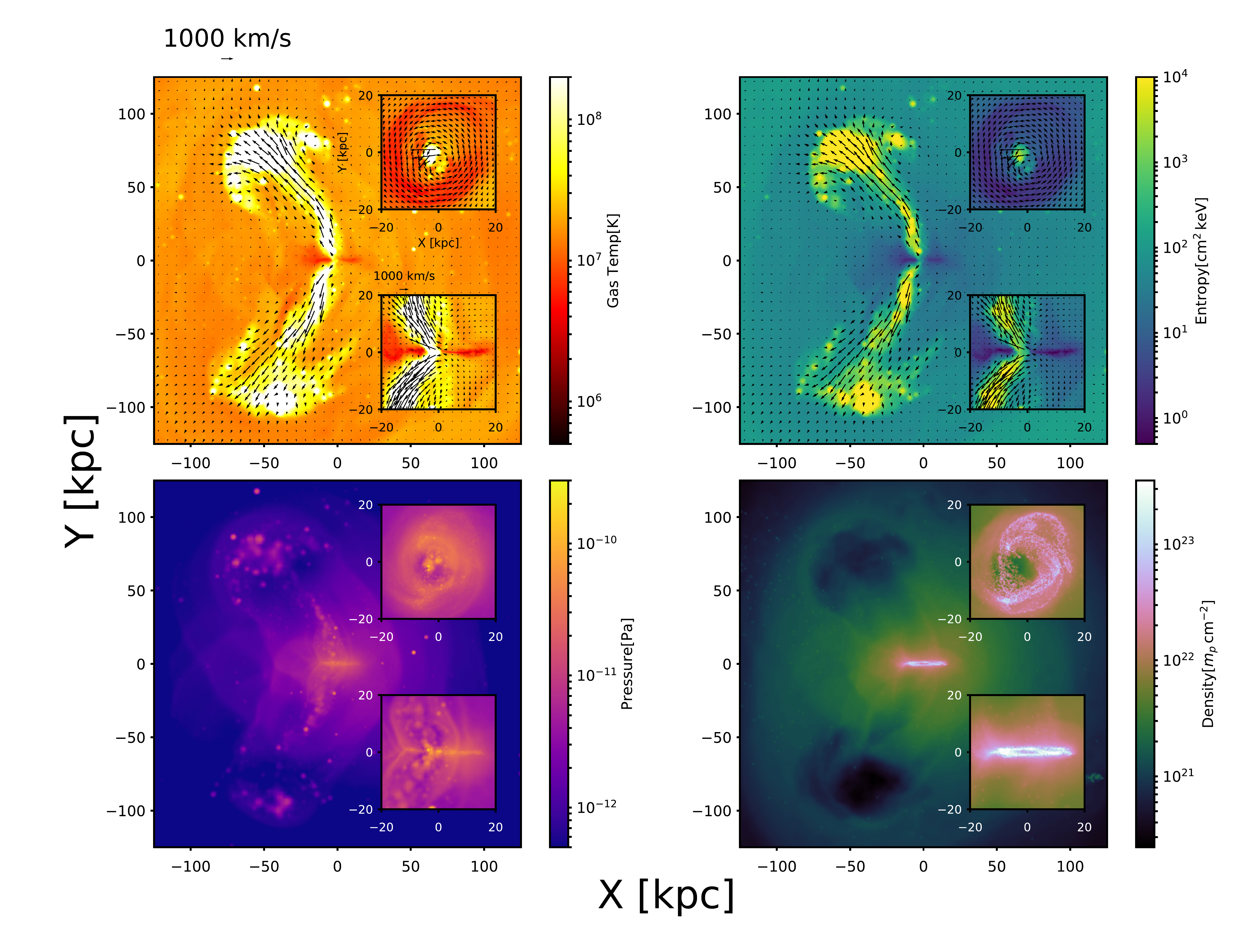}
\caption{{\sc Outflows in Action}. Images of temperature (top left), entropy (top right), pressure (bottom left), and column density (bottom right) of cluster gas at $z = 0.53$ ($t=8.41$). Each large panel is a mass weighted slice along a 30 kpc slice through the center of the cluster. The inset panels are 5 kpc slices zooming into the central regions. The exception to this is the density plot, which is a 100 kpc slice for all panels in order to better show the structure of the gas. A large scale outflow is clearly seen in the temperature and entropy plots, which also overlay the velocity field. The gas in the outflow is moving at 1000s km/s. The structure of the outflow is less clear in the pressure plot; the outflow has a similar pressure as the ambient gas, which is how it is able to maintain its collimation. The tips of the outflow show a high pressure, low density region, which can be seen to create bubbles in the gas, similar to what is observed in X-ray cavities. Although the outflow is powerful, it coexists with low entropy, $10^6$ K, rotationally supported core. Shocks can be seen propagating through the inner regions in the pressure plot, helping to balance the cooling as seen in, e.g., \citet{li15}.}
\label{outflow_image}
\end{figure*}


\begin{figure*}
\centering
\includegraphics[trim=32mm 30mm 65mm 17mm, clip, width=61.5mm]{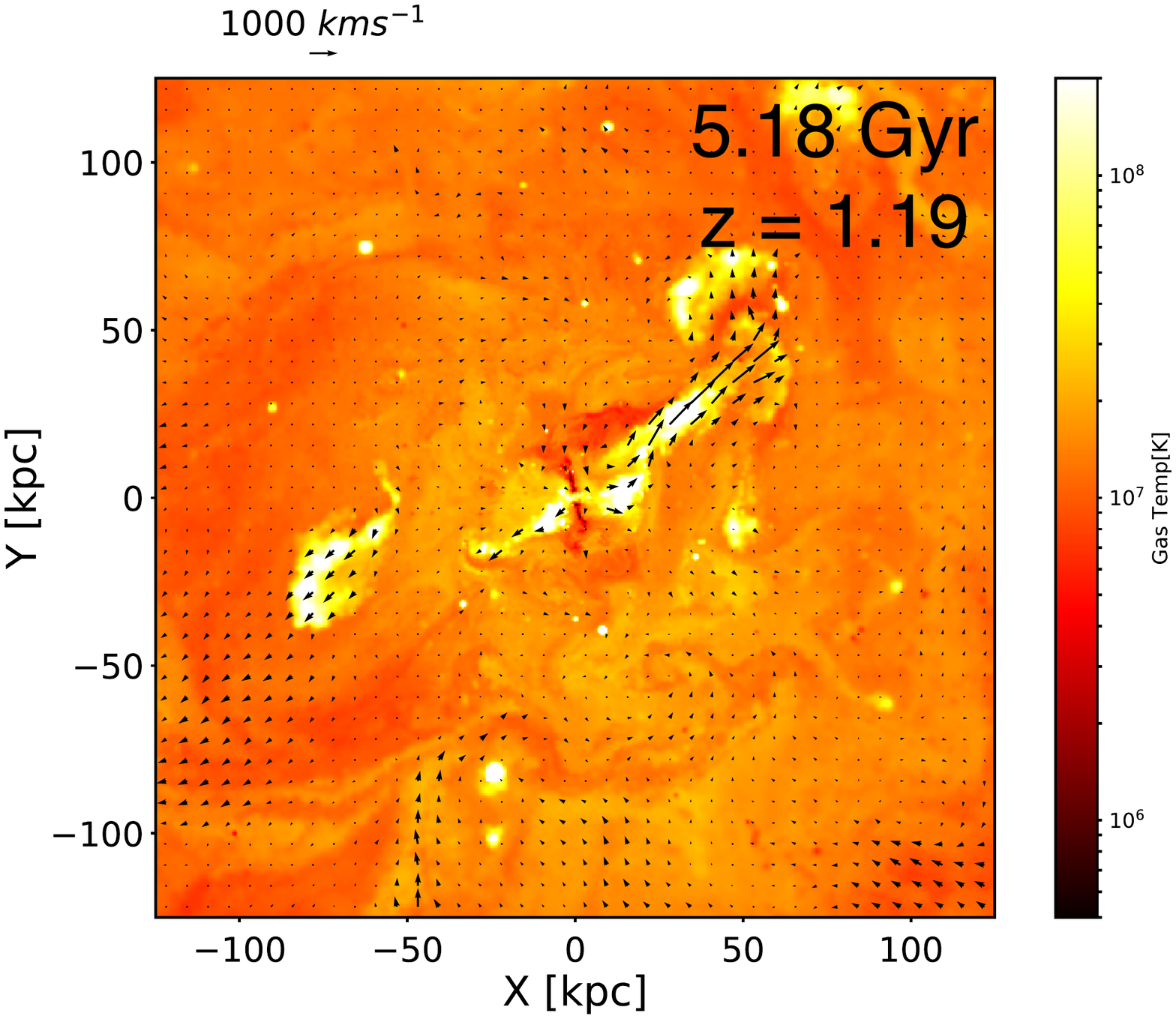}
\includegraphics[trim=67mm 30mm 55mm 15mm, clip, width=53mm]{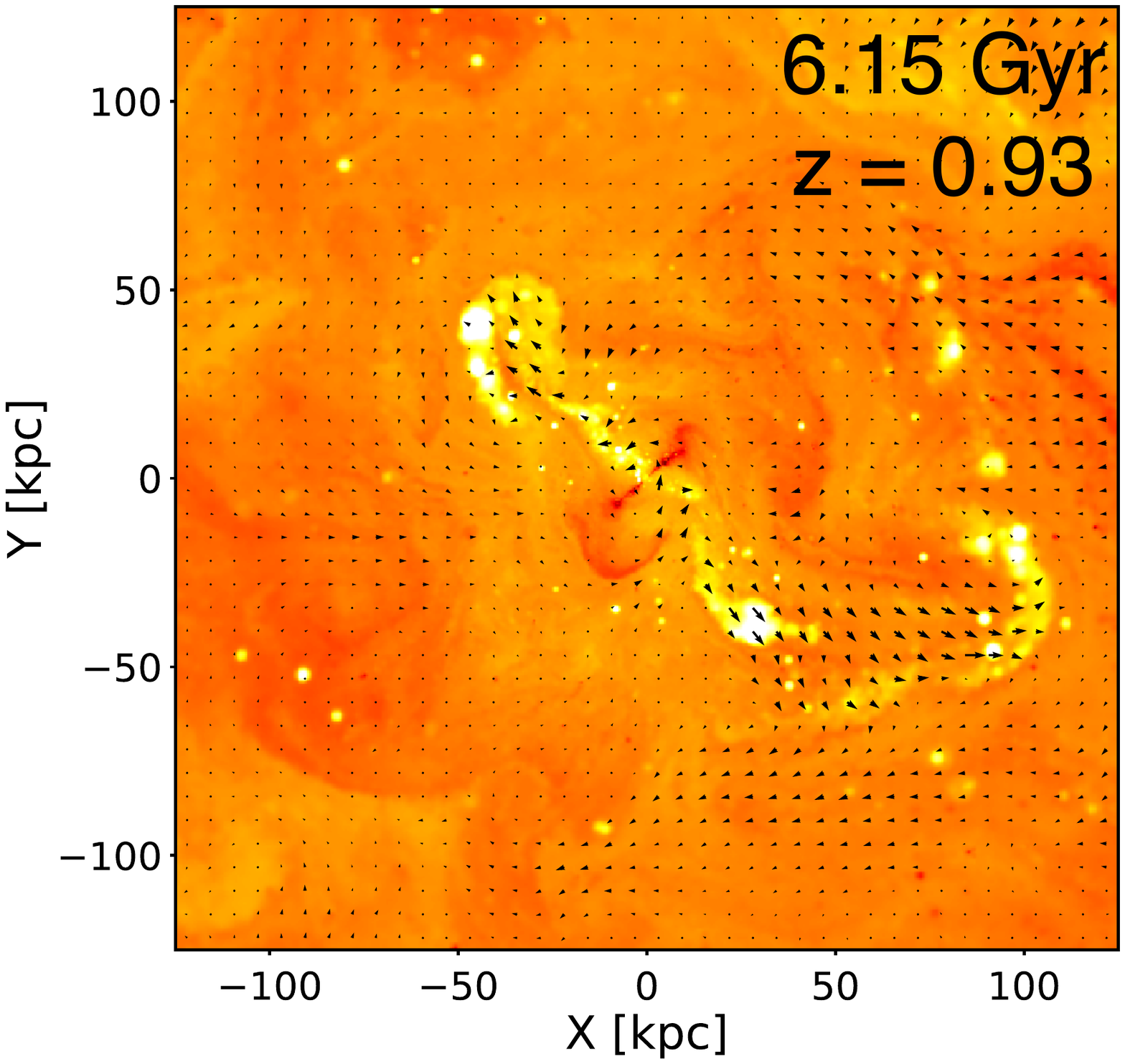}
\includegraphics[trim=67mm 30mm 55mm 15mm, clip, width=53mm]{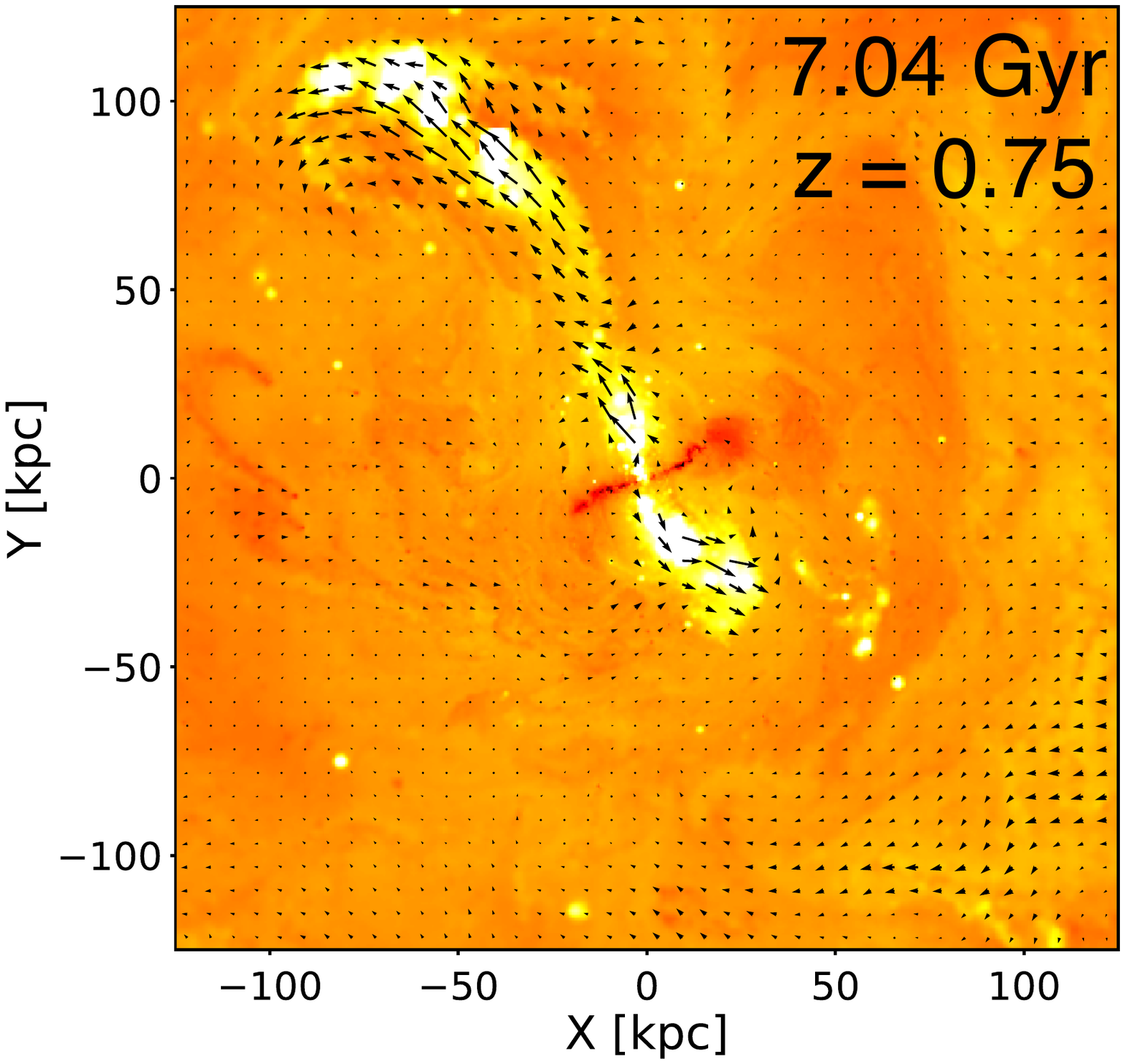}

\includegraphics[trim=41mm 30mm 55mm 30mm, clip, width=61.5mm]{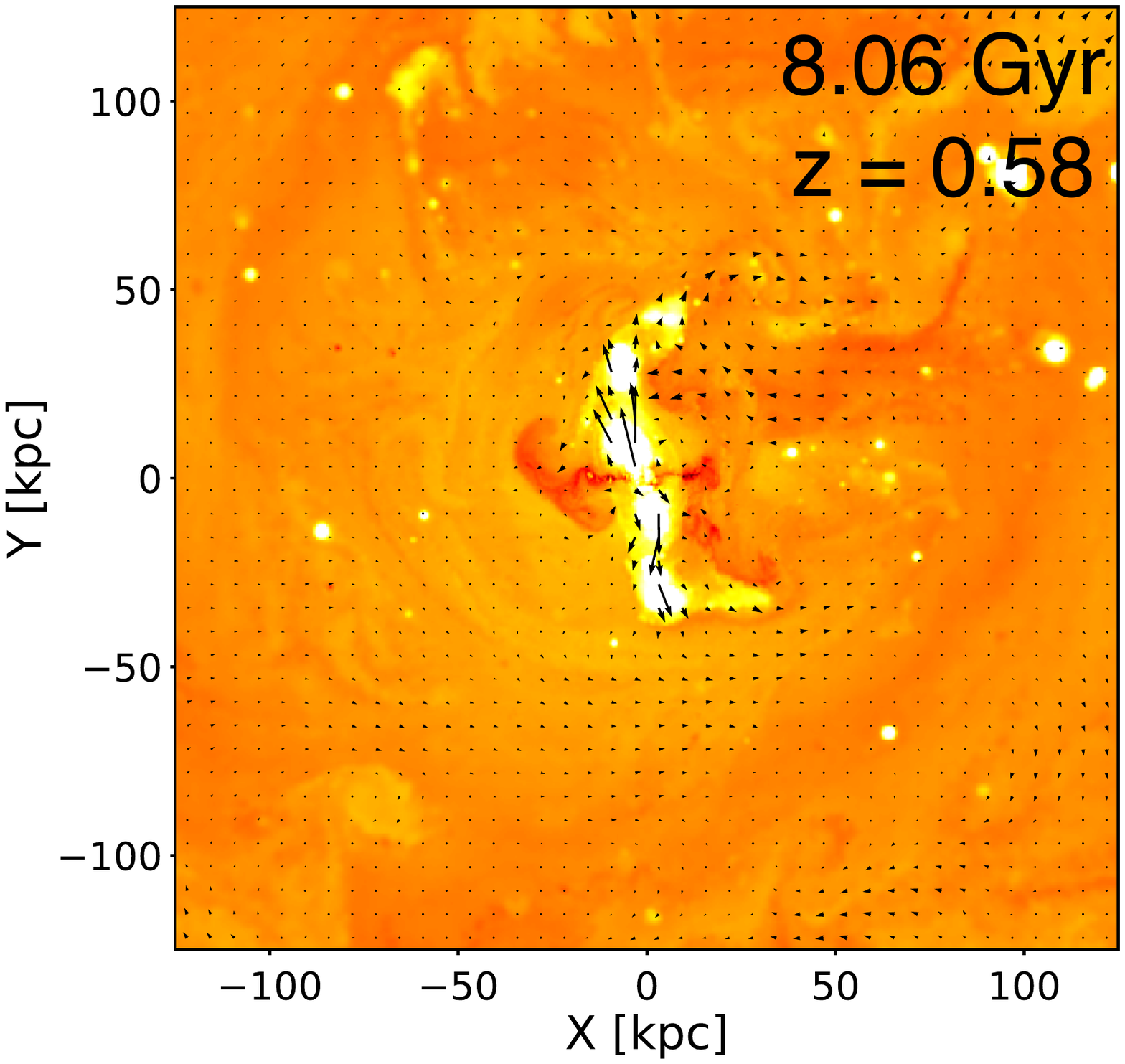}
\includegraphics[trim=67mm 30mm 55mm 30mm, clip, width=53mm]{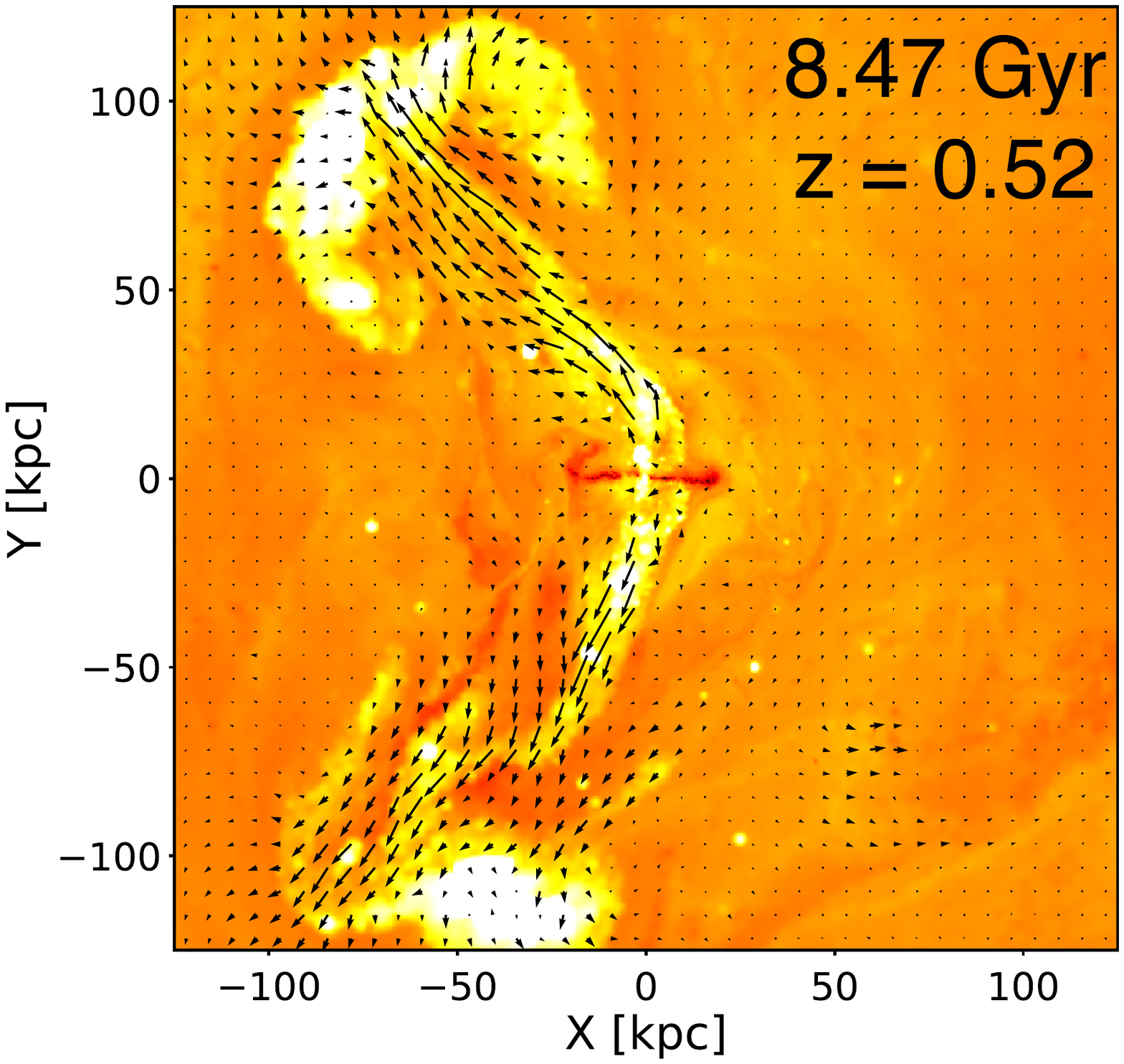}
\includegraphics[trim=67mm 30mm 55mm 30mm, clip, width=53mm]{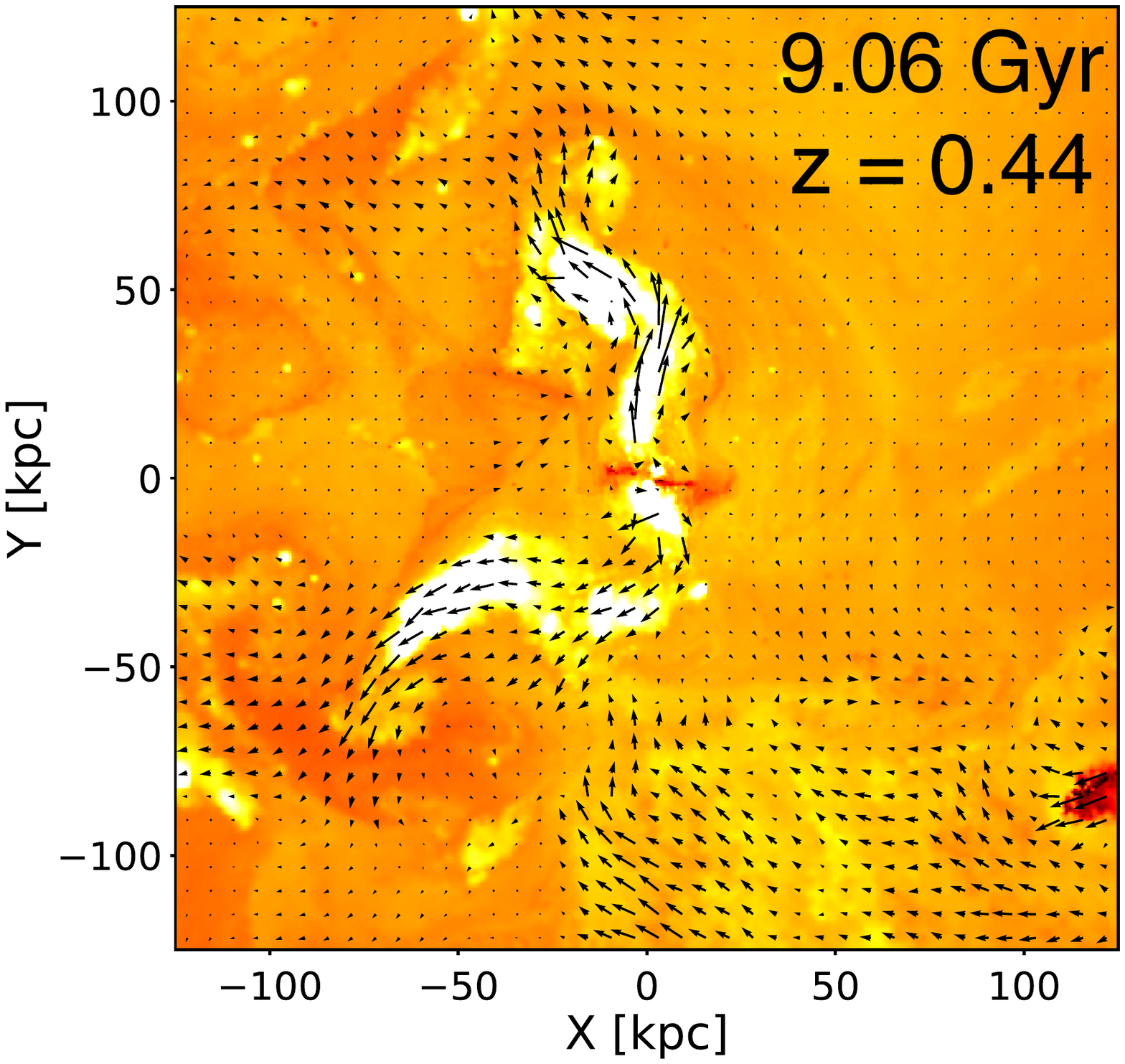}

\includegraphics[trim=40mm 5mm 55mm 30mm, clip, width=61.5mm]{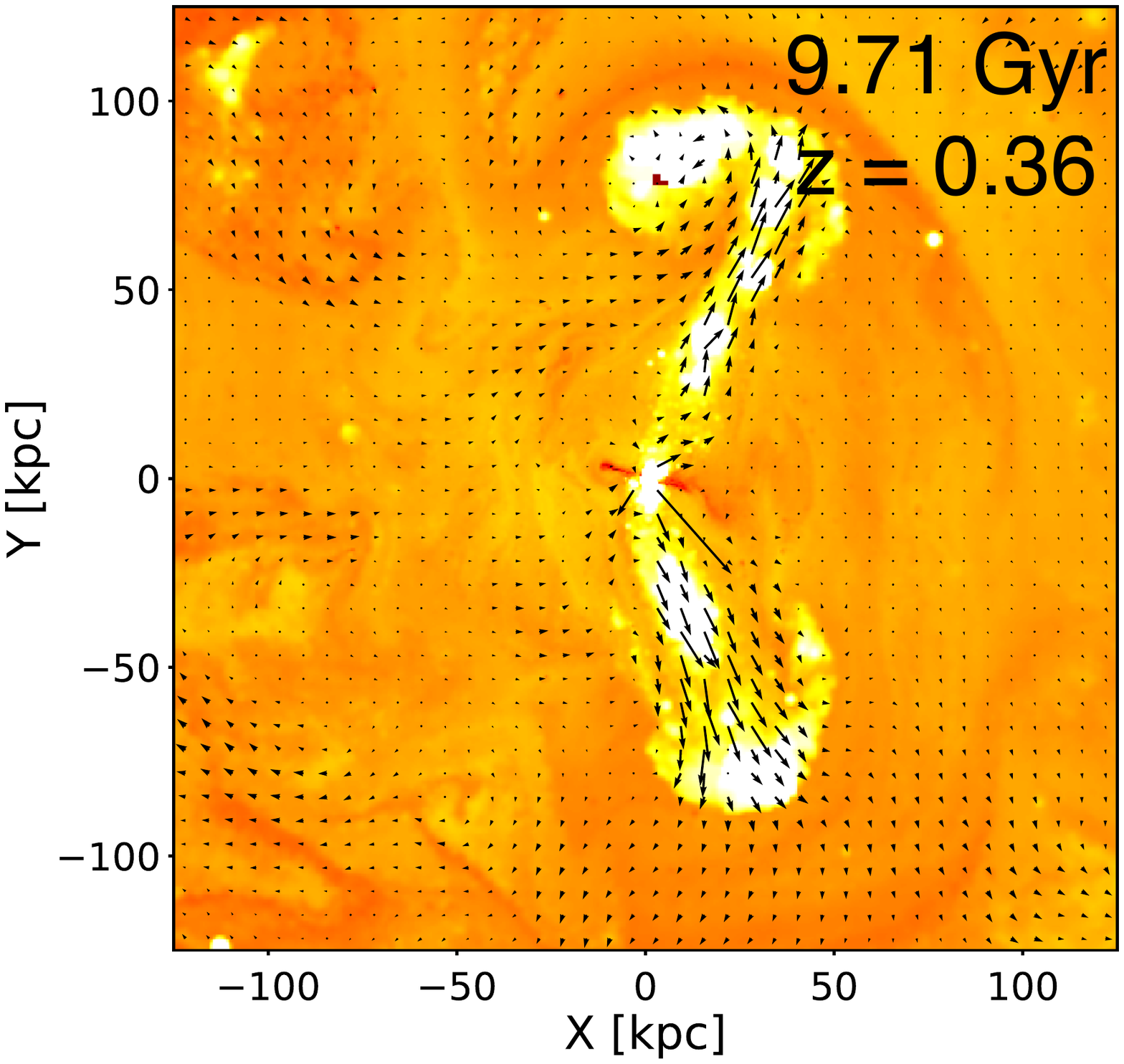}
\includegraphics[trim=67mm 5mm 55mm 30mm, clip, width=53mm]{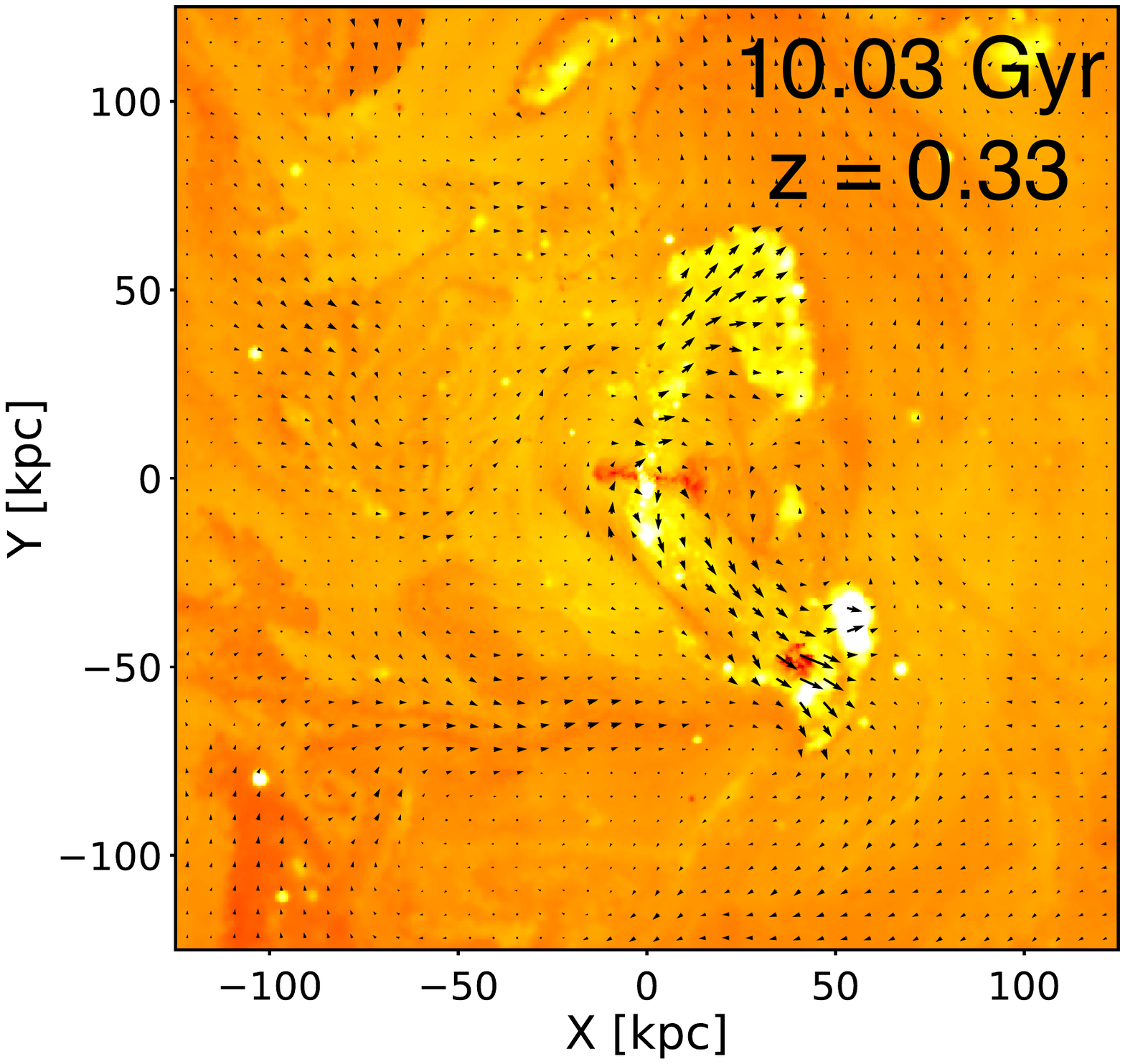}
\includegraphics[trim=67mm 5mm 55mm 30mm, clip, width=53mm]{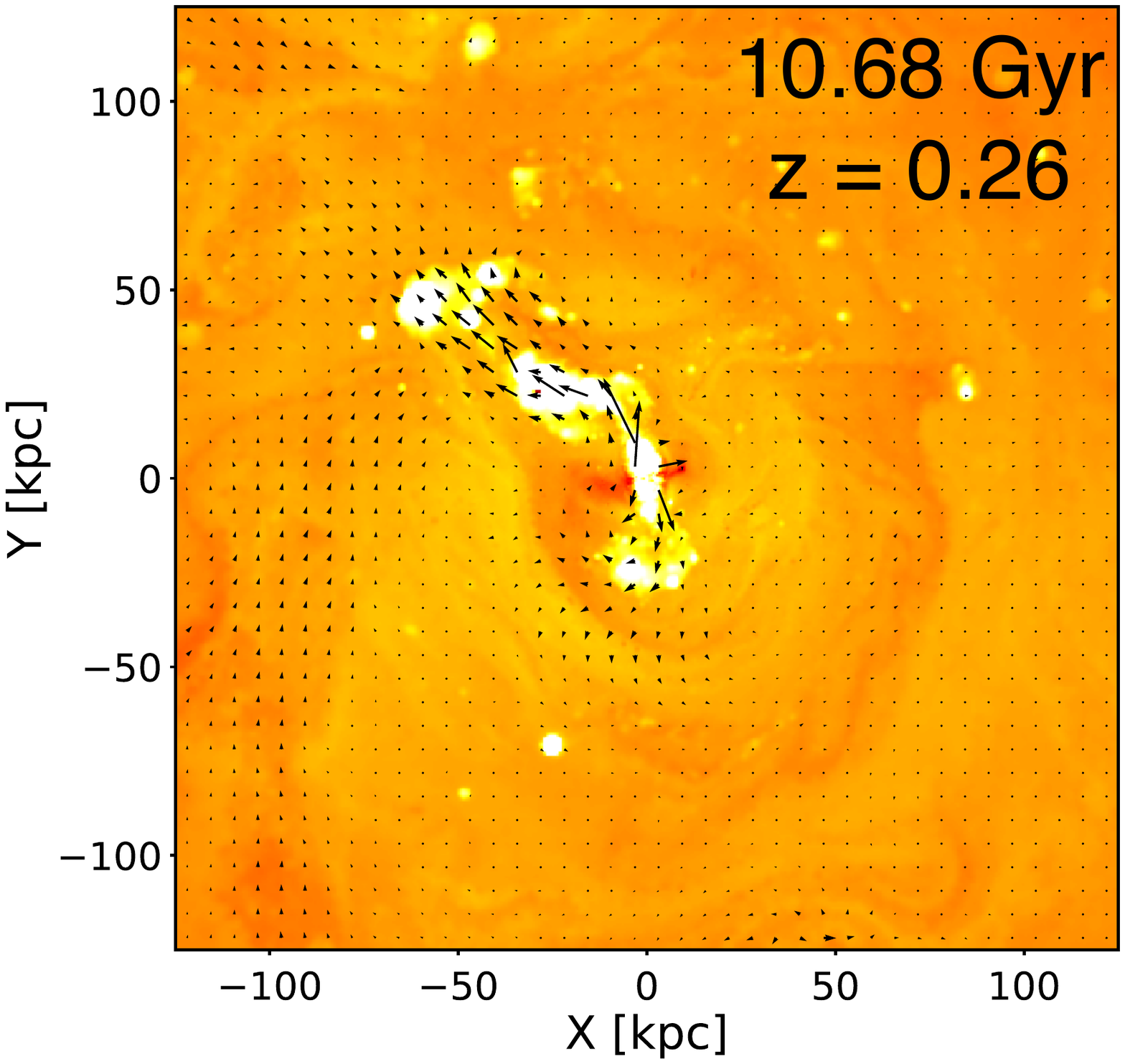}
\caption{{\sc AGN-driven Outflows Over Time}. Snapshots of the gas temperature and velocity field in the inner regions of {\sc RomulusC} over cosmic time. Each image shows the same projection, modulo rotations about the y-axis in order to best image capture any outflow structures present. The colors are the same as Figure~\ref{outflow_image}. The temperatures are mass weighted averages along a 2 kpc slice centered at halo center. It is clear that the winds evolve in direction, shape, and magnitude. The direction is dictated by the rotation of gas in the inner 10-20 kpc of the halo as well as the larger scale motions of the ICM. Between 8 and 10 Gyr, the outflows are particularly strong, corresponding to a period of both high SMBH activity and declining star formation in the BCG (see Figure~\ref{sfh_bhacc}).}
\label{outflow_time}
\end{figure*}

\subsubsection{The Connection between AGN Feedback and BCG Quenching}

 \begin{figure}
\centering
\includegraphics[trim=5mm 0mm -10mm 5mm, clip, width=90mm]{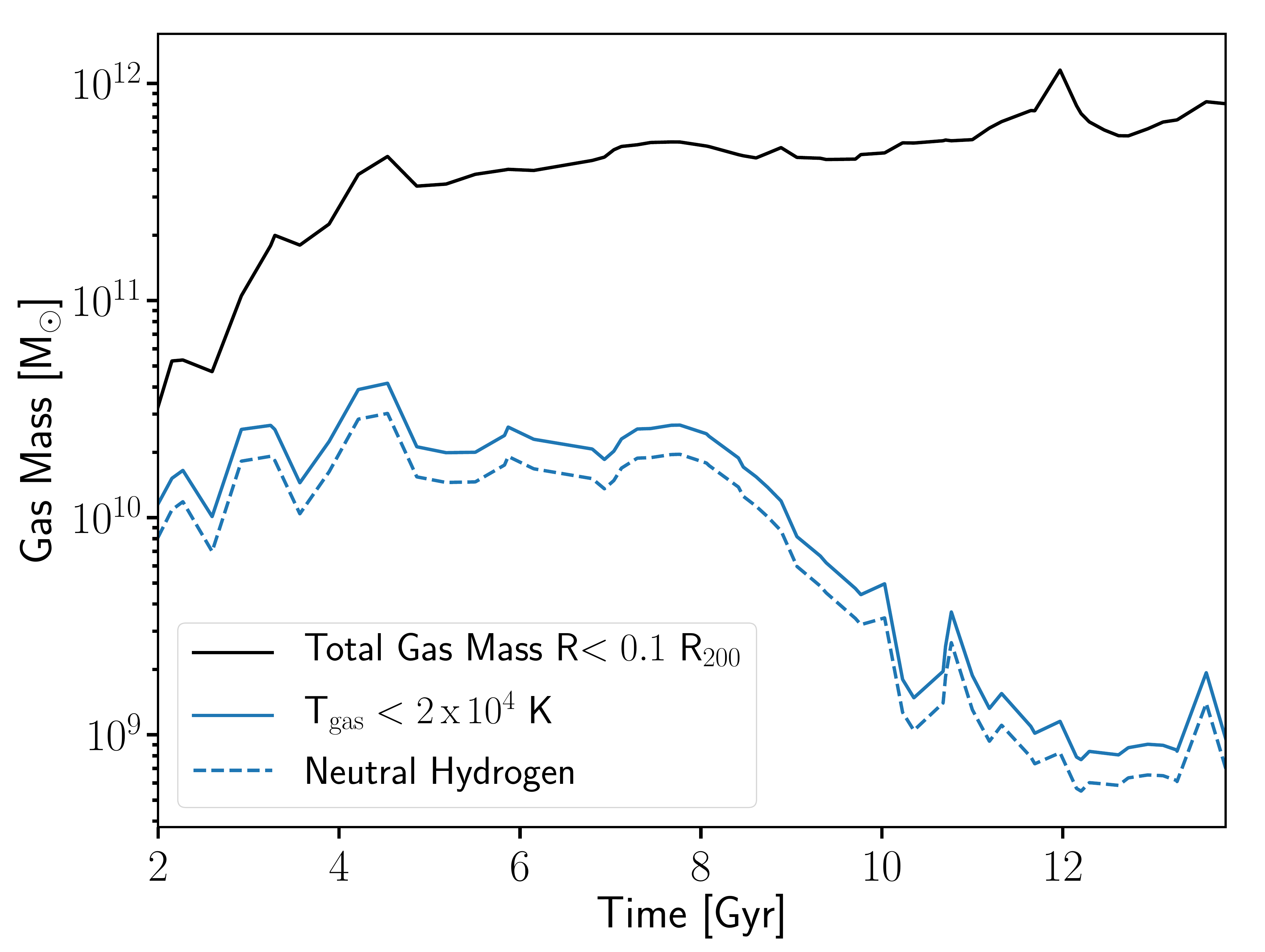}
\caption{{\sc Central Gas Content of the BCG}. The total (black line), cold (blue solid) and HI (blue dashed) gas masses within the central 0.1 R$_{200}$ for {\sc RomulusC}. Large scale feedback from the central SMBH taking place after 8 Gyr results in depletion of cold gas in the central regions without affecting the overall gas supply.}
\label{gasfrac}
\end{figure}

 \begin{figure*}
\centering
\includegraphics[trim=10mm 0mm 10mm 0mm, clip, width=180mm]{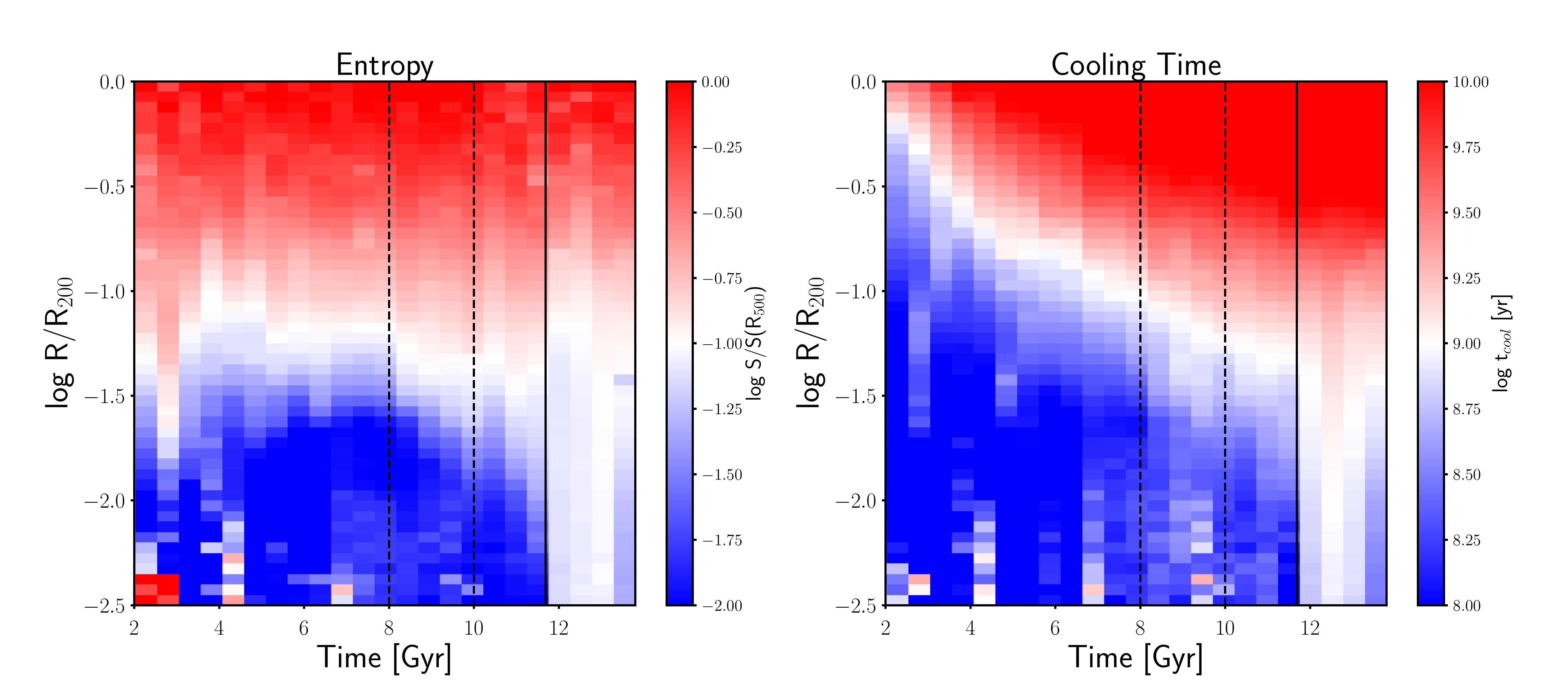}
\caption{{\sc Survival of the Cool Core}. Entropy and cooling time profiles as a function of time for {\sc RomulusC}. Up until the onset of a major merger, marked by the vertical black line, the entropy profile remains steeply declining toward the center of the halo and the cooling times remain below 1 Gyr, both important characteristics of a cool core cluster. Between 8 and 10 Gyr, the star formation and amount of cold gas in the central regions declines due to large scale AGN feedback, but the cluster maintains a low entropy core with sub-Gyr cooling times. Following the major merger, the entropy profile flattens out and the cooling times become several Gyrs, consistent with non-cool core systems.}
\label{Entropy_tcool}
\end{figure*}

Several previous cosmological simulations have demonstrated the importance of AGN feedback in regulating star formation in massive galaxies \citep{dimatteoBH05,Teyssier11,IllustrisBH15,eagle15,beckmann17,pontzen17} and the same is true for {\sc RomulusC}. The resolution of {\sc RomulusC} allows us to examine the interaction between the central AGN, the ICM, and the gas content and star formation history of the central BCG with unprecedented detail. Figure~\ref{sfh_bhacc} shows the specific star formation rate (sSFR) within $0.1\mathrm{R}_{200}$ and the feedback energy ($\epsilon_r \epsilon_f \dot{M} c^2$) imparted by the central SMBH (taken to be the brightest SMBH inside 10 kpc of halo center at any given time). The SMBH feedback rates are presented in 100 Myr bins to reduce the noise and better show overall trends. The star formation rate of the BCG is calculated in 10 Myr bins.

The specific star formation history closely follows the change in SMBH activity. This is particularly evident at early times. For every trough in the sSFR there is a peak in SMBH activity and when there is a period of SMBH quiescence there is a rise in sSFR. There are a few instances with particularly powerful ($\dot{E}>10^{44}$ ergs/s) feedback events, but the one occurring at $\sim8-10$ Gyrs is the longest. There is a shorter period of energetic AGN activity at 6-7 Gyrs, and another phase occurring at 4-6 Gyrs that is less continuous. It is only during this final 2 Gyr long episode that large-scale, powerful outflows persist (Figure 11) and  star formation finally plummets in the BCG.


It is common for cool-core clusters to have non-negligible star formation ranging from several to 10s and sometimes up to 100s M$_{\odot}$yr$^{-1}$ \citep[e.g.][]{bildfell08,loubser16}. Consistent with these observations, the BCG in {\sc RomulusC} maintains a SFR around 1-10 M$_{\odot}$yr$^{-1}$ even after the sSFR drops well below $10^{-11}$yr$^{-1}$ and the central galaxy would be considered quenched. This low level star formation continues until the cool core of the cluster is disrupted by an infalling group at $z\sim0.2$. For the purposes of our analysis here we do not focus on the ICM or BCG evolution during this merger, which is still ongoing at $z=0$, but analysis of the impact of this event will be the focus of future work. The star formation history of the BCG is remarkably similar to the median sSFR values presented in \citet{bonaventura17}, which are derived from IR detections of clusters. The \citet{mcdonald16} results use multiple methods to estimate star formation at various wavelengths, but find that cool core clusters have systematically higher SFRs compared to their overall sample, which could explain why {\sc RomulusC}, which maintains a cool core until $z\sim0.2$, would also have comparatively more star formation.

It is important to remember that the thermal coupling of AGN feedback is a \textit{local} phenomenon in the simulation. Energy is transferred only to the 32 nearest gas particles (generally within $\sim100 pc$ of the SMBH). Any outflows that are generated are the natural consequence of hydrodynamic processes occurring as a result of this thermal heating.

Figure~\ref{outflow_image} shows the central region of {\sc RomulusC} at $z=0.53$, just after the sSFR begins to decline and the AGN begins a prolonged period of activity. Different properties of the gas are shown, averaged by mass along a 30 kpc slab (5 kpc for the inset figures), except for the collumn density plots which are all integrated over 100 kpc, similar to observed radio sources and X-ray cavities that can extend out to 10s to over 100 kpc from the center of the BCG \citep[e.g.][]{mcnamara00,mcnamara09,mcnamara12, osullivan12}. The large-scale, collimated outflow is clearly seen in the temperature and entropy figures, which both have velocity fields overplotted. The outflowing gas is typically traveling at a few thousands of km/s and extends beyond $0.1R_{200}$. Such large-scale outflows dissipate their energy through shocks, as well as turbulent dissipation, and are able to contain cooling within the central regions of the halo. Shocks can be seen in the pressure plot propagating through the innermost core of the halo. Two cavities can be seen in the collumn density figure. The are associated with the end of the outflow and are reminiscent of x-ray cavities (or radio lobes) expanding due to the injection of hot, high pressure gas from the outflows.

Figure~\ref{outflow_time} shows several timesteps between 5 and 11 Gyrs. Large-scale outflows from the AGN are commonplace throughout the simulation, but the powerful outflows taking place at $t>8$ Gyr are able to finally quench star formation. The onset of quenching and these powerful outflows are coincident with a prolonged phase of SMBH activity with feedback rates exceeding $10^{44}$ ergs/s nearly continuously over 2 Gyr.

The outflows also interact with the ICM on larger scales and change directions due to bulk shear flows. Such mechanisms involving `ICM weather' have been suggested as a way to overcome the problems with fixed-direction jets \citep[e.g.][]{heinz06,soker06,morsony10b,mendygral12}. At the time shown in Figure~\ref{outflow_image} ($z=0.53$) we do find a shear velocity of $\sim100$ km/s between the gas at scales below 50 kpc from the cluster center and that between 50 and 100 kpc. We also measure a shear of $\sim60$ km/s between 0-30 kpc and 30-60 kpc. These values are relatively small and consistent with other cosmological simulations \citep{lau17} and slightly lower than results from Hitomi's observations of the Perseus cluster \citep{hitomi16}. We confirmed that the central BCG is not moving significantly with respect to the center of mass of the cluster (as measured from a variety of different radial scales). Rather, the shear observed here is likely due to a recent pericenter passage of a cluster galaxy. Shear velocities as low as 100-300 km/s may be able to have a substantial impact on outflow structure \citep{hardcastle05}. Of course, the structure of the outflow is developed over a long period of time while here we only examine a single snapshot. A deeper analysis is needed to better understand this large-scale evolution of the wind structure, which will be conducted in future work.

\subsubsection{The Effect of AGN Feedback on the ICM}

AGN feedback does not limit star formation by evacuating nor directly heating gas in the inner cluster core. Figure~\ref{gasfrac} plots the mass of gas within $0.1R_{200}$ as a function of time. The overall supply of gas rises at early times while the cluster progenitor is still growing rapidly (see figure 2), then remains nearly constant from $\sim4$ Gyrs onward. However, following the series of very powerful feedback events at $t>8$ Gyr, the cold gas (or, equivalently, the HI gas, which is tracked self-consistently in the simulation) supply declines along with the sSFR. During this time when quenching is in progress or completed (8-11.7 Gyr), the AGN is not drastically increasing the entropy or cooling time of the gas. Figure~\ref{Entropy_tcool} shows the time evolution of the mass weighted entropy and cooling time profiles for {\sc RomulusC}. Within the central $\sim0.05R_{200}$, both entropy and cooling times are low and indicative of a cool-core/relaxed ICM. The time during which the central BCG is becoming quenched is marked between two vertical dashed lines. The cluster core is able to survive with low entropy and cooling timescales of $\sim10^8$ yrs throughout the period where star formation is quenching. This is consistent with observed findings that cool core clusters are more likely to host radio loud AGN \citep[e.g.][]{mittal09}. It is also in agreement with high resolution simulations of isolated galaxies, which show how AGN-driven outflows have little direct effect on the gas within their host galaxy \citep{gabor14}.

The vertical solid line indicates the time where a major merger with a nearby group causes the cool core to be destroyed, an event that is apparent in the nearly flat entropy and cooling time profiles after this time. As stated previously, we will examine this merger event and cool-core destruction in future work, but use it here as an illustrative comparison between the effect of a major merger and that of AGN feedback.

\subsection{Cluster Member Galaxies}

An important advantage of {\sc RomulusC}'s resolution is the ability to better resolve cluster galaxies down to smaller masses than ever before. Our threshold for what is `resolved' in {\sc RomulusC} is very conservative. In the following analysis we only consider halos of total mass at least $3\times10^9$ M$_{\odot}$, corresponding to a minimum dark matter particle count of $\sim10^4$ per halo. We compare galaxy evolution in the cluster and proto-cluster environments simulated in {\sc RomulusC} to galaxy evolution in isolated field galaxies simulated in {\sc Romulus25}. 

The zoom-in Lagrangian region used to model {\sc RomulusC} extends, at $z = 0$, approximately out to 2R$_{200}$. We only include galaxies in our analysis that are `uncontaminated' with low resolution dark matter particles, thereby selecting only those galaxies that lie well within the high resolution region. We still caution that galaxies near the boundaries may still have been affected by the lower resolution regions nearby, but even when we include all uncontaminated galaxies in the zoom region such affected galaxies would be rare. While we do discuss projection effects in the following sections, all results are presented using 3D radial bins relative to the cluster center. Having only a single system makes {\sc RomulusC} more susceptible to spurious results due to the exact choice of projection. Further, because of the limited zoom-in region size, we would still not be able to approximate the full effects of contamination from non-cluster galaxies at large radii.

In order to compare with galaxy evolution in relatively low density environments, we extract central, relatively isolated galaxies from {\sc Romulus25} based on the criteria that they do not exist within R$_{200}$ of any halo hosting a central galaxy of similar or greater stellar mass. For galaxies in {\sc Romulus25} with stellar masses below $10^{10}$ M$_{\odot}$, we apply an additional criteria that they be no closer than 1.5 Mpc from any galaxy with stellar mass greater than $2.5\times10^{10}$M$_{\odot}$ to be considered isolated field galaxies. This is motivated by results from \citet{geha12} that show environmental quenching in low mass galaxies taking place at such scales. Following the results of \citet{munshi13} that account for limitations in observing the total stellar mass of a galaxy, we define the observed stellar mass of a halo's central galaxy to be $0.6$M$_{\star,tot}$, where M$_{\star,tot}$ is the total stellar mass of the halo. We confirm that at the halo masses we examine here, and given our criteria for isolated galaxies, our results would not change were we to explicitly remove satellite galaxy contributions, which do not account for a significant portion of the total stellar mass.

\subsubsection{Environmental Quenching}

\begin{figure*}
\centering
\includegraphics[trim=20mm 5mm 20mm 10mm, clip, width=150mm]{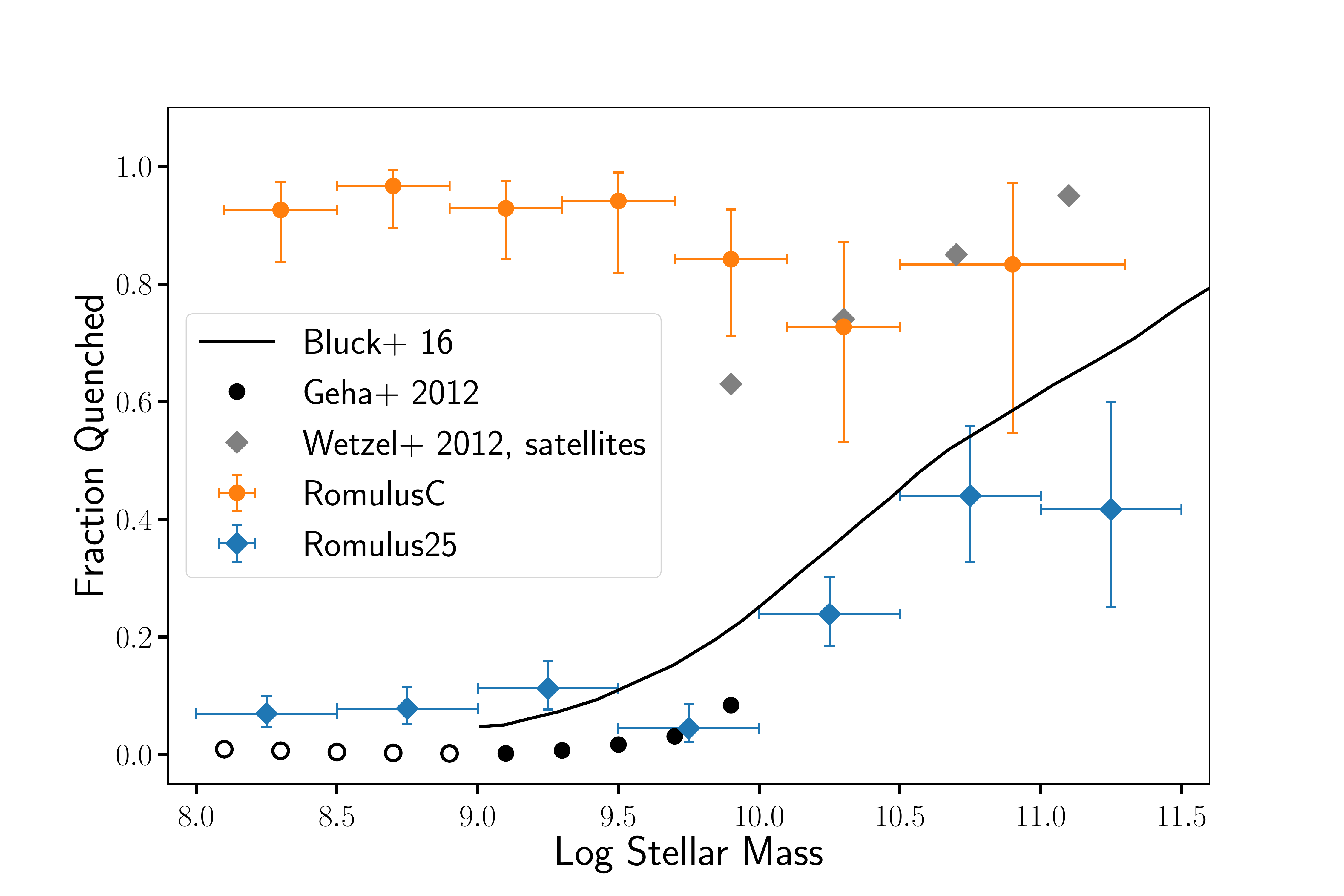}
\caption{{\sc Quenching as a Function of Stellar Mass.} The fraction of quenched galaxies as a function of stellar mass for both {\sc RomulusC} (orange points) and {\sc Romulus25} (blue diamonds). At all but the highest masses we find that satellite galaxies in {\sc RomulusC} have a much higher quenched fraction than isolated galaxies. Our results are consistent with the quenched fraction found by \citet{wetzel12}, but due to small number statistics we cannot confirm that we also see a trend at high mass. At low masses we find the quenched fraction remains nearly constant at 80-90\%. Error bars represent the 68\%  binomial confidence interval \citep{cameron11}. Open points are also from \citet{geha12} and represent upper limits.}
\label{quench_mass}
\end{figure*}

\begin{figure*}
\centering
\includegraphics[trim=20mm 5mm 20mm 10mm, clip, width=150mm]{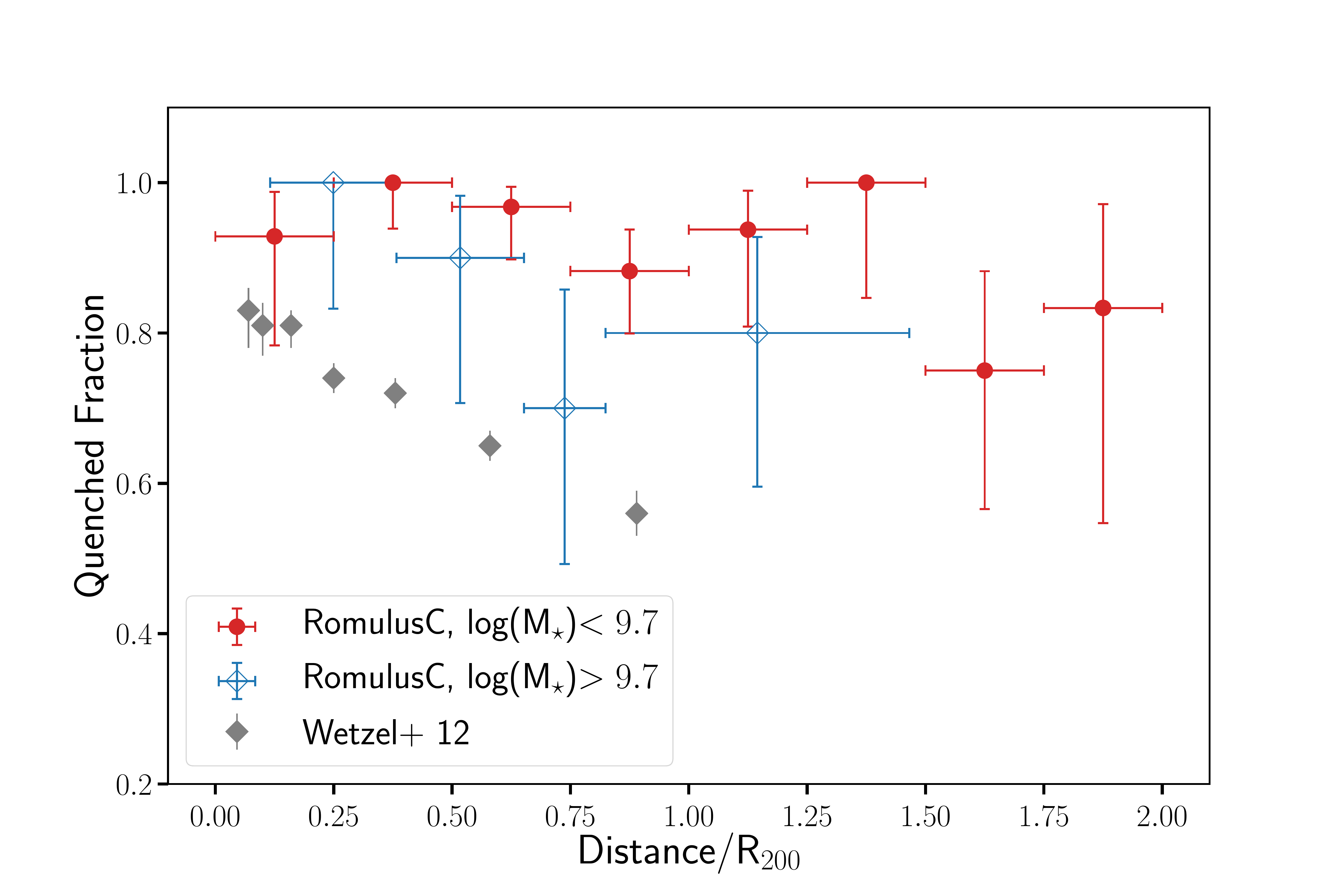}
\caption{{\sc Quenching as a Function of Position in a Cluster}. The quenched fraction as a function of distance from halo center for two mass bins. At high mass (blue, open points; chosen to match the masses of observed cluster galaxies from \citet{wetzel12}) our values are high compared with observations, but as explained in the text this is due to the lowest mass galaxies in this bin. At low masses we predict that no trend exists. At all radial distances low mass galaxies have a quenched fraction of 80-90\%. Error bars represent the 68\%  binomial confidence interval \citep{cameron11}}
\label{quench_distance}
\end{figure*}

 \begin{figure*}
\centering
\includegraphics[trim=0mm 0mm 0mm 0mm, clip, width=150mm]{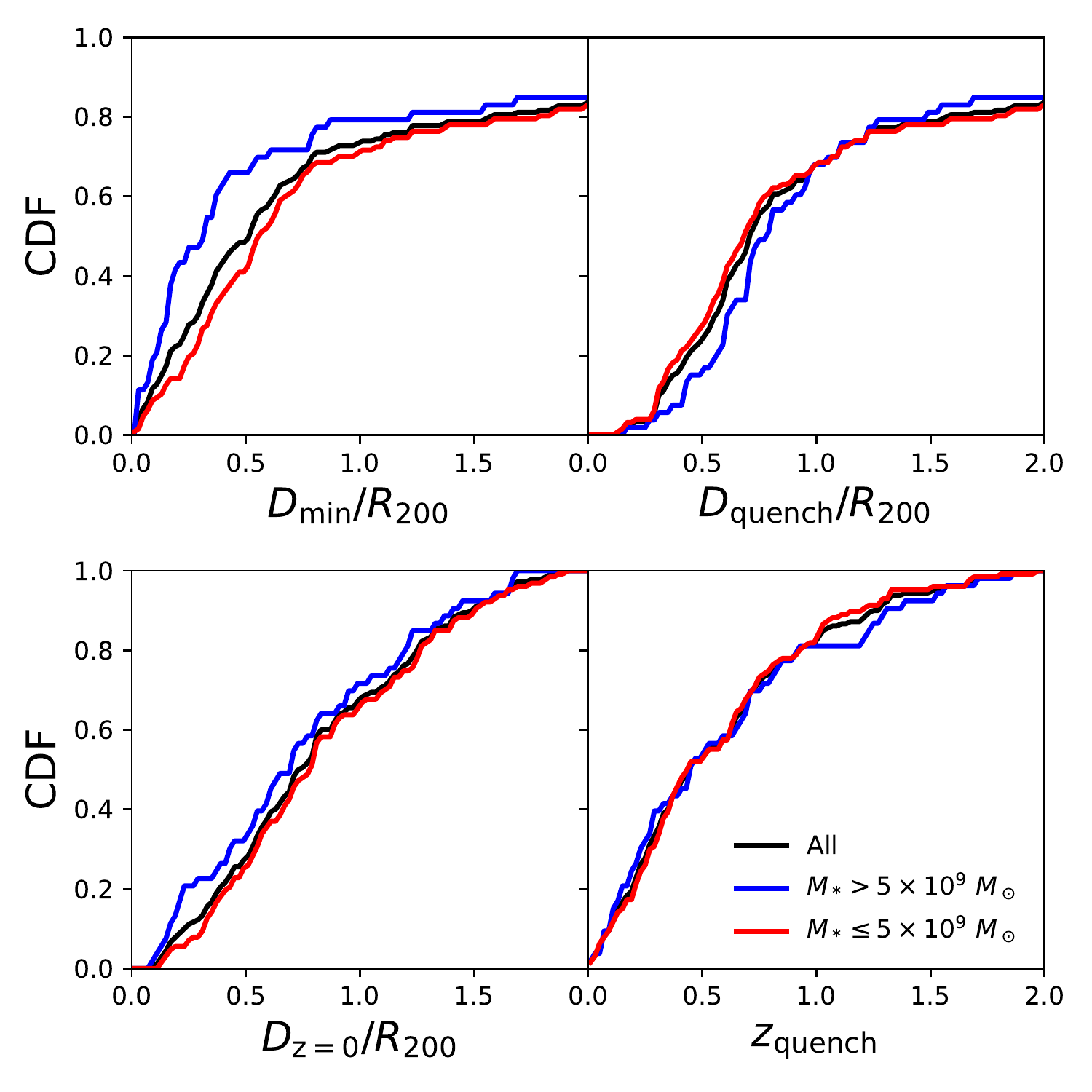}
\caption{{\sc When and Where Cluster Satellites Quench.} Cumulative distribution functions for quenched massive (blue) and low mass (red) cluster galaxies. \textit{Upper Right:} The distribution of minimum distances relative to the cluster center prior to quenching. \textit{Upper Left:} The distance at which the galaxies quench (fall below 1 dex relative to the main sequence). \textit{Lower Left:} The distance of the galaxies at $z=0$. \textit{Lower Right:} The redshift when the galaxies quench. There is little distinction between high and low mass quenched galaxies in all cases, but for the minimum distance to the cluster center achieved prior to quenching. High mass quenched galaxies are more likely to have had orbits that took them closer to the cluster center. This distinction, however, is mostly lost by $z=0$.}
\label{quench_track}
\end{figure*}

In order to obtain a self consistent picture of quenching in both {\sc RomulusC} and {\sc Romulus25}, we define a star forming main sequence based on central, isolated galaxies in {\sc Romulus25}. We follow a similar procedure to observations \citep[e.g.][]{bluck16} and fit the median values of the star formation rate within 0.1 dex bins of stellar mass between $10^8$ and $10^{10}$ M$_{\odot}$. We find a best fit main sequence of $\log(\mathrm{SFR}) = 1.206\times \log(M_{\star}) - 11.7$ for $z = 0$. While this differs from main sequence definitions derived from observations in both slope and normalization, for stellar masses between $10^8$ and $10^{10}$ M$_{\odot}$ the fit lies within the 0.5 dex scatter associated with the observed main sequence. We take any galaxy whose star formation is a factor of 10 below our fitted main sequence to be quenched at $z = 0$. To calculate star formation, we use the formation times of star particles within each halo and calculate the average formation rate in the previous 25 Myr. However, because of limited mass resolution, the smallest possible star formation rates are quantized and therefore subject to numerical noise. For halos that formed only 2 or fewer particles in the previous 25 Myr (corresponding to a SFR of 0.064 M$_{\odot}$ yr$^{-1}$), we calculate the SFR averaged over the previous 250 Myr instead. We confirm that our results are insensitive to the choice of timescale over which to measure SFR except for the lowest mass galaxies.

This definition of `quenched' does differ from some observations, including \citet{wetzel12}, who adopt a flat threshold of $10^{-11}$ yr$^{-1}$. Based on our main sequence definition, quenched galaxies are defined on a mass dependent specific SFR threshold between $10^{-11}$ and $10^{-10.5}$ yr$^{-1}$ at $z=0$ across the stellar mass range we cover. \citet{wetzel12} find what would be considered a nearly constant main sequence in specific SFR, while our main sequence has a slight evolution with mass. As discussed briefly in \citet{bluck16}, because the simulations are not equipped to fully mimic the SFR diagnostics that observers use, such a definition allows us to define a quenched threshold that is fully self consistent for our simulated galaxies while still maintaining the ability to compare with observations that have a different distribution of (inferred) SFRs. Our definition is also not reliant on the factor of $\sim2$ difference between the total stellar masses of our simulated galaxies and those that would be inferred by observations \citep{munshi13}, which would add further uncertainty to our results. Finally, this threshold definition allows us to derive self consistent quenched fractions at different redshifts. We confirm that changing our threshold to a flat sSFR value of $10^{-11}$ yr$^{-1}$ will affect the $z=0$ classification of a handful of the most massive galaxies in both {\sc Romulus25} and {\sc RomulusC}, bringing the quenched fraction lower for the highest mass bin in both simulations but still within the 68\% confidence interval for our fiducial definition (see Figure~\ref{quench_mass}). We also confirm that our definition of quenched is consistent with a definition based on UVJ colors \citep[e.g.][]{whitaker11}.

Figure~\ref{quench_mass} plots the quenched fraction of galaxies as a function of mass for both {\sc Romulus25} (blue) and {\sc RomulusC} (orange). In both simulations there are $>20$ galaxies per bin ($>100$ for the lowest masses) except for the highest mass bins which contain 7 and 12 galaxies for {\sc RomulusC} and {\sc Romulus25} respectively. For {\sc RomulusC} galaxies, here we only consider those within R$_{200}$ of the cluster center in order to compare directly with observations from \citet{wetzel12}. The quenched fractions from {\sc Romulus25} are mostly consistent with results from SDSS \citep{bluck16} at higher masses, following the same increasing trend with stellar mass, although the quenched fractions from {\sc Romulus25} are biased low, particularly in the highest mass bin. However, the fact that our predictions are roughly consistent with observations and follow the same trend in gradually increasing with mass is encouraging. More work is needed to explain this discrepancy which is beyond the scope of this paper.

At the low mass end, {\sc Romulus25} predicts very low quenched fractions, similar to observations. However, at the lowest masses our fractions of several percent are still significantly higher than observations from \citet{geha12}. Exploring this difference in more detail is beyond the scope of this paper, but can be due to several factors. Our definition of quenched is based on direct star formation rates from the simulation, as opposed to \citet{geha12}, where quenching is defined on the basis of H$\alpha$ emission as well as old stellar ages derived from from D$_{4000}$. It is possible that low level and/or recent star formation in these small galaxies would make them non-quenched in the \citet{geha12} definition. There is also evidence that a subset of low mass galaxies have low gas fractions and emission consistent with AGN \citep{bradford18}, which could provide enough H$\alpha$ to make a galaxy appear star forming. In future work we plan on doing a more in-depth analysis of the dwarf galaxy population in {\sc Romulus25} and comparing it in a more self consistent way to observations. For the purposes of this work, the important point is that we predict very low quenched fractions for low mass galaxies in isolation.

\begin{figure*}
\centering
\includegraphics[trim=20mm 10mm 20mm 10mm, clip, width=180mm]{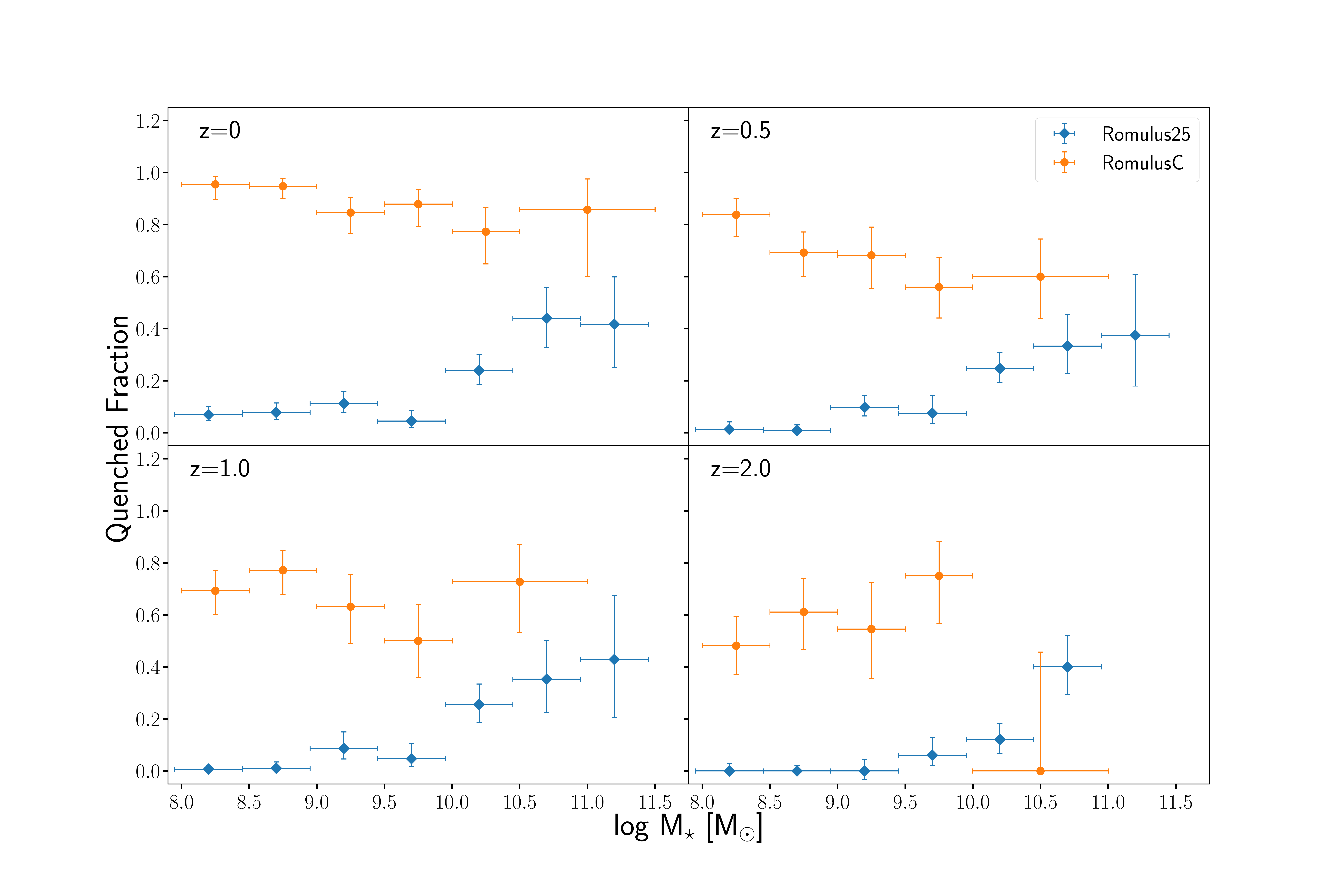}
\caption{{\sc Quenching Over Cosmic Time}. The fraction of quenched galaxies as a function of stellar mass in {\sc RomulusC} (orange) and {\sc Romulus25} (blue) at four different redshifts. All galaxies within $2R_{200}$ are shown. The enhanced quenching seen in high density environments like {\sc RomulusC} is in place even at high redshift and for low mass galaxies. Error bars represent the 68\%  binomial confidence interval \citep{cameron11}}
\label{quench_multiz}
\end{figure*}

\begin{figure*}
\centering
\includegraphics[trim=20mm 10mm 20mm 10mm, clip, width=180mm]{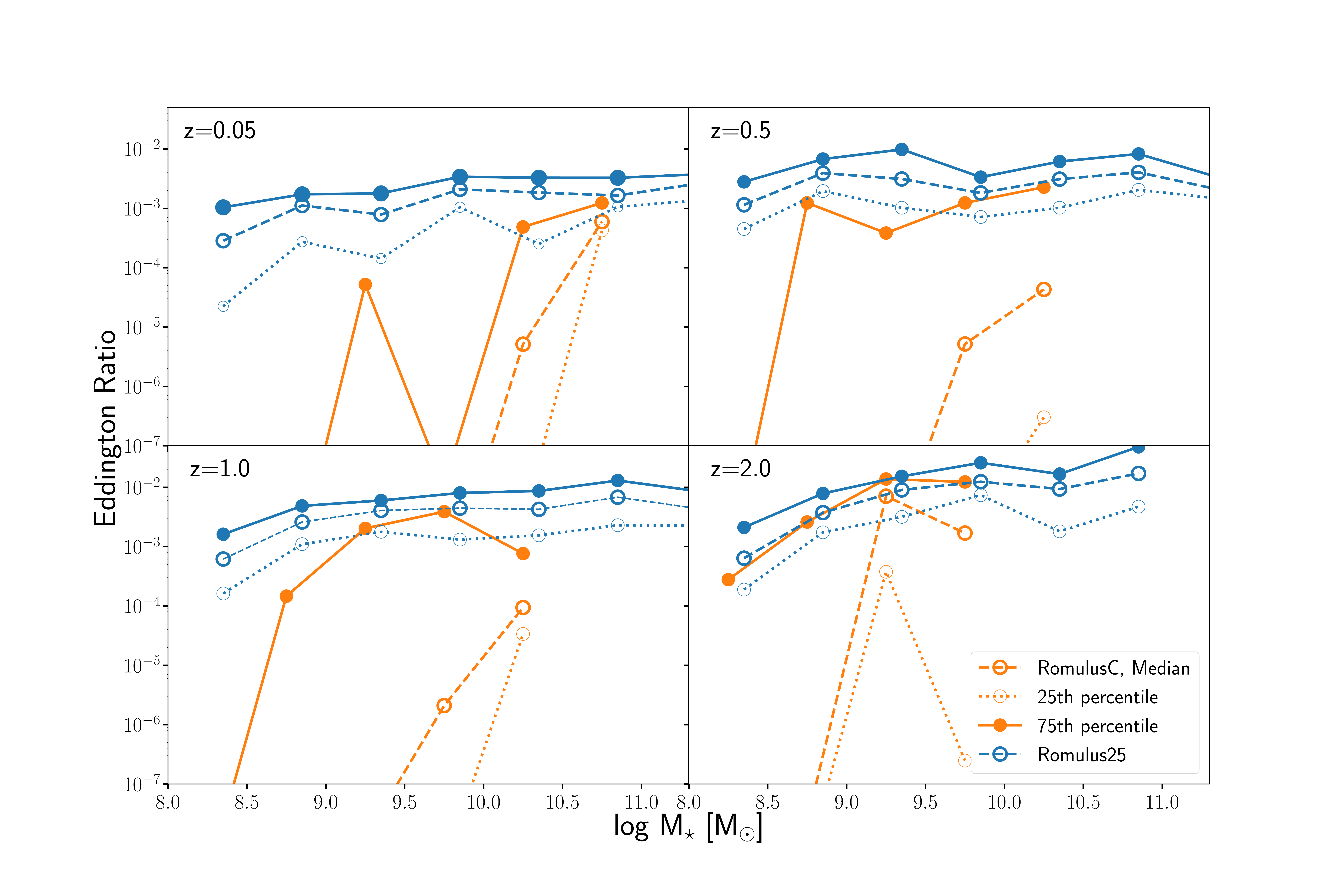}
\caption{{\sc SMBH Growth Over Cosmic Time}. The distribution of Eddington ratios for SMBHs in different environments. The 25th, 50th, and 75th percentiles are shown for SMBHs in different mass galaxies in high density environments ({\sc RomulusC}, orange) as well as relatively isolated galaxies in the field ({\sc Romulu25}, blue). In the cluster environment, there is a much more significant fraction of SMBHs that are growing at extremely low rates and in some cases not going at all. This occurs at all mass scales and at all redshifts. The highest Eddington ratio SMBHs are most similar between the different environments, especially at higher redshift.}
\label{edd_multiz}
\end{figure*}

\begin{figure*}
\centering
\includegraphics[trim=20mm 10mm 20mm 10mm, clip, width=180mm]{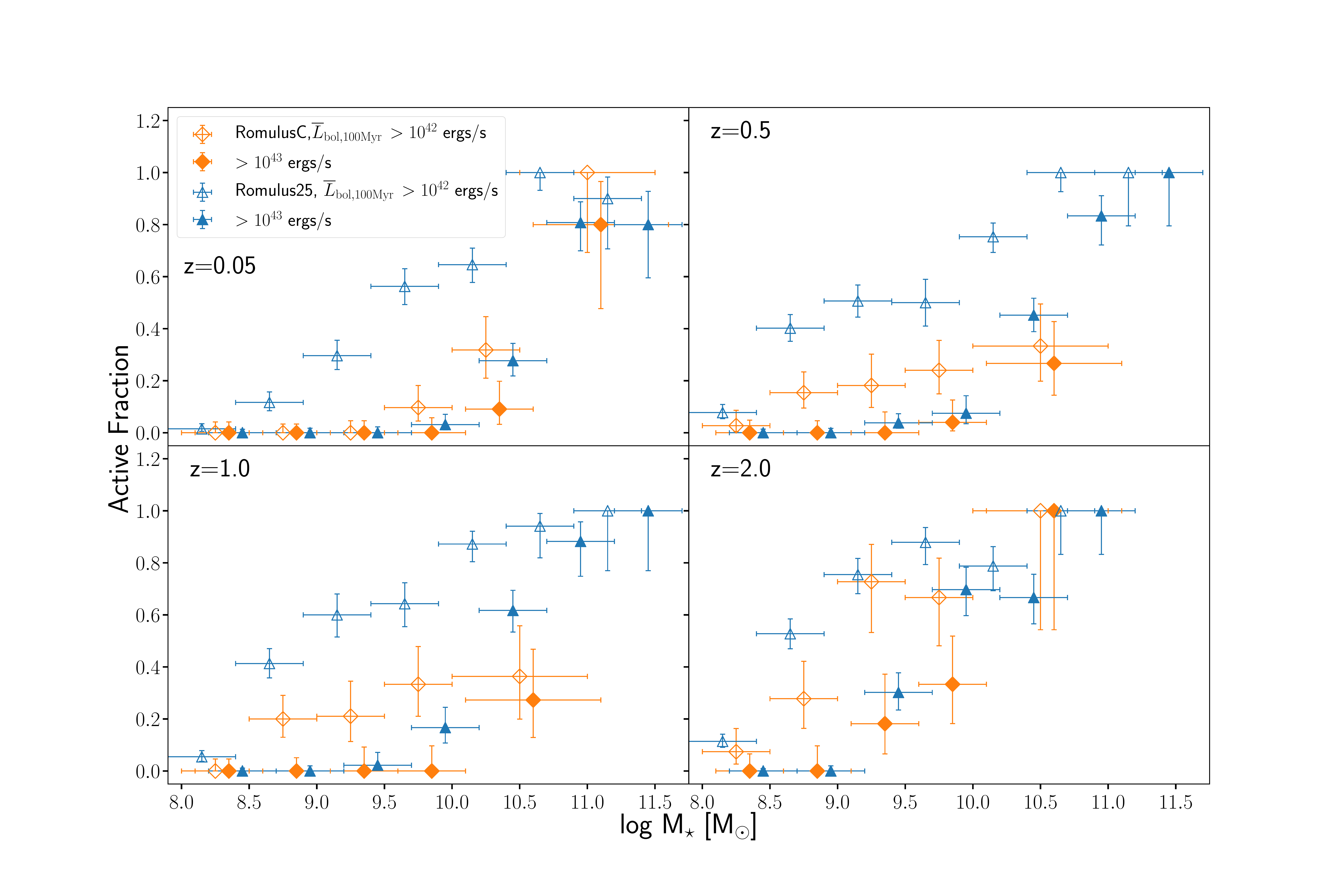}
\caption{{\sc The Population of Luminous AGN}. The fraction of galaxies hosting SMBHs that have a luminosity above a threshold of $10^{42}$ (open points) and $10^{43}$ (closed points) ergs/s as a function of stellar mass at four different redshifts. Luminosities are averaged over the previous 100 Myr. Note that we do not include $z=0$ because the black hole properties are only followed to $z=0.05$ in {\sc Romulus25}. Low luminosity AGN are much more affected by their environment while high luminosity AGN, which occur only in more massive galaxies, seem relatively insensitive to being in a cluster versus relatively isolated galaxy. Both AGN fractions evolve significantly with redshift such that, by $z=2$, both are very similar to the field. This is in contrast with Figure~\ref{edd_multiz}, which shows a substantial population of very low Eddington ratio SMBHs. When only focusing on high luminosity (which means high accretion rate) much of the difference in environment seen in the Eddington ratio vanishes, leading to claims that observed high redshift AGN population in clusters are similar to or even enhanced compared with the field. Error bars represent the 68\%  binomial confidence interval \citep{cameron11}}
\label{agn_multiz}
\end{figure*}

For {\sc RomulusC}, the highest mass bins are overall comparable to the results from \citet{wetzel12}, but we do not see the same trend with stellar mass. Of course, given our relatively low number statistics at the higher masses, such a trend would be difficult to resolve. Here we have also tailored our bins to match those used in \citet{wetzel12} at masses greater than $10^{9.7}$ M$_{\odot}$, but we combine the final two in order to have enough galaxies in the bin. We do predict significantly higher quenched fractions at M$_{\star} = 10^{9.7-10.1}$ M$_{\odot}$.  We see a significant difference in the quenched fraction at high masses between cluster and field environments, unlike observations \citep{wetzel12,bluck16}. This is likely due to the fact that {\sc Romulus25} under-produces quenched galaxies at higher masses, making the environmental effects of the cluster more apparent.


For low mass galaxies, {\sc RomulusC} predicts a nearly constant quenched fraction as stellar masses get smaller. The trend with stellar mass observed at high masses by \citet{wetzel12}, therefore, does not continue to lower mass galaxies, with 80-100\% of galaxies below $10^{10}$ M$_{\odot}$ predicted to be quenched within R$_{200}$ independent of their stellar mass. This is in stark contrast to galaxies  in the field from both observations and {\sc Romulus25}. While the ability to observe such low mass galaxies is limited, there have been detailed observations within a small number of nearby clusters that show a significant population of quenched dwarf galaxies \citep[e.g.][]{drinkwater01,weinmann11, balogh16,roediger17, habas18}. \citet{weinmann11} find varying results as to how quenched fractions change with galaxy mass. For Virgo and Coma they find little dependence on luminosity, but clear dependence in Perseus. In all three cases, the quenched fractions presented in \citet{weinmann11} range from $\sim70-90\%$ for dwarf galaxies. The quenched fractions in {\sc RomulusC} are on the high end of these observed clusters, though not inconsistent. These results are also consistent with \citet{geha12}, who find that the quenched population of low mass galaxies quickly increases with proximity to higher mass galaxies, even outside of R$_{200}$. Finally, we also note that these results are consistent with lower resolution cluster simulations \citep[e.g.][]{bahe17} that will be discussed further in \S5.2.


Figure~\ref{quench_distance} plots the fraction of quenched galaxies as a function of radial distance from the cluster center at $z=0$ for two stellar mass bins. The high mass bin was chosen to match that probed by \citet{wetzel12} and we find that our results are biased high in comparison. According to Figure~\ref{quench_mass}, this is due to the $10^{9.7-10.1}$M$_{\odot}$ galaxies being more quenched in {\sc RomulusC} than in the \citet{wetzel12} observations. For these higher mass galaxies we choose the radial bins such that each bin contains 10 galaxies. We confirm that, were we to only include galaxies with M$_{\star}>10^{10}$ M$_{\odot}$, our values would be much more similar to the \citet{wetzel12} results. Again we note that projection effects can affect the results from observations such that these points should be considered lower limits, particularly at large separations where contamination from the field can be important. Indeed we find that the quenched fractions in several radial bins decreases for {\sc RomulusC} if we were to use the projected distances, though the exact effect depends on the line of sight. 

While at high masses the radial dependence is difficult to estimate based on the large error bars, we do see some evidence for a decreasing quenched fraction with radius, though possibly not as steep as that presented in \citet{wetzel12}. For the low mass bin, we find with more confidence that there is no radial dependence, and the quenched fraction remains steady at $\sim80-100\%$ out beyond R$_{200}$. There is some evidence that the quenched fraction does fall off beyond $\sim1.5R_{200}$, but we would need a larger sample of galaxies and a larger zoom-in region in order to know for sure.


Our results indicate that the processes causing galaxy quenching in high density environments are much more efficient compared to the field for all but possibly the highest mass galaxies. For low mass galaxies, this process is particularly efficient and seems to act evenly at all distance scales. It has been suggested that the main process leading to galaxy quenching in cluster environments is ram pressure stripping and has been supported by observations \citep[e.g.][]{smith10,merluzzi13,boselli14,roedinger15, haines13,haines15} as well as both hydrodynamic simulations and analytic models \citep[e.g.][]{murakami99,hester06,bahe15,mistani16,zinger18}. This would explain why observations of higher mass cluster member galaxies have a quenched fraction that negatively correlates with distance, but we predict such radial dependence does not exist for lower mass galaxies. Ram pressure is more efficient at stripping a small galaxy. For a higher mass galaxy with a larger gas disk, such processes would take longer and require higher ram pressure, thus becoming more efficient for orbits passing closer to halo center. 

In order to examine the process of quenching in more detail, we follow the evolution of each galaxy that is quenched at $z=0$ in {\sc RomulusC} within $2 R_{200}$, the extent of our high resolution region at $z=0$. Figure~\ref{quench_track} plots cumulative distribution functions for various properties of quenched galaxies in {\sc RomulusC}: their final distance from the cluster center, the distance at which they become quenched, their minimum distance prior to quenching, and the redshift at which they become quenched. In order to determine when a galaxy is quenched, we compare to the main sequence fitted to {\sc Romulus25} data as described above. To account for an evolving main sequence, we perform fits at $z = 1, 2, \,\mathrm{and}\, 3$ in addition to our $z=0$ fit and compare galaxies to the main sequence by interpolating between each main sequence. To see whether quenching occurs differently at different masses, we split our sample into high and low mass bins around M$_{\star}=10^{9.7}$M$_{\odot}$. We choose this splitting because it corresponds to the lower mass limit of the \citet{wetzel12} study and represents a rough boundary below which observations of galaxies in cluster environments is limited to a handful of nearby systems. Note that our time resolution for tracking galaxy evolution is limited to $\sim100$ Myr, the time between saved snapshots.

We see little difference between high mass and lower mass galaxies in terms of their quenching redshift, their $z=0$ final distance to the cluster center, and the distance at which they become quenched. We do, however, see a significant difference in the minimum distance relative to the cluster center achieved prior to quenching. Massive, quenched galaxies are more likely to have orbits that take them closer to halo center before they quench. This supports the idea of ram pressure stripping being the dominant process acting on in-falling cluster galaxies, at least at higher masses. At lower masses multiple processes affect the galaxies' gas supply. First, 36\% of lower mass galaxies have been satellites of halos prior to cluster in-fall. These galaxies therefore experienced a phase of `pre-processing' by hot halo gas in another halo. At lower masses, galaxies are also more susceptible to ram pressure stripping because of their shallower potential and often explosive stellar feedback processes \citep{murakami99,bahe15}. This difference may also be due to massive galaxies/halos sinking faster due to dynamical friction, potentially combined with ram pressure taking longer to strip more massive disks. We will examine the role of ram pressure stripping as a driving force of galaxy quenching in more detail in future work.

Importantly, Figure~\ref{quench_track} also shows how the dependence of quenching processes on the position within a cluster can be quickly erased simply by orbital dynamics. The upper left panel implies that quenching in high mass galaxies on average requires closer approaches to the cluster center compared with lower mass galaxies. However, by the time the galaxies quench, this difference is no longer apparent. This is expected if these galaxies are on more eccentric orbits when they quench, as would be the case if they were to quench soon after their initial in-fall before they have virialized, in agreement with results from the Magneticum simulations \citep{lotz18}. These radially plunging orbits would take the galaxies near the cluster center, where they experience a large amount of ram pressure, and then quickly take them back out toward the cluster outskirts. If this first passage is enough to quench star formation, and if the orbit is radial enough, then the galaxy can quickly be taken very far from the cluster center by the time it quenches. As time goes on, the orbits  evolve, particularly as in-falling galaxies virialize and feel the effects of dynamical friction. 


\subsubsection{The Evolution of Star Formation in Cluster Galaxies}

We follow the evolving population of galaxies within $2R_{200}$ of the cluster's main progenitor halo in {\sc RomulusC} in order to examine how the population changes over time relative to field galaxies. Figure~\ref{quench_multiz} shows the quenched fraction of this population of galaxies in {\sc RomulusC} (orange) as a function of their stellar mass at $z=0, 0.5, 1.0$, and $2.0$. At each redshift we compare to the population of central, non-interacting galaxies from {\sc Romulus25} (blue). We find that the enhanced quenched fraction for low mass galaxies is well established at high redshift. Note that Figure~\ref{quench_track} does not indicate that any $z = 0$ cluster galaxy quenches prior to $z=2$. This is because all of the galaxies that are quenched at higher redshifts have since merged with another galaxy, most commonly the BCG, or have been otherwise disrupted or stripped so as to no longer be considered resolved substructure in the simulation (based on our very conservative definition of what is adequately resolved).

\subsubsection{Black Hole Growth in Cluster Galaxies}

{\sc RomulusC} is the first cosmological cluster simulation to include all three of the following: 1) realistic SMBH dynamics \citep{tremmel15,tremmel18}, 2) SMBH accretion that accounts for the kinematics of gas within the galaxy \citep{tremmel17}, and 3) SMBH formation criteria that requires no \textit{a priori} assumptions about halo occupation of SMBHs, seeding them in small halos ($10^8-10^{10} M_{\odot}$) at early times \citep{tremmel17}. All of this, combined with the high resolution of {\sc RomulusC} means that we can accurately follow SMBH evolution within cluster member galaxies, whose gas, as well as overall morphology, is undergoing tremendous evolution through interactions with both the ICM and other galaxies. Our model allows SMBHs to exist and dynamically evolve within low mass and high mass galaxies alike. They are also seeded within those galaxies early enough such that they experience the full effects of the dense cluster environment along with their host galaxy.

Taking galaxies once again within $2 R_{200}$ at different redshifts, we examine their SMBH activity compared with the field. Figure~\ref{edd_multiz} shows the 25th (dotted), 50th (dashed), and 75th (solid) percentiles in average SMBH Eddington ratio over the previous 100 Myr for the SMBH with the highest accretion rate in each galaxy as a function of galaxy mass. The average Eddington ratio is calculated by solving the following equation for $f_{edd}$.

\begin{equation}
M_0 + \Delta M = M_0 \exp \left[ \frac{f_\mathrm{Edd} \cdot (1-\epsilon)}{\epsilon}\left(\frac{100\mathrm{Myr}}{t_\mathrm{Edd}}\right)\right]
\end{equation}

\noindent $M_0$ is the initial mass of the SMBH at $t_0 = t(z) - 100 Myr$ ($t(z)$ is the time at any given redshift) and $\Delta M$ is the amount of growth that took place during the previous 100 Myr. The characteristic timescale for Eddington limited accretion, t$_\mathrm{Edd}$, is given by the following equation given a Thomson scattering cross section, $\sigma_T$, and the mass of a hydrogen atom, $m_h$.

\begin{equation}
t_{Edd} = \frac{\sigma_T c}{4\pi G m_h}
\end{equation}

\noindent Using this relation allows us to account for the fact that the Eddington accretion rate is evolving continuously as the SMBH mass grows. This relation for exponential mass growth is equivalent to what is actually used to calculate mass growth during each SMBH timestep, $\Delta t$, in the simulation ($\Delta M = \dot{M}\times\Delta t$) in the limit as $\Delta t/t_{edd}$ goes to zero \citep{volonteri13}. Given that the average SMBH timestep in {\sc RomulusC} is at most $\sim10^5$ yr (compared with t$_{edd} \sim5\times 10^8$ yr) this is not a bad assumption to make. Nevertheless we confirm that our results remain the same were we to calculate f$_{Edd}$ using $<\dot{M}_{BH}>/\dot{M}_{Edd}$, averaged still over the previous 100 Myr and $\dot{M}_{Edd}$ calculated using the black hole mass at the end of the time bin.

In Figure~\ref{edd_multiz} we see that at lower redshifts there is a dearth of actively growing SMBHs compared to the field which is especially drastic for lower mass galaxies. Looking out to higher redshift, the population of cluster SMBHs with the highest Eddington ratios become more similar to that of the field. Still, out to $z=2$ we see there remains a much more significant population of SMBHs in cluster member galaxies that experience very little growth or none at all. 

Observations of AGN in clusters, particularly those that are X-ray selected, generally can only pick out the highest accretion rates. We find that, while the distribution of Eddington ratios in cluster galaxies is significantly different from the field at all redshifts, the population of bright AGN may not show such a stark difference at all redshifts. In Figure~\ref{agn_multiz} we plot the fraction of galaxies hosting SMBHs that would result in luminous AGN as a function of galaxy stellar mass for two bolometric luminosity thresholds similar to observational limits for AGN detection in the X-ray \citep[e.g.][]{rosario13b,rosario15}. As with the average Eddington ratios, the average luminosity over the previous 100 Myrs is used. At the highest redshifts, the population of luminous AGN is similar between the field and the cluster environments. At lower redshifts, the higher luminosity AGN remain mostly similar to the field, but the low luminosity AGN population, particular those in low mass galaxies, becomes significantly lower in {\sc RomulusC}. 


\section{Discussion}

\subsection{The Success of Thermally Coupled AGN Feedback}

The ability of a thermally coupled feedback prescription for SMBHs to create large-scale, \textit{naturally collimated} winds is an important result of {\sc RomulusC} with critical implications for the cosmological simulation community. It demonstrates that previous failures of thermal coupling of AGN feedback to limit star formation in massive galaxies, often used as justification for implementing more complicated mechanical feedback models \citep[e.g.][]{ragone-figueroa13, choi15}, are not intrinsic to thermal feedback. Rather, we argue that it is the result of limited resolution. With high resolution in mass, space, and time, we are able to more effectively model the interaction between the AGN and surrounding \textit{nearby} gas because that gas is able to appropriately respond to the influx of thermal energy (in part due to our brief cooling shutoff). The inference that limited resolution leads to an over-cooling of gas is not a new concept and has previously been examined in relation to stellar feedback \citep{katz92} and galaxy clusters \citep{lewis00}, but {\sc RomulusC} demonstrates that the solution to the problem for energy input from AGN does not have to lie in a kinetically coupled feedback (i.e. where particles are pushed rather than heated) or a very high coupling efficiency. This is not to say that there aren't important physical motivations behind implementing kinetic or variable efficiency models (e.g. outflows and jets observed to exist in massive galaxies), but this must be properly separated from the issue of overcoming a numerical limitation.

The fact that the thermal feedback naturally forms collimated outflows is an interesting result. This is likely due to both the morphology of the gas near the SMBHs (the outflowing gas will take the path of least resistance) and the  angular momentum of the gas that receives feedback and drives the base of the outflow. The transformation of an initially isotropic outflow into a collimated outflow on large scales is a well studied phenomena. Various works on stellar winds have shown that when an initially isotropic outflow interacts with a medium with anisotropic density and pressure, it will elongate and form a jet-like structure along the steepest pressure gradient \citep{konigl82,canto88,raga89}. As the jet propagates, it can remain collimated due to external pressure support when certain conditions are met in both the ambient medium and the velocity of the jet relative to its internal sound speed \citep{konigl82,begelman84}. The jet-like structures we see in {\sc RomulusC} are therefore the direct consequence of an initially spherical outflow interacting with the denser gas in the cluster core that has a disk morphology. Indeed when this structure is destroyed by the merger at $z\sim0.2$, the black hole's activity declines by almost two orders of magnitude and the bipolar outflows are no longer present. More idealized, high resolution simulations of AGN feedback also have shown that a purely thermal model can drive asymmetric, large-scale outflows in Milky Way-mass galaxies \citep{gabor14}. 


It is important to note that our sub-grid models for stellar and SMBH physics have all been optimized based on reproducing empirical relations at Milky Way and dwarf mass scales. {\it The results presented in this work are therefore purely a prediction of our model and have been in no way constrained to produce a realistic BCG or ICM.} From this perspective, the results presented here from {\sc RomulusC} are fully emergent from our sub-grid prescriptions. That {\sc RomulusC} reproduces many key observed properties of clusters and cluster galaxies is a significant success of our sub-grid model and our optimization process, but it also implies that the physics of star formation and, in particular, SMBHs, does not have to be particularly different in cluster environments, as it is in many other cosmological simulations through the implementation of `two-mode` AGN feedback \citep[e.g.][]{weinberger17,pillepich18}

A logical next step in exploring the physics of AGN feedback would be to make the coupling and/or radiative efficiency variable over the lifetime of the SMBH. This would affect the overall balance between SMBH growth and feedback. For a higher efficiency, less accretion is required for the same energetic effect. Less accretion means that the SMBH grows less over time and that gas near the SMBH does not need to be as cool or dense and so there may also be less star formation. Possibly, the fact that we see a relatively high stellar mass in our BCG compared with observations is an indication that a more efficient feedback is required at high masses. The final mass of the BCG's central SMBH is 10$^{10}$ M$_{\odot}$, which is a factor of a few above the black hole mass stellar mass relation given our BCG's stellar mass. While there is mounting observational evidence in support of overly massive black holes in BCGs and groups \citep[e.g.][]{mcconnell13,mezcua18}, this could be additional evidence that higher feedback efficiency is needed. Such a model could be justified by observations of different feedback mechanisms, such as jets and radio bubbles in the centers of clusters. However, there remains a lot of uncertainties and adopting such a model would require the addition of several free parameters.

As discussed in \S2.1, {\sc RomulusC} does not include metal line cooling, an important coolant for warm/hot ICM gas in the centers of massive halos. This will affect the accretion history of the central galaxy in massive halos. \citet{vandevoort11} show that metal line cooling will impact the accretion history of gas in the central galaxies hosted in the most massive halos at both early and late times. However, AGN feedback also has a significant effect on gas inflow onto galaxies in massive halos \citep{vandevoort11b,vandevoort11}. The relative roles of metal line cooling and AGN feedback in regulating the cooling of the ICM and star formation in the central galaxy is uncertain. Exploring this further will require simulations that self consistently model the ISM and CGM/ICM with full metal line cooling and molecular hydrogen physics, as well as significantly higher resolution than even what we have attained with {\sc RomulusC}.

\subsection{Star Formation in Cluster Member Galaxies}

The lack of star formation we see in cluster member galaxies is consistent with both observations \citep[e.g][]{drinkwater01, vandenbosch08, weinmann11,haines13,haines15,boselli14,balogh16,bluck16, habas18} as well as recent simulation work \citep[e.g.][]{bahe15,oman16,zinger18,shao18}. It is interesting that we do not see any radial dependence for low mass galaxy quenching. This is however not very surprising. Observations of such low mass galaxies are generally limited to only a few clusters in the local Universe. \citet{weinmann11} do find evidence of radial dependence for dwarf galaxy quenching in Virgo, which is somewhat similar in mass to {\sc RomulusC} \citep{urban11}, but not in the more massive Perseus cluster. We do still find some evidence for radial dependence at the higher mass end of our low mass bin ($10^9 < M_{\star} < 10^{9.7} M_{\odot}$, not plotted), but any such dependence seems to go away at the lowest masses, which are both hard to observe and dominate our low mass bin shown in Figure~\ref{quench_distance}. Quenching rates that depend little on radial distance are consistent with theoretical results from \citet{zinger18}, who find that the virial shock extends out to $2R_{200}$ and can cause significant ram presure stripping of smaller galaxies even before they in-fall beyond $R_{200}$, in addition to pre-processing. While \citet{bahe17} also find significant quenching at low masses in the Eagle simulation, they discuss how this might be a resolution effect, as they also see significant quenching of low mass galaxies in lower density environments \citep{eagle15} due to stellar feedback creating disks that are more unstable to stripping. As explored in \citet{eagle15}, the quenching at the low mass end is resolution dependent, but in the highest resolution EAGLE simulation, which is of similar resolution to {\sc Romulus}, they find reasonable quenched fractions down to stellar masses of $10^{8}$ M$_{\odot}$, as we find for isolated field dwarfs in {\sc Romulus25} (see Figure~\ref{quench_mass}).

Analysis of cluster galaxies in the Virgo cluster by \citet{boselli14} indicate that at lower masses ram pressure stripping is likely the dominant source of quenching  in the inner regions of the halo. \citet{zinger18} show that although ram pressure is crucial in stripping away hot halo gas even at large radii, much closer approaches to the cluster core are required to destroy star forming disks. This process often leads to strangulation of star formation by stripping away the supply of gas from galaxies but not directly destroying their disks. However, it has been shown that the inclusion of feedback processes, which in isolation create galactic outflows and fountains, make ram pressure more effective by making the ISM more susceptible to being stripped \citep{murakami99,bahe15}. Thus, it is not surprising that we see an enhanced quenched population of low mass galaxies extending out to large radii.


This picture of ram pressure dominated quenching is also consistent with our result that more massive galaxies tend to quench after falling closer to the cluster center. Ram pressure stripping will be less efficient in these galaxies due to their deeper potential well and massive gaseous disk. Feedback processes that can enhance this stripping are also less effective for the same reason and so higher amounts of ram pressure are needed to destroy the gaseous disk. It is also possible that this difference is due to a combination of ram pressure stripping taking longer in more massive galaxies and shorter orbital decay timescales for more massive sub-halos. We will conduct a more detailed analysis relating ram pressure to both star formation and SMBH activity (see \S4.2.3 and \S5.3).

The majority (71\%) of the galaxies that quench beyond $R_{200}$ in {\sc RomulusC} have been pre-processed as satellites of another in-falling halo. Such pre-processing has been used to explain observations of quenched galaxies at large clustercentric radii \citep{fujita04,haines15}. Still, a significant fraction of these quenched galaxies in {\sc RomulusC} that quench far from the cluster center have never been within $R_{200}$ of another galaxy, including the main halo (i.e. they are not backsplash galaxies). This could imply that cluster galaxies can experience unique interactions with their environment without requiring them to be a satellite prior to cluster infall, as shown in previous simulations \citep{bahe13,zinger18}.

\subsection{AGN Activity in Cluster Member Galaxies}


Like star formation, we find that SMBH activity is significantly decreased in cluster environments, in agreement with low redshift observations of luminous AGN in clusters compared to the field \citep{haines12,ehlert13,ehlert14}. The fact that only the most massive cluster galaxies host more luminous AGN with high accretion rates at lower redshift is also consistent with observations \citep{pimbblet13}. Observations have also indicated that the AGN population in clusters evolves significantly with redshift, eventually meeting or even surpassing the AGN fractions in the field \citep{martini13, lehmer13}. {\sc RomulusC} also shows a significant evolution with redshift, where higher luminosity, higher Eddington ratio AGN become more common at all masses out to $z = 2$. Importantly, while the fraction of luminous AGN in cluster galaxies matches closely with the field values at $z=2$, consistent with observed clusters, it is clear that looking only at the SMBHs with the highest accretion rates does not give the full picture. Rather, we predict a much more substantial population of extremely low Eddington ratio (or even completely dormant) SMBHs in the cluster environment at all stellar masses and redshifts. This population will not be observed in most  AGN surveys, which will generally be limited to sources with higher accretion rates. One potential way to distinguish this affect is to examine the dynamics of galaxies hosting AGN. \citet{haines12} show that luminous AGN in clusters tend to reside in in-falling galaxies, indicating that SMBH activity does indeed decline as the galaxies interact with the ICM.

It is important to note that with our limited volume and only a single cluster we do not completely sample the rarer, much higher luminosity SMBH growth events ($L_{bol}>10^{44}$ ergs/s). However, our results indicate that the picture of how the cluster environment affects SMBH growth may be quite different if one only focuses on high luminosity AGN compared with the much more common low and intermediate luminosity sources. The fact that we find the brightest AGN mostly in higher mass galaxies indicates that the extent to which ram pressure has been able to strip the galaxy's gas supply may be the deciding factor. It may also be possible that ram pressure can drive some SMBH activity in these higher mass galaxies \citep{poggianti17,marshall18}. We will examine the connection between SMBH activity, galaxy gas supply, and ram pressure stripping in more detail in future work.

\section{Summary}

We have presented first results from {\sc RomulusC}, the highest resolution cosmological simulation of a galaxy cluster to date. The simulation is able to resolve cluster member galaxies with unprecedented detail down to dwarf galaxy mass scales. With a novel approach to SMBH physics \citep{tremmel15,tremmel17} and spatial resolution on the order or 100 pc, {\sc RomulusC} is able to resolve the internal processes of galaxies from cluster dwarfs to BCGs with unprecedented detail. While {\sc RomulusC} lacks metal line cooling, as discussed in \S2.2.4 and \S5.1 this choice was made based on previous studies demonstrating the problems with including it in simulations that are unable to resolve multiphase gas. This should be considered a limitation of our model to the same extent as including metal line cooling without the necessary resolution should be considered a limitation in other (generally lower resolution) simulations.  Further, {\sc RomulusC} represents an important test to our sub-grid models, particularly those related to SMBH growth and feedback, as they have only been optimized to produce realistic galaxies for halos 100 times smaller than the main halo of {\sc RomulusC}. We demonstrate that {\sc RomulusC} is consistent with observations in terms of baryonic content, bulk properties and structure of the ICM, stellar mass and star formation history of the BCG, and quenched fractions for higher mass cluster member galaxies.

We show that the central BCG in {\sc RomulusC} has a star formation history consistent with the average sSFR of BCGs observed out to high redshift, finding that local maxima and minima in the star formation history are closely associated with lower and higher levels of SMBH activity respectively. Large-scale, collimated outflows are ubiquitous throughout the simulation and the longest period of sustained SMBH activity (8 Gyr - 10 Gyr) is associated with the final quenching of star formation in the BCG as well as particularly powerful outflows. The final stellar mass of the BCG is within a factor of 2 of the median of the stellar masses observed in similar mass halos. Importantly, the large-scale outflows that are critical to this quenching co-exist with a low entropy core that maintains a short ($<1$ Gyr) cooling time. The effect of the AGN feedback is to limit the ability for halo gas to cool onto the central galaxy, rather than directly heat the gas to high temperatures (and entropy) or blow it away \citep[e.g.][]{mccarthy11,pontzen17}. 

The fact that a simple AGN feedback model with thermally coupled, isotropic energy injection is able to drive powerful outflows that are \textit{naturally} collimated in morphology is a major success and demonstrates that the failure of thermal AGN sub-grid feedback prescriptions have been limited not by inherent physics, but by poor resolution in time, space, and mass. The strength of this simple approach is that the large-scale nature of the outflows are purely a prediction of the model and not something placed by hand in the simulation through explicit kinetic feedback. The model also does not assume any characteristic mass scale at which different feedback modes become dominant. The motions of gas are driven by both local morphology (i.e. the morphology and angular momentum of the gas in the center of the cluster) as well as larger scale cluster `weather' driven by cluster galaxies and overall turbulence in the ICM. Our approach does include a (brief) cooling shutoff for gas that receives feedback in order to avoid any spurious overcooling. While this is physically motivated by the fact that feedback does not occur instantaneously, higher resolution simulations are still needed to self consistently model the coupling of AGN feedback to gas on 100 pc scales. The high BCG stellar mass and black hole mass in {\sc RomulusC} might be indicative that a higher feedback efficiency is required for SMBHs, a natural consequence if, as observations indicate, additional physical processes such as coupling to AGN jets are occurring on small scales preferentially for more massive galaxies/SMBHs.

Beyond the BCG and its central SMBH, we examine the evolving population of cluster member galaxies. We demonstrate the success of both {\sc RomulusC} and our uniform volume simulation, {\sc Romulus25}, in reproducing the observed fraction of quenched galaxies as a function of stellar mass in cluster environments and the field respectively. Taking advantages of the high resolution of {\sc RomulusC} we predict that 80-100\% of low mass (M$_{\star} < 10^{10}$M$_{\odot}$) galaxies are quenched at $z=0$ regardless of stellar mass or distance from the cluster center, a fraction that is more than ten times higher than that of isolated galaxies that almost never quench at low mass. More massive quenched galaxies in {\sc RomulusC} have orbits that have taken them systematically closer (D$_{min}\sim0.4$ R$_{200}$) to the cluster center prior to quenching while less massive galaxies tend to quench farther out (D$_{min}\sim0.6$ R$_{200}$). A significant fraction of galaxies ($\sim25\%$) quench before falling beyond R$_{200}$ and 71\% of these galaxies have previously been satellites of another in-falling galaxy. This means that $\sim7\%$ of cluster galaxies quench beyond R$_{200}$ yet have not been preprocessed as a satellite of another halo. These galaxies may have been quenched by the larger-scale cluster environment, consistent with results from \citet{zinger18}. The enhanced fraction of quenched galaxies is in place even at $z=2$ within $2R_{200}$ at all but the very highest masses.

Finally, we examine the population of SMBHs in {\sc RomulusC} and find that overall SMBH growth is significantly suppressed in the cluster environment at all masses and redshifts. While the pouplation of the highest Eddington ratio SMBHs, particularly those in more massive galaxies, become closer to field SMBHs at higher redshifts, the distribution of Eddington ratios in the cluster environment remains quite different. Thus, we predict, consistent with observations, that the most luminous AGN population evolves significantly with redshift and by $z=2$ even relatively low luminosity AGN ($L_{bol} > 10^{42}$ergs/s) become similar to the field. The population of AGN most affected by the cluster environment is low luminosity AGN, which are generally missed in observational studies of AGN.


\section{Future Work and Simulations}

As mentioned throughout the Paper, significant follow-up analysis is planned to better understand the detailed consequences of ram-pressure stripping and pre-processing on star formation and SMBH accretion in cluster member galaxies. Also, a more detailed analysis of the structure and evolution of AGN winds and how they affect star formation and cooling in the BCG will take place in future work. The nature of SPH allows us to easily track the evolution of gas which directly receives AGN feedback and that which becomes entrained in large-scale winds, allowing us to better understand the morphology of these outflows and why star formation shuts off when it does.

{\sc RomulusC} is the first of a planned suite of zoom-in simulations of massive halos. Currently, we have several more zoom-in simulations planned including group-scale halos (M$_{200} = 3-5\times 10^{13}$M$_{\odot}$) as well as more massive clusters (M$_{200} = 2\times10^{14}-1\times10^{15}$ M$_{\odot}$). These simulations, combined with the small galaxy groups in {\sc RomulusC} (M$_{200} \sim10^{12.5-13}$ M$_{\odot}$) will allow us to further explore galaxy evolution and SMBH feedback in different environments and in the most massive galaxies. The success of our fiducial simulation that we present here is very encouraging and indicates that our sub-grid physics implementations are very much up to the task of modeling galaxy and SMBH evolution correctly within these unique environments.

 \section*{Acknowledgments}
 
The Authors thank the anonymous referee for their careful reading and useful comments and suggestions for this manuscript. MT gratefully acknowledges support from the YCAA Prize Postdoctoral Fellowship. AB, TQ and MT were partially supported by NSF award AST-1514868. AR acknowledges support from NASA Headquarters under the NASA Earth and Space Science Fellowship Program Grant 80NSSC17K0459. PN acknowledges useful discussions in her weekly research group meeting. AP is supported by the Royal Society. MV acknowledges funding from the European Research Council under the European Community's Seventh Framework Programme (FP7/2007-2013 Grant Agreement no.\ 614199, project ``BLACK''). This work is supported in part by NSF grant AST-141276 and the UCL Cosmoparticle Initiative. This research is part of the Blue Waters sustained-petascale computing project, which is supported by the National Science Foundation (awards OCI-0725070 and ACI-1238993) and the state of Illinois. Blue Waters is a joint effort of the University of Illinois at Urbana-Champaign and its National Center for Supercomputing Applications. This work is also part of a PRAC allocation support by the National Science Foundation (award number OAC-1613674). The analysis presented in this Paper was done using the publicly available software packages, pynbody \citep{pynbody} and TANGOS \citep{tangos}.

\bibliography{bibref_mjt.bib}
\bibliographystyle{mn2e}

\label{lastpage}
\end{document}